\documentclass[prb, superscriptaddress,
reprint,
amsmath,
longbibliography]{revtex4-2}

\usepackage{graphicx}
\usepackage{dcolumn}
\usepackage{bm}
\usepackage{enumerate}
\usepackage{refcount}
\usepackage{fmtcount}
\usepackage{amssymb}
\usepackage{hyperref}
\usepackage[usenames,dvipsnames]{color}
\usepackage[normalem]{ulem}
\usepackage[capitalise]{cleveref}
\usepackage{physics}
\usepackage{dsfont}
\usepackage{tikz}
\usetikzlibrary{calc}
\newcommand{\tikznode}[2]{%
\ifmmode%
\tikz[remember picture,baseline=(#1.base),inner sep=0pt] \node (#1) {$#2$};%
\else
\tikz[remember picture,baseline=(#1.base),inner sep=0pt] \node (#1) {#2};%
\fi}
\usepackage{cancel}
\usepackage[justification=justified]{subcaption}
\usepackage[format=plain,justification=raggedright]{caption}

\Crefname{equation}{Eq.}{Eqs.}
\Crefname{figure}{Fig.}{Figs.}
\Crefname{tabular}{Tab.}{Tabs.}
\Crefname{table}{Tab.}{Tabs.}
\Crefname{section}{Sec.}{Secs.}
\hypersetup{pdfstartview=FitH,pdfpagemode=UseOutlines,colorlinks,
citecolor=blue,linkcolor=blue,urlcolor=blue}

\begin{document}

\title{Schwinger boson study of the \texorpdfstring{$J_1$}{J1}-\texorpdfstring{$J_2$}{J2}-\texorpdfstring{$J_3$}{J3} kagome Heisenberg antiferromagnet with Dzyaloshinskii-Moriya interactions}
\author{Dario Rossi}
    \email[]{dario.rossi@unige.ch}
    \affiliation{Department of Theoretical Physics, University of Geneva, Quai Ernest-Ansermet 24, 1211 Geneva, Switzerland}
\author{Johannes Motruk}
    \affiliation{Department of Theoretical Physics, University of Geneva, Quai Ernest-Ansermet 24, 1211 Geneva, Switzerland}
\author{Louk Rademaker}
	\affiliation{Department of Quantum Matter Physics, University of Geneva, Quai Ernest-Ansermet 24, 1211 Geneva, Switzerland}
 \author{Dmitry A. Abanin}
    \affiliation{Department of Theoretical Physics, University of Geneva, Quai Ernest-Ansermet 24, 1211 Geneva, Switzerland}
    \affiliation{Google Research, Mountain View, CA, USA}

\date{\today}
\begin{abstract}
Schwinger boson mean field theory is a powerful approach to study frustrated magnetic systems which allows to distinguish long-range magnetic orders from quantum spin liquid phases, where quantum fluctuations remain strong up to zero temperature. In this work, we use this framework to study the Heisenberg model on the kagome lattice with up to third nearest neighbour interaction and Dzyaloshinskii-Moriya (DM) antisymmetric exchange. This model has been argued to be relevant for the description of transition metal dichalcogenide bilayers in certain parameter regimes, where spin liquids could be realized. By means of the projective symmetry group classification of possible ans\"atze, we study the effect of the DM interaction at first nearest neighbor and then compute the $J_2$-$J_3$ phase diagram at different DM angles. We find a new phase displaying chiral spin liquid characteristics up to spin $S=0.5$, indicating an exceptional stability of the state.
\end{abstract}
\maketitle

\section{Introduction}
The Heisenberg model on the kagome lattice has been the subject of extensive investigations due to its strong geometric frustration which makes it a prime candidate for the realization of a quantum spin liquid state~\cite{Norman2016,Mendels2016,Savary2016,Knolle2019}. This interest is not only theoretical in nature, but various materials are believed to be approximately described by the model~\cite{Norman2016,Mendels2016}. The most intensely studied compound is Herbertsmithite (ZnCu$_3$(OH)$_6$Cl$_2$)~\cite{Shores2005,Helton2007,Bert2007,Mendels2007,Khuntia2020}, but alternative material realizations have been investigated, and found not to exhibit any signs of ordering much below the temperature scales associated to the respective spin coupling~\cite{Aidoudi2011,Faak2012,Clark2013a,Chen2020,Zeng2022,Liu2022,Lu2022}. 

On the theoretical side, the problem has a long history and has been approached with every conceivable method ranging from mean-field theory of partons~\cite{Sachdev1992, Ran2007,Hermele2008,Lu2011,Messio2012}, variational~\cite{Tay2011,Iqbal2011b,Iqbal2013,Clark2013b,Iqbal2014} and renormalization techniques~\cite{Buessen2016,Hering2019,Thoenniss2020} to numerically exact algorithms~\cite{Chalker1992,Leung1993,Lecheminant1997,Waldtmann1998,Jiang2008,Nakano2011,Lauchli2011,Yan2011,Depenbrock2012,Jiang2012,He2017,Changlani2018,Zhu2018,Lauchli2019} and tensor networks~\cite{Xie2014,Mei2017,Liao2017,Jiang2019}. It is widely believed that the nearest-neighbor-only model hosts a spin liquid ground state, but its nature is under ongoing debate.
Early density matrix renormalization group (DMRG) studies pointed to a gapped ($\mathds{Z}_2$) ground state~\cite{Yan2011,Depenbrock2012,Jiang2012}, whereas variational Monte Carlo~\cite{Iqbal2011b,Iqbal2013,Iqbal2014}, more recent DMRG~\cite{He2017,Zhu2018} and two-dimensional tensor network studies~\cite{Liao2017,Jiang2019} favor a gapless $U(1)$ Dirac spin liquid instead.  While the nearest-neighbor-only model has always been in the focus of attention, various perturbations to the system have been considered as well. Both longer range interactions~\cite{Suttner2014,Iqbal2015,Kolley2015,Gong2014,Gong2015,Iqbal2021,Sun2022} and $SU(2)$ breaking Dzyaloshinskii-Moriya (DM)~\cite{Dzyaloshinsky1958,Moriya1960,Cepas2008,Rousochatzakis2009,Messio2010,Messio2017,Mondal2017,Lee2018,Hering2017,Buessen2019} terms have been studied. With varying parameters, it has been shown that other states such as a chiral spin liquid (CSL)~\cite{Kalmeyer1987,Schroeter2007,Thomale2009} and valence bond orders can be ground states or closely competing states~\cite{Iqbal2011a,Messio2012,Gong2014,He2014,He2015,Gong2015,Iqbal2015,Messio2017,Wietek2020,Sun2022}.

Recently, we proposed moir\'e bilayers of transition metal dichalcogenides (TMDs)~\cite{Wu2018,Wu2019,Xu2020,Huang2021} as a platform for realizing spin models on the kagome lattice that feature both long-range and DM interactions~\cite{Motruk2022,Pan2020_1}. Motivated by this proposal,
we study the phase diagram of the antiferromagnetic Heisenberg model on the kagome lattice with up to third-nearest neighbor and additional DM interactions
by Schwinger boson mean field theory (SBMFT). This approach is particularly useful in this context as it allows to distinguish gapped spin liquid phases from competing magnetic orders. We derive the projective symmetry group (PSG) classification of both time reversal (TR) symmetric and TR symmetry breaking chiral $\mathds{Z}_2$ spin liquid ans\"atze on the kagome lattice for the given interactions. When minimizing the parameters of these ans\"atze, we find phase diagrams that are mostly consistent with previous SBMFT~\cite{Messio2012,Messio2017,Mondal2017} and DMRG studies~\cite{Gong2014,Gong2015,Motruk2022}. By varying the spin size, we can strengthen or weaken quantum fluctuations. Upon revisiting the phase diagram of Ref.~\cite{Messio2012}, we find an additional chiral spin liquid phase that exists up to a spin value of 1/2. This is to the best of our knowledge the first spin liquid phase found to be stable at such high spin in SBMFT and may indicate a remarkable stability of the CSL for the $J_1$-$J_2$-$J_3$ kagome Hamiltonian.

The outline of the paper is as follows. In \cref{sec:model}, we describe the model under consideration, review SBMFT and provide details about the numerical procedure used to reach the ground state. We derive the PSG classification of the model with and without DM interactions and introduce the classical long-range magnetic orders we expect to appear in \cref{sec:PSG}. Readers interested in the results can directly skip to \cref{sec:results}, where we report the phase diagrams obtained by SBMFT for different values of interactions and DM angles, commenting on the new phases and spin structure factors. Finally, we discuss our results in \cref{sec:conclusion}.

\section{Model and methods} \label{sec:model}
\begin{figure}
	\includegraphics[width=0.3\textwidth]{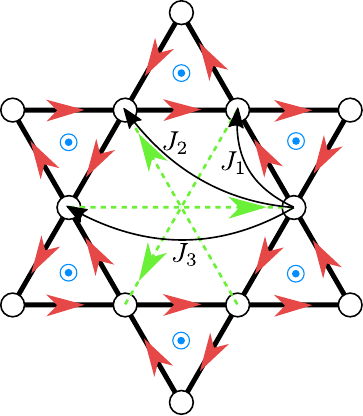}
	\caption{DM directions for 1st (red) and 3rd (green) nearest neighbor. Blue circles indicate the sign of the DM vector in the bonds surrounding the triangles. Black arrows indicate the 1st, 2nd and 3rd nearest neighbours' exchange interaction considered in the model.}\label{fig:DM_directions}
\end{figure}

We consider the $XXZ$ model with Dzyaloshinskii–Moriya (DM) interactions on the kagome lattice, with up to third (across hexagons) nearest-neighbor (n.n.) terms (\cref{fig:DM_directions}). The Hamiltonian can be written as
\begin{equation}\label{eq:H_initial}
\begin{split}
	\mathcal{H}=\sum_\gamma J_\gamma\sum_{ij\in \left\{\gamma-nn\right\}} \biggl( &S_i^zS_j^z + \cos(2\phi_{ij}^\gamma)\left(S^x_iS^x_j + S^y_iS^y_j\right)\\
    &+ \sin(2\phi_{ij}^\gamma)\left[\hat{z}\cdot\left(\vec{S}_i\times\vec{S}_j\right)\right]\biggr),
 \end{split}
\end{equation}
where $\gamma$ denotes first, second and third n.n., and the second summation runs over all bonds at distance $\gamma$. $S_i$ are the usual Pauli spin operators on site $i$, and $\phi_{ij}^\gamma$ is the DM phase of bond $\{i-j\}$. The model reduces to the $SU(2)$ symmetric Heisenberg model if all $\phi_{ij}^\gamma=0$. In general, there would be three different independent phases for first, second and third n.n. bonds, but in this work instead we consider the following regime
\begin{subequations}\label{eq:DM_phases}
	\begin{align}
		\phi^2 &= 0,\\
		\phi^3 &= 2\phi^1.
	\end{align}
\end{subequations}
This choice is inspired by the description of twisted bilayer TMDs where symmetry arguments lead to such a DM phase dependence~\cite{Pan2020_2,Zang2021,Wietek2022,Kiese2022,Rademaker2022,Motruk2022}.

From now on let us denote $\phi^1=\phi$ which we will later tune to study the emerging $J_1$-$J_2$-$J_3$ phase diagram. The Hamiltonian term describing general DM interactions is usually written as 
\begin{equation}\label{eq:DM}
    \mathcal{H}_{DM}\propto \Vec{D}_{ij}\cdot\left(\Vec{S}_i\times\Vec{S_j}\right).
\end{equation}
In our case, the DM vector $\Vec{D}_{ij}$ is pointing uniformly in the $\hat{z}$ direction. Its orientation is shown in \cref{fig:DM_directions}. The red arrows indicate the direction over which to take the cross product of \cref{eq:DM}. We will refer to this kind of DM interaction as ``uniform'' DM interaction since all the DM vectors point in the same direction \footnote{There is a DM vector for each bond, the blue dot in \cref{fig:DM_directions} indicates the direction for all the bonds neighboring the triangle.}, as opposed to the ``staggered'' case which has been considered in many other works in the literature~\cite{Messio2017,Messio2010,Mondal2017,Huh2010}. 
To study the phase diagram, we will use Schwinger boson mean field theory (SBMFT), which is well suited to distinguish between gapped spin liquids (SL) and gapless long-range ordered states (LRO).

The DM interaction changes the symmetry of the original Hamiltonian. The $SU(2)$ symmetry of the pure Heisenberg model is reduced to a $U(1)$ rotation symmetry around $\hat{z}$. Nevertheless, our choice of the DM angles in~\cref{eq:DM_phases} is such that at particular angles $\phi=n\pi/3$, the $SU(2)$ symmetry is restored. It has been shown~\cite{Motruk2022} that such DM phases can be eliminated by a gauge transformation of the spins (a local spin rotation in the $xy$-plane) following a $\sqrt{3}\times\sqrt{3}$ type of pattern. For this reason, we expect to see a periodicity in the phase diagram as a function of $\phi$ with period $2\pi/3$.
\subsection{Schwinger-Boson Mean Field Theory} \label{sec:SBMFT}
The main idea of Schwinger boson mean field theory is to replace the spin operators $\mathbf{S}$ with boson operators $a,b$. This allows a mean field theory treatment of both symmetric and symmetry broken phases. Concretely, the spin operators are substituted by Schwinger bosons as follows,
\begin{equation}\label{eq:SB_def}
	\vec{S}_i=\frac{1}{2}\left(\hat{a}^\dagger_i\hat{b}^\dagger_i\right)\vec{\sigma}
	\begin{pmatrix}
		\hat{a}_i	\\
		\hat{b}_i
	\end{pmatrix},
\end{equation}
with the constraint $\hat{a}^\dagger \hat{a} + \hat{b}^\dagger \hat{b} = 2S$, in order to remain in the physical sector after enlarging the Hilbert space. 

In this subsection, we review the derivation of the mean field Hamiltonian in order to explain our notation and highlight the key steps of the procedure. For a thorough explanation of the method see Refs.~\cite{Auerbach1998,Wen2002}. By substituting the Schwinger bosons into the Hamiltonian we obtain the familiar form 
\begin{equation}        \label{eq:H_schw}
\begin{split}
	\mathcal{H}=&\sum_\gamma J_\gamma \sum_{ij\in \left\{\gamma-nn\right\}} \left(:\hat{B}^{\gamma\dagger}_{ij}\hat{B}^\gamma_{ij}:-\hat{A}^{\gamma\dagger}_{ij}\hat{A}^\gamma_{ij}\right) \\
    &+ \sum_i\lambda_i\left(\hat{a}^\dagger_i\hat{a}_i+\hat{b}^\dagger_i\hat{b}_i - 2S\right).
 \end{split}
\end{equation}
The double dots denote normal ordering, $\lambda_i$ are the Lagrange multipliers and $S$ is the effective value of the spin. 

The pairing $\hat{A}_{ij}^\gamma$ and hopping operators $\hat{B}_{ij}^\gamma$ are defined as
\begin{subequations}\label{eq:bond_ops}
	\begin{align}
		\hat{A}_{ij}^\gamma &= \frac{1}{2}(\tau_{ij}^{\gamma*}\hat{a}_i\hat{b}_j-\tau_{ij}^\gamma\hat{a}_j\hat{b}_i),\\
		\hat{B}_{ij}^\gamma &= \frac{1}{2}(\tau_{ij}^\gamma\hat{a}^\dagger_i\hat{a}_j+\tau_{ij}^{\gamma*}\hat{b}^\dagger_i\hat{b}_j),
	\end{align}
\end{subequations}
where $(\tau_{ij}^\gamma)^2=e^{-i2\phi_{ij}^\gamma}$. In this way, we expressed the quartic Hamiltonian in terms of products of $U(1)$ invariant bond operators that explicitly preserve the rotational invariance of the interactions in our Hamiltonian. In the Heisenberg case without DM interaction, they are $SU(2)$ invariant. The last term in \cref{eq:H_schw} enforces the local constraint of having $2S$ Schwinger bosons per site.

In SBMFT, the spin $S$ is a free parameter not necessarily restricted to $1/2$. A lower value produces more quantum fluctuations, while for $S\rightarrow\infty$, the classical limit is recovered. A value frequently used arises from fixing $\langle \mathbf{S}^2\rangle = 3S(S+1)/2 = 3/4$, which leads to $S=(\sqrt{3}-1)/2 \approx 0.366$ (see \cite{Auerbach1988a,Auerbach1998,Messio2010}). In this work, we vary $S$ to evoke transitions from long-range magnetic orders to quantum spin liquids.

Up to this point we merely rewrote the original Hamiltonian and if the single occupation constraint is strictly respected at each site, the two models are equivalent and will have exactly the same ground state. We note that the operators appearing in the Hamiltonian have changed from 2-spin operators to 4-boson operators. The next step is to implement a mean field approximation of the form
\begin{equation}
	\hat{A}^{\gamma\dagger}_{ij}\hat{A}_{ij} \simeq A^{\gamma*}_{ij}\hat{A}^\gamma_{ij} + \hat{A}^{\gamma\dagger}_{ij}A^\gamma_{ij} - |A^\gamma_{ij}|^2,
\end{equation}
and same for the hopping operators $\hat{B}_{ij}^\gamma$. Then, we perform a Fourier transformation of the spinon operators
\begin{equation}
    \hat{a}_i=\frac{1}{\sqrt{\Omega}}\sum_{k\in\text{BZ}}e^{i\vec{k}\cdot\vec{x}_i} a_{\mu_i,k},
\end{equation}
where we split the site index $i$ into the unit cell (UC) coordinates $\vec{x}_i$ and the index within the unit cell $\mu_i$. We obtain the mean field Hamiltonian
\begin{widetext}
\begin{equation}\label{eq:superlong}
	\begin{split}
		\frac{\mathcal{H}_{MF}}{N_s}=&\sum_\gamma\frac{1}{2}z_\gamma J_\gamma\left(|A^\gamma_{ij}|^2-|B^\gamma_{ij}|^2\right) + \lambda\left(2S+1\right)+\frac{\lambda}{\Omega m}\sum_k\sum_{i\in UC}\left(\hat{a}^\dagger_{\mu_i,k}\hat{a}_{\mu_i,k}+\hat{b}_{\mu_i,-k}\hat{b}^\dagger_{\mu_i,-k}\right)
		+\sum_\gamma\frac{J_\gamma}{2\Omega m}\sum_k\sum_{i\in UC}\\
  \sum_{j\in \gamma-nn(i)}\Biggl\{&\left[B^{\gamma*}_{ij}\left(e^{i\vec{k}\cdot\vec{\delta}}\tau^\gamma_{ij}\hat{a}^\dagger_{\mu_i,k}\hat{a}_{\mu_j,k}+e^{-i\vec{k}\cdot\vec{\delta}}\tau^{\gamma*}_{ij}\hat{b}^\dagger_{\mu_i,-k}\hat{b}_{\mu_j,-k}\right)+B^{\gamma}_{ij}\left(e^{-i\vec{k}\cdot\vec{\delta}}\tau^{\gamma*}_{ij}\hat{a}_{\mu_i,k}\hat{a}^\dagger_{\mu_j,k}+e^{i\vec{k}\cdot\vec{\delta}}\tau^\gamma_{ij}\hat{b}_{\mu_i,-k}\hat{b}^\dagger_{\mu_j,-k}\right)		\right.\\
		&-\left.A^{\gamma*}_{ij}\left(e^{-i\vec{k}\cdot\vec{\delta}}\tau^{\gamma*}_{ij}\hat{a}_{\mu_i,k}\hat{b}_{\mu_j,-k}-e^{i\vec{k}\cdot\vec{\delta}}\tau^\gamma_{ij}\hat{b}_{\mu_i,-k}\hat{a}_{\mu_j,k}\right)-A^{\gamma}_{ij}\left(e^{i\vec{k}\cdot\vec{\delta}}\tau^{\gamma}_{ij}\hat{a}^\dagger_{\mu_i,k}\hat{b}^\dagger_{\mu_j,-k}-e^{-i\vec{k}\cdot\vec{\delta}}\tau^{\gamma*}_{ij}\hat{b}^\dagger_{\mu_i,-k}\hat{a}^\dagger_{\mu_j,k}\right)		\right]\Biggr\},
	\end{split}
\end{equation}
\end{widetext}
with $z_\gamma$ the coordination number, $m$ the number of sites in the unit cell, $\Omega$ the number of points in the Brillouin zone (see \cref{App:gap} for the relation with $N_s$ and $m$) and $\vec{\delta}=\vec{x}_j-\vec{x}_i$ the distance between unit cells. In \cref{eq:superlong} we have summations over $\gamma$ which labels the neighbour distance (from first n.n. to third), over $k$ which spans the Brillouin zone, over the sites of a unit cell $i$ and finally over $j$ which is the $\gamma$-th nearest neighbor of site $i$.

In this step, we considered $\lambda_i=\lambda$ in order to have a single Lagrange multiplier to enforce the occupation constraint. The constraint is thus only imposed on average. To increase accuracy one could consider a different $\lambda$ for each site in the unit cell.

The mean field Hamiltonian can be rewritten in a compact form by introducing the vectors 
\begin{equation}
    \hat{\psi}_k^\dagger=\left(\hat{a}^\dagger_{1,k},\hat{a}^\dagger_{2,k}, \dots \hat{a}^\dagger_{m,k},\hat{b}_{1,-k},\hat{b}_{2,-k},\dots \hat{b}_{m,-k}\right),
\end{equation}
such that~\cref{eq:superlong} becomes
\begin{equation}
\begin{split}
	\frac{\mathcal{H}_{MF}}{N_s}=&\sum_\gamma\frac{1}{2}z_\gamma J_\gamma\left(|A^\gamma_{ij}|^2-|B^\gamma_{ij}|^2\right) + \lambda\left(2S+1\right)+\\
 &\frac{1}{\Omega m}\sum_k\hat{\psi}_k^\dagger\mathcal{N}_k\hat{\psi}_k.
 \end{split}
\end{equation}
Here $\mathcal{N}_k$ is a $(2m,2m)$ matrix. The elements of $\mathcal{N}_k$ are found in the last summation of \eqref{eq:superlong}, which gives this matrix a general structure
\begin{equation}
\mathcal{N}_k = \text{diag}(\lambda) + 
	\begin{pmatrix}
		\tikznode{a}{} & \alpha & 	\tikznode{c}{} & \tikznode{e}{} & \eta \\
		\gamma & \ &  	\ & \theta & \tikznode{f}{} \\
        \hline
		\tikznode{g}{} & \zeta & 	\ & \ & \delta \\
		\epsilon & \tikznode{h}{} &	\tikznode{d}{} & \beta & \tikznode{b}{} 
	\end{pmatrix}
\end{equation}

\begin{tikzpicture}[remember picture, overlay]
  \draw[dashed] (a)+(-0.04,0.19) -- ($(b) + (0.05,-0.02)$);
  \draw ($ (c) + (0,0.25) $) -- ($ (d) + (0,-0.1) $);
  \draw[dashed] ($(e)+(-0.25,0.19)$) -- ($(f) + (0.1,-0.02)$);
  \draw[dashed] ($(g)+(-0.06,0.17)$) -- ($(h) + (0.28,-0.04)$);
\end{tikzpicture}
where the Greek letters from $\alpha$ to $\theta$ refer to the terms in \eqref{eq:superlong}. They are upper/lower triangular matrices which overlap on the diagonals. For example,
\begin{equation}
	\alpha = \sum_\gamma\frac{J_\gamma}{2}\sum_{i\in UC}\sum_{j\in \gamma-nn(i)} B^{\gamma*}_{ij}e^{i\vec{k}\cdot\vec{\delta}}\tau^\gamma_{ij},
\end{equation}
and so forth. Since $\mathcal{N}_k$ is Hermitian, these terms are related as $\gamma=\alpha^\dagger,\; \zeta=\theta^\dagger,\; \epsilon=\eta^\dagger$, and $\beta=\delta^\dagger$.

In order to diagonalize the mean field Hamiltonian, we perform a Bogoliubov transformation. The energy per site then reads
\begin{equation}    \label{eq:min_H}
    \begin{split}
	\mathcal{E}_{MF}= &\sum_\gamma\frac{1}{2}z_\gamma J_\gamma\left(|A^\gamma_{ij}|^2-|B^\gamma_{ij}|^2\right) + \lambda\left(2S+1\right) + \\
    &\frac{1}{\Omega m}\sum_{k,\mu}\epsilon^\mu(k),
 \end{split}
\end{equation}
where $\mu = 1,\dots,m$ and $\epsilon^\mu$ are the positive eigenvalues. This transformation for bosons is reviewed in \cite{Messio2013} (Appendix A) and more generally in \cite{Colpa1978,Sachdev1992}. To be more specific, we need to find a matrix $M_k$ which transforms
\begin{equation} \label{eq:bog}
	\psi_k=M_k\tilde{\psi}_k,
\end{equation}
such that two conditions are satisfied: the final matrix has to be diagonal and the vectors $\tilde{\psi}_k$ in addition have to satisfy the canonical commutation relations $\left[\tilde{\psi}_k^\dagger,\tilde{\psi}_k\right]=J$, where $J$ is a diagonal $(2m,2m)$ matrix with $-1$ on the first $m$ terms and $1$ on the others. These two conditions can be written as
\begin{subequations}	\label{eq:bogo}
	\begin{align}
		&M^\dagger_k\mathcal{N}_kM_k=\omega_k,\\
		&M^\dagger_kJM_k=J.
	\end{align}
\end{subequations}
The second condition makes the Bogoliubov transformation different from the normal diagonalization, where $J=\openone$. In order to perform this transformation we first have to verify that $\mathcal{N}_k$ is positive definite. In fact, while performing the minimization there might be parameter choices that do not yield positive eigenvalues, which then have to be discarded as non-physical. Then, we find an upper-triangular matrix $\mathcal{C}_k$ such that $\mathcal{N}_k=\mathcal{C}^\dagger_k\mathcal{C}_k$ through a Cholesky decomposition. Finally, we diagonalize $\mathcal{G}_k=\mathcal{C}^\dagger_kJ\mathcal{C}_k$. This is a Hermitian matrix whose first $m$ eigenvalues are positive and the others negative.

Since the kagome lattice has a three-site unit cell, it is not well suited for an analytic treatment \footnote{This is due to the fact that there is no analytic expression for the eigenvalues of a $6\times6$ matrix.}. Therefore, we rely on numerical simulations. The ground state of the model can be found by extremizing the MF energy with respect to all the mean field parameters and the Lagrange multiplier. This procedure involves a large number of mean field parameters, so, in order to be able to find a solution we rely on the \emph{projective symmetry group} classification of possible ans\"atze. This assumes that some symmetries, such as translational invariance, will be respected by the solution, thus reducing the number of free parameters. The classification for our model will be laid out in the next section. 

There are two equivalent ways of finding the ground state. The first one consists of looking for the the saddle point where the derivative of the free energy with respect to all the mean field parameters is zero. We call this method \emph{gradient descent}. An important aspect of this gradient descent minimization is that we are not minimizing the energy with respect to each of the parameters. This is due to the fact that the solution lies actually at a saddle point: it is maximal with respect to the Lagrange multiplier. In addition to this, there is a fundamental difference between pairing $|A^\gamma|$ and hopping $|B^\gamma|$ mean field parameters: the solution will be at the minimum with respect to $|A^\gamma|$ and maximum of $|B^\gamma|$ for positive couplings $J_\gamma$. This can be seen by computing the second derivative of $\mathcal{E}_{MF}$: 
\begin{subequations}
	\begin{align}
		\frac{\partial^2\mathcal{E}_{MF}}{\partial^2|A^\gamma|}\propto J_\gamma,	\\
		\frac{\partial^2\mathcal{E}_{MF}}{\partial^2|B^\gamma|}\propto -J_\gamma.	
	\end{align}
\end{subequations}
Hence, while performing the minimization, we also need to consider the sign of the Hessian for the various parameters~\footnote{Since the mean field parameters are in general complex valued, we also need to determine their phase. It turns out that in most cases they respect the same Hessian sign as their corresponding amplitude.}.

The gradient descent method requires computation of derivatives of the free energy with respect to the mean field parameters, which when done with a finite difference method, can introduce numerical inaccuracies. Furthermore, the method becomes computationally expensive under a growing number of minimization parameters. Another way of reaching the ground state is that of iteratively solving the \emph{self-consistency} relations. These are of the form
\begin{equation} \label{eq:self-consistency}
    A^\gamma_{ij}=\langle\hat{A}^\gamma_{ij}\rangle.
\end{equation}
In order to do so, we need to write down the hopping and pairing operators in terms of Bogoliubov bosons using \cref{eq:bog} and exploit the definition of ground state as the vacuum of such excitations. We decompose the transformation matrix $M_k$ as
\begin{equation}
    M_k=\begin{pmatrix}
	U_k&X_k	\\
	V_k&Y_k
\end{pmatrix},
\end{equation}
where each component is a $m\times m$ matrix, with $m$ the size of the (ansatz) unit cell. The final form of pairing and hopping operators then is
\begin{subequations}\label{eq:sc}
    \begin{align}
        \langle \hat{A}^\gamma_{ij}\rangle=&\frac{1}{2\Omega}\sum_k\Big(\tau_{ij}^{\gamma*} e^{i\vec{k}\cdot(\vec{x}_i-\vec{x}_j)}U_{\mu\nu}(k)V_{\lambda\nu}^*(k) \nonumber\\&-\tau_{ij}^\gamma e^{-i\vec{k}\cdot(\vec{x}_i-\vec{x}_j)}Y_{\mu\nu}^*(k)X_{\lambda\nu}(k)\Big),   \\
        \langle \hat{B}^\gamma_{ij}\rangle=&\frac{1}{2\Omega}\sum_k\Big(\tau_{ij}^{\gamma} e^{-i\vec{k}\cdot(\vec{x}_i-\vec{x}_j)}X_{\mu\nu}^*(k)X_{\lambda\nu}(k)\nonumber\\&+\tau_{ij}^{\gamma*} e^{i\vec{k}\cdot(\vec{x}_i-\vec{x}_j)}V_{\mu\nu}(k)V_{\lambda\nu}^*(k)\Big),
    \end{align}
\end{subequations}
where $\mu,\lambda$ are the unit cell indexes of respectively sites $i,j$ and summation over repeated indexes is implied.
The procedure of this method of solution is the following: starting from a set of mean field parameters $\left\{O\right\}$, we maximize the free energy with respect to the Lagrange multiplier $\lambda$ in order to fulfill the occupation constraint. Then, using $\lambda$ and $\left\{O\right\}$ we compute a new set of mean field parameters using \cref{eq:sc}. We iterate this procedure until convergence to a stationary point. This procedure is more efficient than the gradient descent and allows to consider a larger set of mean field parameters, thus making it best suited for considering a large number of different ans\"atze. In both approaches, one has to pay attention to the fact that they are quite dependent on the initial set of parameters. Hence, we need to repeat the procedure many times with different initial parameters in order to be sure to find all the saddle points of the free energy. 

\section{Symmetry classification}       \label{sec:PSG}
\subsection{Derivation of the algebraic projective symmetry group}
\begin{figure}
	\centering
	\includegraphics[width=0.3\textwidth]{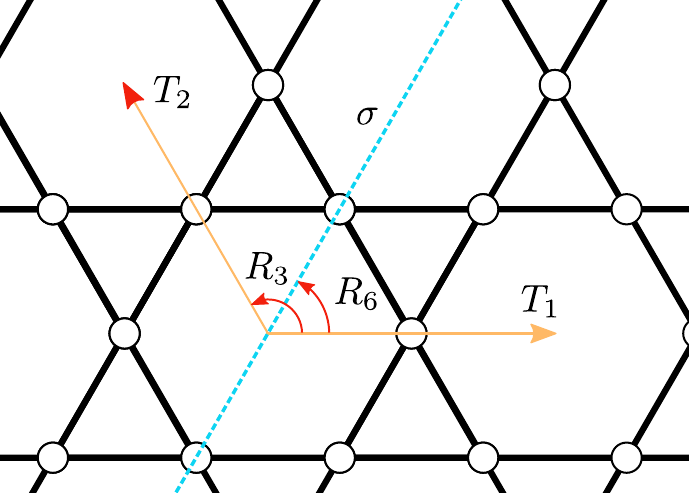}
	\caption{Symmetries of the kagome lattice. The lattice directions considered in the text coincide with $T_1$ and $T_2$.}\label{fig:symmetries}
\end{figure}

In this section, we review the derivation of the projective symmetry group (PSG) classification presented in~\cite{Messio2013} and adapt it to our specific model. The mean field parameters' manifold grows exponentially in system size. In addition, the Lagrange multipliers have to be optimized for each free energy evaluation, making the numerical convergence of the problem a demanding task. This was done for relatively small system sizes in \cite{Misguich2012} and it was shown that in almost all cases the mean field solution was highly symmetric. The idea then is to restrict our search to solutions of the self-consistency equations that respect some of the symmetries of the model. A set of mean field parameters $\left\{A_{ij},B_{ij},\lambda\right\}$ is called an ansatz. We demand that our ans\"atze respect some symmetries of the original Hamiltonian. The symmetries of the Heisenberg interaction on the kagome lattice are: global spin rotation, time reversal (TR) symmetry and lattice symmetries (translations $T_1$ and $T_2$, rotations $R_6$ and reflections $\sigma$ for the kagome lattice, shown in \cref{fig:symmetries}). In our case the ansatz automatically respects the spin rotation symmetry since we are considering bond operators of the form \eqref{eq:bond_ops}. Furthermore, we want to consider ans\"atze which respect the lattice symmetries and, eventually, also time-reversal symmetry. In addition, our Hamiltonian contains DM interactions which break some of these symmetries. The resulting PSG will thus be different from the one of the pure Heisenberg model.

First of all, let us note that the introduction of bosonic operators imbues the theory with a $\mathcal{G} = U(1)$ gauge freedom. A gauge transformation (GT) acts on the Schwinger bosons as
\begin{equation}
	\hat{b}_{j,\sigma} =\exp(i\theta(j))\hat{b}_{j,\sigma}.
\end{equation}
The effect on the bond operators \eqref{eq:bond_ops} is thus
\begin{subequations}\label{eq:gaugeAB}
	\begin{align}
		\hat{A}_{ij}&\rightarrow e^{i(\theta(i)+\theta(j))}\hat{A}_{ij},\\
		\hat{B}_{ij}&\rightarrow e^{-i(\theta(i)-\theta(j))}\hat{B}_{ij},
	\end{align}
\end{subequations}
and since it is just a GT, the Hamiltonian described by $\left\{A_{ij},B_{ij}\right\}$ will remain unaffected by the action of the gauge transformation. If two mean field Hamiltonians $\mathcal{H}_{MF}$ have the same physical properties, then their ans\"atze are related by a GT. Hence, a GT modifies the ansatz, but not the physical quantities.

Let us now consider the lattice symmetries. We call $\chi$ the lattice symmetry group, then the spinon (bosonic) operators will transform under the action of $X\in\chi$ as
\begin{equation}
	\hat{a}_{i}\rightarrow\hat{a}_{X(i)},
\end{equation}
in other words, the site $i$ is transformed to site $X(i)$. The same happens for $\hat{b}_i$ and by extension for the bond operators \eqref{eq:bond_ops}. If an ansatz respects a symmetry, then the physical quantities are the same before and after the application of that symmetry transformation. But we also know that if two systems have the same physical quantities, then they are related by a GT. This means that there exists at least one GT $\hat{G}_X\in\mathcal{G}$ such that $\hat{G}_XX$ leaves the ansatz invariant. The set of transformations $\mathcal{G}\times\chi$ which do not change the ansatz is called the projective symmetry group (PSG) \emph{of that particular ansatz}. This group only depends on the lattice symmetries $\chi$ and on the ansatz, not on the Hamiltonian. An important subgroup of the PSG is the so-called invariant gauge group (IGG) which is the group of gauge transformations related to the identity $X=I$. The IGG is the group of gauge transformations that leaves the ansatz invariant. From the PSG of an ansatz we therefore know which symmetries it preserves and how these symmetries are realized projectively in the gauge group.

For our purpose of restricting the space of relevant ans\"atze, we would like to choose a symmetry group $\chi$ and find all the possible ans\"atze compatible with it. In order to do so we have to find the so-called \emph{algebraic} PSG (A-PSG), which are constructed by constraining the possible gauge group representations using the symmetry relations of $\chi$. The idea is that any symmetry group will have to respect some consistency relations. These can be written in the form of products of symmetry operations on the lattice yielding the identity. If a lattice symmetry can be written in several ways, also the associated gauge transformations should be compatible, with the identity in that case being the IGG. For example, if we have a relation $X_1X_2=X_2X_1$ for $X_i\in\chi$, then we can re-write this as $X_1X_2X_1^{-1}X_2^{-1}=1$. In terms of the associated GT, this becomes $G_1X_1G_2X_2X_1^{-1}G_1^{-1}X_2^{-1}G_2^{-1}\in \text{IGG}$, and leads to a constraint on the possible realizations of $G_\chi$. 

In the non-TR symmetry breaking classification, we consider $\chi$ as the group of all lattice symmetries, and at the end impose TR symmetry by considering only ans\"atze which are real up to a gauge transformation. This was done before in \cite{Wang2006}. Chiral solutions (that break TRS) have been shown to be competitive ground states, as shown for example in \cite{Messio2012}. To also include chiral ans\"atze, we will follow the procedure outlined in \cite{Messio2013}. The idea is to consider ans\"atze that respect all the lattice symmetries only up to a TR transformation. In order to do so, we need to distinguish between \emph{odd} and \emph{even} lattice symmetries: they are characterized by having odd or even parity under TR, respectively. Even symmetries include for example all squares of the elements of $\chi$. Once we know $\chi_e$ we can construct the chiral A-PSG of $\chi$ as the A-PSG of $\chi_e$. Then, we consider the odd symmetries $\chi_o$, which leads to two types of constraints: first, same type (pairing or hopping) mean field parameters on bonds related by such symmetries will have the same modulus. Second, fluxes are physical quantities and thus gauge independent, and are sent to their opposite by TR. Thus, they are unchanged by even transformations. The constraints arise from considering all non-trivial fluxes on the lattice and consider all possible cases of parities for the transformations of $\chi_o$.

We start by deriving the A-PSG of a triangular Bravais lattice, which has $\chi_e=\left\{T_1,T_2,R_3\right\}$. These are all even symmetries for the kagome lattice. Since $R_6$ is also a symmetry in the kagome lattice, $R_3=R_6^2$ must have even parity. By considering the lattice relation $T_2R_6=R_6T_1T_2$ we can see that $T_1\in\chi_e$ and in the same way the relation $T_1R_6=R_6T_2^{-1}$ yields $T_2\in\chi_e$. Thus, the odd symmetry group will be $\chi_o=\left\{R_6,\sigma\right\}$. The symmetries are reported in \cref{fig:symmetries}.
This A-PSG was computed in \cite{Messio2013} and it reads 
\begin{subequations}
	\begin{align}
		\theta_{T_1}(r_1,r_2)&=0,		\\
		\theta_{T_2}(r_1,r_2)&=p_1\pi[r_1],		\\
		\theta_{R_3}(r_1,r_2)&=p_1\pi[r_1]\left([r_2]-\frac{[r_1]+1}{2}+[r_2^*-r_1^*]\right), \nonumber\\
            &\quad+g_{R_3}(r_1^*,r_2^*),	\label{seq:orrible}
	\end{align}
\end{subequations}
with $p_1\in\{ 0,1\}$ labelling the two different ans\"atze and $r_1,r_2$ are coordinates in the two lattice directions, with an integer $[r_i]$ and fractional $r_i^*$ part. One solution is that of taking $g_{R_3}=0$. 

\begin{figure}
	\includegraphics[width=0.45\textwidth]{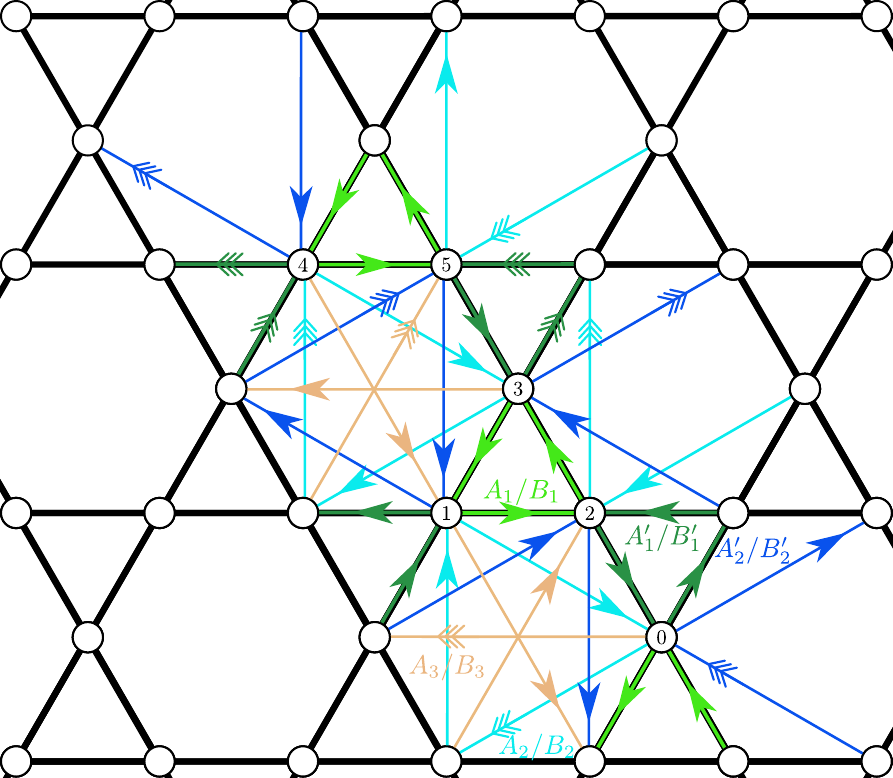}
	\caption{Bond relations of hopping and pairing operators on the kagome lattice. These relations come from the A-PSG derived using only translations and $2\pi/3$ rotations. For first and second neighbour distances there are two inequivalent bonds, colored in light and dark green for first n.n. and in light and dark blue for second n.n.. A double arrow on the bond indicate that the phase of that $A_{ij}/B_{ij}$ has an additional phase $p_1\pi$. Numbers on the sites indicate the used convention for ordering the $6$-sites unit cell.}\label{fig:dir_PSG}
\end{figure}

With the A-PSG at hand we first of all note that it only depends on one parameter $p_1$, meaning that there are only two equivalence classes. In particular, since $p_1$ enters in the definition of $\theta_{T_2}$, we see that the ansatz will have a $3$-site unit cell for $p_1=0$, while it is doubled for $p_1=1$. For each type of bond, we need to consider a starting point to cover all the lattice using the symmetries of $\chi_e$. For first and second nearest neighbours, there are two inequivalent sets of bonds not related by any symmetry. Starting from a bond, we can obtain the values of the others by the procedure
\begin{subequations}
	\begin{align}
		A_{i\rightarrow j}=e^{i(\theta_\chi(i)+\theta_\chi(j))}A_{\chi(i)\rightarrow\chi(j)},	\\
		B_{i\rightarrow j}=e^{i(\theta_\chi(i)-\theta_\chi(j))}B_{\chi(i)\rightarrow\chi(j)}.
	\end{align}
\end{subequations}
This will yield the bond relations in picture \cref{fig:dir_PSG} for first, second and third nearest-neighbors.

Let us now make a counting of the mean field parameters we are left with so far. There is a total of $20$ mean field parameters: $10$ moduli $A_1$, $A_1'$, $A_2$, $A_2'$, $A_3$, $B_1$, $B_1'$, $B_2$, $B_2'$, $B_3$ and the corresponding $10$ phases. This is still a large parameter space in which to look for solutions. Luckily, we can restrict this number of free parameters by exploiting the remaining symmetries. First of all, we can fix $\phi_{A_1}=0$. The relevant additional symmetries not yet considered will be $R_6$ ($\pi/3$-rotations) and $\sigma$ (reflection) for the Heisenberg model, only $R_6$ in the case of the staggered DM interactions considered in \cite{Messio2017,Halimeh2016,Messio2010,Mondal2017,Huh2010} and finally only $\sigma$ for the uniform DM interaction used in our model \footnote{Note the ambiguity in the definition of staggered and uniform DM interactions \cite{GomezAlbarracin2018}. In this work we stick to the notation of \cite{Messio2010}.}.  By looking at the interaction directions in \cref{fig:DM_directions}, we can see that reflection symmetry is respected while $\pi/3$ rotations are not. This can be illustrated by applying a rotation: the link changes direction, but the DM vector stays the same, therefore the energy changes in the symmetry transformed bond. When applying a reflection instead, the link as well as the DM vector stay the same. The completely opposite scenario happens if we consider instead a staggered DM vector on all triangles. Then, it is clear that $\sigma$ is no longer a symmetry, while $R_6$ is restored.

\begin{figure}
	\centering
	\includegraphics[width=0.45\textwidth]{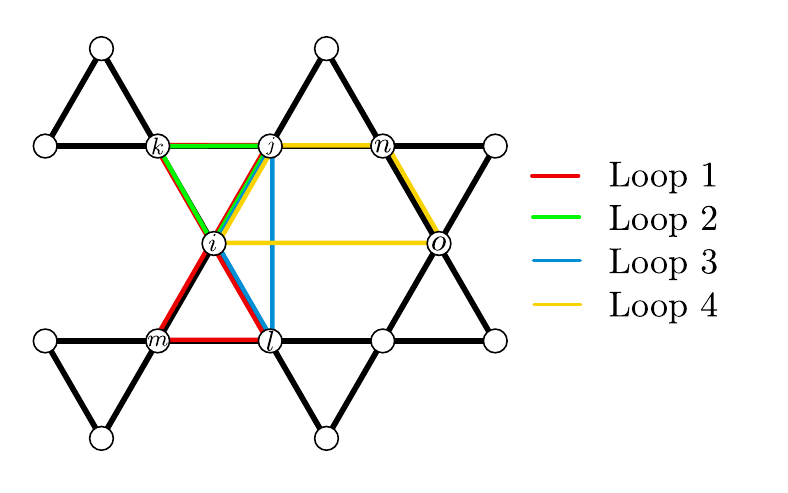}
	\caption{Loops that need to be considered in order to constrain hopping and pairing mean field phases. The list of loops in terms of operators is given in the text.}\label{fig:loops_PSG}
\end{figure}

\begin{table*}
	\centering
	\begin{tabular}{|c|c|c|ccc|cccc|cc|c|}
		\hline
		$N^\circ$ 	& $(\epsilon_R,\epsilon_\sigma)$ & $p_1$	& $\phi_{A_1}'$ & $\phi_{B_1}$ & $\phi_{B_1}'$ & $\phi_{A_2}$ & $\phi_{A_2}'$ & $\phi_{B_2}$ & $\phi_{B_2}'$ & $\phi_{A_3}$ & $\phi_{B_3}$ & compatible order(s)\\
		\hline\hline
		1 	& $(1,1)$ & 0	& $0$	& $p_2\pi$& $p_2\pi$	&free&$\phi_{A_2}$&$p_3\pi$&$p_3\pi$&-&$p_4\pi$&$\mathbf{Q}=0$\\
		
		2 	& $(1,1)$ & 0	& $\pi$	& $p_2\pi$& $p_2\pi$	&-&-&$p_3\pi$&$p_3\pi$&free&$p_4\pi$&$\sqrt{3}\times\sqrt{3}$\\
		
		3 	& $(1,1)$ & 1	& $0$	& $p_2\pi$& $p_2\pi$	&free&$\phi_{A_2}$&$p_3\pi$&$p_3\pi$&free&-&?\\
		
		4 	& $(1,1)$ & 1	& $\pi$	& $p_2\pi$& $p_2\pi$	&-&-&$p_3\pi$&$p_3\pi$&-&-&?\\
		\hline
		5 	& $(1,-1)$ & 0	&$0$	& free	& $\phi_{B_1}$	&$p_2\pi$&$p_2\pi$	&free	&$\phi_{B_2}$&-&$p_3\pi$&$\mathbf{Q}=0$\\
		
		6 	& $(1,-1)$ & 0	&$\pi$	& free	& $\phi_{B_1}$	&$p_2\pi+\frac{\pi}{2}$&$p_2\pi-\frac{\pi}{2}$&free&$\phi_{B_2}$&$p_3\pi$&$p_4\pi$&$\sqrt{3}\times\sqrt{3}$\\
		
		7 	& $(1,-1)$ & 1	&$0$	& free	& $\phi_{B_1}$	&$p_2\pi$&$p_2\pi$&free&$\phi_{B_2}$&$p_3\pi$&$\frac{\pi}{2}+p_4\pi$&?\\
		
		8 	& $(1,-1)$ & 1	&$\pi$	& free	& $\phi_{B_1}$	&$p_2\pi+\frac{\pi}{2}$&$p_2\pi-\frac{\pi}{2}$&free&$\phi_{B_2}$&-&$\frac{\pi}{2}+p_3\pi$&\textit{octahedral}\\
		\hline
		9 	& $(-1,1)$ & 0	&$0$	& free	& $-\phi_{B_1}$	&free	&-$\phi_{A_2}$	&$p_2\pi$	&$p_2\pi$&-&free&$\mathbf{Q}=0$\\
		
		10 	& $(-1,1)$ & 0	&$\pi$	& free	& $-\phi_{B_1}$	&-&-				&$p_2\pi$&$p_2\pi$&$p_3\pi$&free&$\sqrt{3}\times\sqrt{3}$\\
		
		11 	& $(-1,1)$ & 1	&$0$	& free	& $-\phi_{B_1}$	&free&-$\phi_{A_2}$&$p_2\pi$&$p_2\pi$&$p_3\pi$&-&\textit{cuboc-2}\\
		
		12 	& $(-1,1)$ & 1	&$\pi$	& free	& $-\phi_{B_1}$	&-&-&$p_2\pi$&$p_2\pi$&-&-&?\\
		\hline
		13	&	$(-1,-1)$	&0	&	$\phi$	&	$p_2\pi$	&	$p_2\pi$	&	$\frac{\phi}{2}+p_3\pi$	&	$\frac{\phi}{2}+p_3\pi$	&	free	&$-\phi_{B_2}$	&$\frac{\phi+\pi+2p_4\pi}{2}$	&	$p_5\pi$&$\mathbf{Q}=0,\sqrt{3}\times\sqrt{3}$\\
		14	&	$(-1,-1)$	&1	&	$\phi$	&	$p_2\pi$	&	$p_2\pi$	&	$\frac{\phi}{2}+p_3\pi$	&	$\frac{\phi}{2}+p_3\pi$	&	free	&$-\phi_{B_2}$	&$\frac{\phi+2p_4\pi}{2}$	&	-&\textit{cuboc-1}\\
        \hline\hline
		15 	& $(\cdot,1)$ & 0	& $0$	& free	& $-\phi_{B_1}$	&free&free&	$p_2\pi$ &$p_3\pi$&-&free&$\mathbf{Q}=0$\\
		
		16 	& $(\cdot,1)$ & 0	& $\pi$	& free	& $-\phi_{B_1}$	&-&-&	$p_2\pi$ &$p_3\pi$&free&free&$\sqrt{3}\times\sqrt{3}$\\
		
		17 	& $(\cdot,1)$ & 1	& $0$	& free	& $-\phi_{B_1}$	&free&free&	$p_2\pi$ &$p_3\pi$&free&-&\textit{cuboc-2}\\
		
		18 	& $(\cdot,1)$ & 1	& $\pi$	& free	& $-\phi_{B_1}$	&-&-&	$p_2\pi$ &$p_3\pi$&-&-&?\\
		\hline		
		19 	& $(\cdot,-1)$ & 0	& $\phi$	& free	& $\phi_{B_1}$	&$\frac{\phi}{2}+p_2\pi$&$\frac{\phi}{2}+p_3\pi$&	free &free&$\frac{\phi+\pi+2p_4\pi}{2}$&$p_5\pi$&$\mathbf{Q}=0,\sqrt{3}\times\sqrt{3}$\\
		
		20 	& $(\cdot,-1)$ & 1	& $\phi$	& free	& $\phi_{B_1}$	&$\frac{\phi}{2}+p_2\pi$&$\frac{\phi}{2}+p_3\pi$&	free &free&$\frac{\phi+2p_4\pi}{2}$&$p_5\pi-\frac{\pi}{2}$&\textit{cuboc-1},\textit{octahedral}\\
		\hline
	\end{tabular}
	\caption{All possible ans\"atze for the Heisenberg model on the Kagome lattice without ($1$ to $14$) and with uniform ($15$ to $20$) DM interaction. $p_i=\pm1$ and an empty spot means that the corresponding bond has to vanish because of symmetry constraints. All amplitudes at the same distance (1st, 2nd, 3rd n.n.) are equal (separately for $A$ and $B$ parameters) without DM. For uniform DM interaction this is not true for the 2nd nn parameters, i.e. $|A_2|\neq|A_2'|$ and same for $B_2$. The last column reports the regular orders compatible with each ansatz. The dot in ans\"atze $15$ to $20$ refers to the fact that $\epsilon_R$ is not needed in these ans\"atze since $R_6$ symmetry is not considered.
    }	\label{tab:allHeisTMD}	
\end{table*}
Let us now take into consideration both $R_6$ and $\sigma$, in order to classify all the ans\"atze of the pure Heisenberg model. The loops that need to be considered in the lattice are reported in \cref{fig:loops_PSG}. 
The loop operators are
\begin{subequations}	\label{seq:loops}
	\begin{align}
		\text{Loop 1:}&\qquad	\hat{A}_{ij}^\dagger\hat{A}_{jk}\hat{A}_{ki}^\dagger\hat{A}_{il}\hat{A}_{lm}^\dagger\hat{A}_{mi},	\\
		\text{Loop 2:}&\qquad	\hat{A}_{ij}^\dagger\hat{A}_{jk}\hat{B}_{ki},	\\
		\text{Loop 3-a:}&\qquad	\hat{A}_{lj}^\dagger\hat{A}_{ji}\hat{B}_{il},	\\
		\text{Loop 3-b:}&\qquad	\hat{A}_{ji}^\dagger\hat{A}_{il}\hat{B}_{lj},	\\
		\text{Loop 4-a:}&\qquad	\hat{A}_{io}^\dagger\hat{A}_{on}\hat{B}_{nj}\hat{B}_{ji},	\\
		\text{Loop 4-b:}&\qquad	\hat{A}_{on}^\dagger\hat{A}_{nj}\hat{B}_{ji}\hat{B}_{io},	
		\end{align}
\end{subequations}
where each bond has to be taken with the correct sign and phase, as defined in \cref{fig:dir_PSG}. In order to implement a symmetry $\chi$, it needs to equate the phase of a loop to $\epsilon_\chi$ times the phase of the $\chi$-transformed loop.

If we consider both $R_6$ and $\sigma$ we obtain the following constraints
\begin{subequations}\label{eq:constraints}
	\begin{align}
		\phi_{A_1}'&=-\epsilon_{R_6}\phi_{A_1}',		\label{seq:cA1_R}				\\
		\phi_{A_1}'&=-\epsilon_\sigma\phi_{A_1}',	\label{seq:cA1_s}				\\
		\phi_{B_1} &=\epsilon_{R_6}\phi_{B_1}',		\label{seq:cB1_R}				\\
		\phi_{B_1} &=-\epsilon_\sigma\phi_{B_1}',	\label{seq:cB1_s}				\\
		\phi_{A_2}'&=\epsilon_{R_6}(\phi_{A_2}-\phi_{A_1}'),		\label{seq:cA2_R}			\\
		\phi_{A_2}'&=\epsilon_\sigma(\phi_{A_2}'-\phi_{A_1}'),		\label{seq:cA2_s}			\\
		\phi_{B_2}'&=\epsilon_{R_6}\phi_{B_2},			\label{seq:cB2_R}				\\
		\phi_{B_2}'&=-\epsilon_\sigma\phi_{B_2}',		\label{seq:cB2_s}			\\
		\phi_{A_3}+p_1\pi&=\epsilon_{R_6}\phi_{A_3}+\pi+\phi_{A_1}',		\label{seq:cA3_R}			\\
		\phi_{A_3}+p_1\pi&=\epsilon_\sigma\phi_{A_3}+\pi+\phi_{A_1}',		\label{seq:cA3_s}			\\
		\phi_{B_3}+p_1\pi&=-\epsilon_{R_6} \phi_{B_3},			\label{seq:cB3_R}		\\
		\phi_{B_3}+p_1\pi&=\epsilon_\sigma\phi_{B_3}.			\label{seq:cB3_s}
	\end{align}
\end{subequations}
There is one main difference between Heisenberg and staggered DM, which respect $R_6$, and uniform DM, which does not. 

\begin{table*}
	\centering
	~\clap{
	\begin{tabular}{|l|c|c|cccccc|ccccccccc|}
		\hline
		$Order$ 	& $(\epsilon_R,\epsilon_\sigma)$ & $p_1$ & $A_1$ & $B_1$ & $A_2$ & $B_2$ & $A_3$ & $B_3$ &$\phi_{A_1}'$ & $\phi_{B_1}$ & $\phi_{B_1}'$ & $\phi_{A_2}$ & $\phi_{A_2}'$ & $\phi_{B_2}$ & $\phi_{B_2}'$ & $\phi_{A_3}$ & $\phi_{B_3}$\\
		\hline\hline
        $\mathbf{Q}=0$      &   $(\pm1,\pm1)$     &   $0$     &   $\sqrt{3}/4$  &   $1/4$   &   $\sqrt{3}/4$    &   $1/4$   &   $0$ &   $1/2$   &   $0$ &   $\pi$&   $\pi$&   $\pi$&   $\pi$&   $\pi$&   $\pi$&   -     &   $0$     \\
        \hline
        $\sqrt{3}\times\sqrt{3}$      &   $(\pm1,\pm1)$     &   $0$     &   $\sqrt{3}/4$  &   $1/4$   &   $0$    &   $1/2$   &   $\sqrt{3}/4$ &   $1/4$   &   $\pi$ &   $\pi$&   $\pi$&   - &   - &   $0$&   $0$&   $0$     &   $\pi$     \\
                \hline
        \textit{cuboc-1}      &   $(-1,-1)$     &   $1$     &   $\sqrt{3}/4$  &   $1/4$   &   $1/4$    &   $\sqrt{3}/4$   &   $1/2$ &   $0$   &   $2\theta_0$ &   $\pi$&   $\pi$&   $\pi+\theta_0$&   $\pi+\theta_0$&   $-\theta_0$&   $\theta_0$&   $\theta_0$     &   -     \\
                \hline
        \textit{cuboc-2}      &   $(-1,1)$     &   $1$     &   $1/4$  &   $\sqrt{3}/4$   &   $\sqrt{3}/4$    &   $1/4$   &   $1/2$ &   $0$   &   $0$ &   $\pi+\theta_0$&   $\pi-\theta_0$&   $\pi-\theta_0$&   $\pi+\theta_0$&   $0$&   $0$&   $0$     &   -     \\
                \hline
        \textit{octahedral}      &   $(1,-1)$     &   $1$     &   $1/(2\sqrt{2})$  &   $1/(2\sqrt{2})$   &   $1/(2\sqrt{2})$    &   $1/(2\sqrt{2})$   &   $0$ &   $1/2$   &  $\pi$&   $5\pi/4$&   $5\pi/4$&   $3\pi/2$&   $\pi/2$&   $\pi/4$&   $\pi/4$     &   -   &   $3\pi/2$     \\
        \hline
	\end{tabular}
}
	\caption{Pairing and hopping amplitudes and phases for classical $O(3)$ regular orders on the Kagome lattice. The value of $\theta_0$ is $\arctan(\sqrt{2})\approx 0.95$.}	\label{tab:classical_orders}	
\end{table*}

In fact, when $R_6$ is a symmetry, all bonds of the same distance have the same amplitude. In the uniform DM case instead, they can in principle assume different values. But since $\sigma$ is a symmetry, the difference arises only at the second-nearest neighbor level. We report the full set of ans\"atze for the Heisenberg model and the uniform DM interaction in \cref{tab:allHeisTMD}. Parameters $p_i$ can take values $0,1$ and differentiate the ans\"atze. In the table, we numbered the ans\"atze for different choices of $\epsilon_\chi$ and $p_1$. 
\begin{figure*}
	\includegraphics[width=\textwidth]{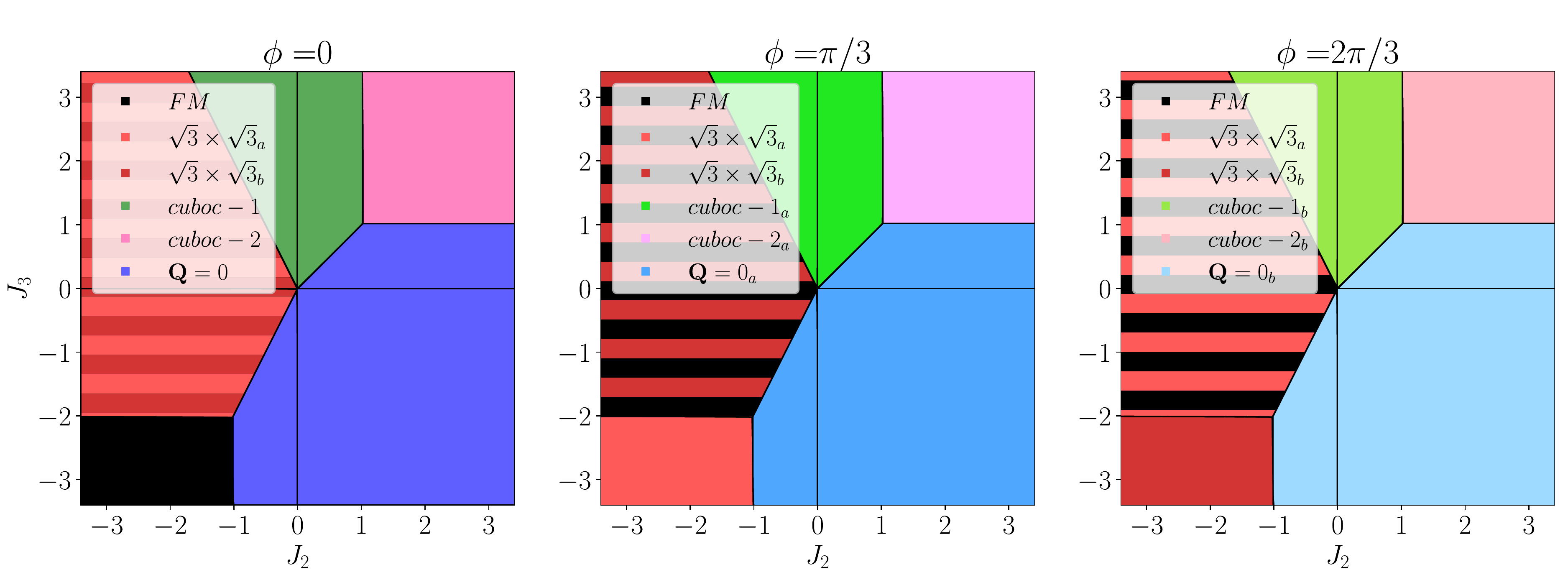}
	\caption{Classical phase diagram of the Heisenberg model on the kagome lattice with up to third n.n. interactions and DM interactions as defined in \cref{sec:model}. Choosing the value of the DM (first n.n.) angle to $n\pi/3,  n=0,1,2$, yields the same phase diagram with gauged transformed orders. $\sqrt{3}\times\sqrt{3}$ comes in two vector chiralities denoted by subscript $a,b$. Gauged transformed versions of $\mathbf{Q}=0,$ \textit{cuboc-1},  \textit{cuboc-2}, \textit{octahedral} are also referred to in this way. $FM$ stands for ferromagnetic order. } \label{fig:classical_PD_SU2}
\end{figure*}
Counting all the possible combinations of $p_i$, there are many more different ans\"atze for the pure Heisenberg model than for the model with DM interactions. This is due to the fact that in the latter case, there are less symmetry restrictions leading to more free parameters in general, both phases and amplitudes.

\subsection{Classical orders}
Once the ground state of the mean field Hamiltonian has been obtained, we determine from the spinon dispersion whether it features gap or not. SBMFT is particularly well fitted for distinguishing long-range orders from gapped spin liquids. When the gap closes, the spinons are allowed to condense in the ground state. This condensate breaks the spin rotational invariance and results in a long-range ordered arrangement of spins on the lattice. In general, the long-range order to which an ansatz can condense is given by the form of the condensate. However, there are also examples of ans\"atze which do not allow condensation to any order as we will see later in the results' section. The set of classical orders that can be formed is restricted to states respecting the remaining symmetries up to a global spin rotation. These are the so-called regular orders, which have been classified, also for chiral orders, in \cite{Messio2011}. In this classification, all possible classical orders on the kagome lattice that are $O(3)$-regular have been taken into account. These are the orders which for any lattice symmetry $X$ allow for  a global spin rotation $S_X\in O(3)$ such that the state is invariant under $S_XX$. The group $O(3)$ has been considered since it is the spin symmetry of the Heisenberg model. 

The important quantities that will be useful for our analysis are the values of the amplitudes and phases of the pairing and hopping parameters in these classical orders, which then can be used as starting parameters for the minmization procedure. These values can be found in \cref{tab:classical_orders} for $O(3)$ regular orders on the kagome lattice with all lattice symmetries.

All phases and amplitudes have been obtained by considering the same loops used to evaluate the constraints \eqref{eq:constraints}. We start by taking the product of pairing and hopping operators along a loop and subsequently expanding them as in \eqref{eq:bond_ops}. This results in a sum of terms containing an even number of single-site boson operators. By considering the expression of the spins in terms of Schwinger bosons \eqref{eq:SB_def}, we can then derive
\begin{subequations}
	\begin{align}
		\hat{a}^\dagger_i\hat{a}_i &= \frac{1}{2} + S^z,		\\
		\hat{b}^\dagger_i\hat{b}_i &= \frac{1}{2} - S^z,		\\
		\hat{a}^\dagger_i\hat{b}_i &= S^x+iS^y.		
	\end{align}
\end{subequations}
Inserting these relations into the loop expressions and considering the real space values of the spin directions for the different orders, we obtain the results in \cref{tab:classical_orders}. 

\section{Numerical results}       \label{sec:results}
\subsection{Classical phase diagram}
It is instructive to first determine the classical ground state phase diagram corresponding to our model. As candidate states, we consider the regular orders defined in \cite{Messio2011}. In addition, we take into account the orders obtained by gauge transforming the regular orders through our $\sqrt{3}\times\sqrt{3}$-type transformation mentioned in the introduction. A third set of possible ground states is given by the generalized spiral orders. These are very general orders which are obtained by considering only translations as symmetries in the method of \cite{Messio2011}. They are defined by the directions of the spins in a $6$-site unit cell \footnote{The general unit cell has $6$ sites because of the parameter $p_1$ found in the PSG classification. By considering only a $3$-site unit cell, orders like \textit{cuboc-1}, \textit{cuboc-2} and \textit{octahedral} are excluded.} and by two angles defining the rotations of the spins when moving to the neighboring unit cell along the two inequivalent lattice directions. In addition to the rotation, there can be also an inversion leading to two additional parameters and a total of $15$ parameters. By minimizing the energy of such spiral states, we find the most probable candidate ground state at the classical level. With this most general choice of parameters, spiral states include all regular orders, gauged orders and umbrella states. The phase diagrams for different DM angles at $SU(2)$-invariant values are shown in \cref{fig:classical_PD_SU2}. 

We can see that the two $\sqrt{3}\times\sqrt{3}$ orders with opposite vector chirality are degenerate at zero DM angle, while at DM angle $\phi=n\pi/3, n=1,2$, the degeneracy is lifted and one of them becomes degenerate with the in-plane ferromagnetic order at each point. This is due to the fact that the ferromagnetic order is transformed into one of the $\sqrt{3}\times\sqrt{3}$ orders under the gauge transformation, which in turn is transformed into its other vector chirality and finally back to the ferromagnetic order. As with the $\sqrt{3}\times\sqrt{3}$/ferromagnetic state, all the orders come in sets of three states that are generated by acting with the gauge transformation once or twice on the original order. We denote the once (twice) gauge transformed versions with the subscript $a$ ($b$). Since the points with the DM phases $\phi=n\pi/3,\quad n=1,2$ are all effectively $SU(2)$ invariant, the structure of the phase diagram is the same.

\begin{figure}
    \centering
    \includegraphics[width=0.5\textwidth]{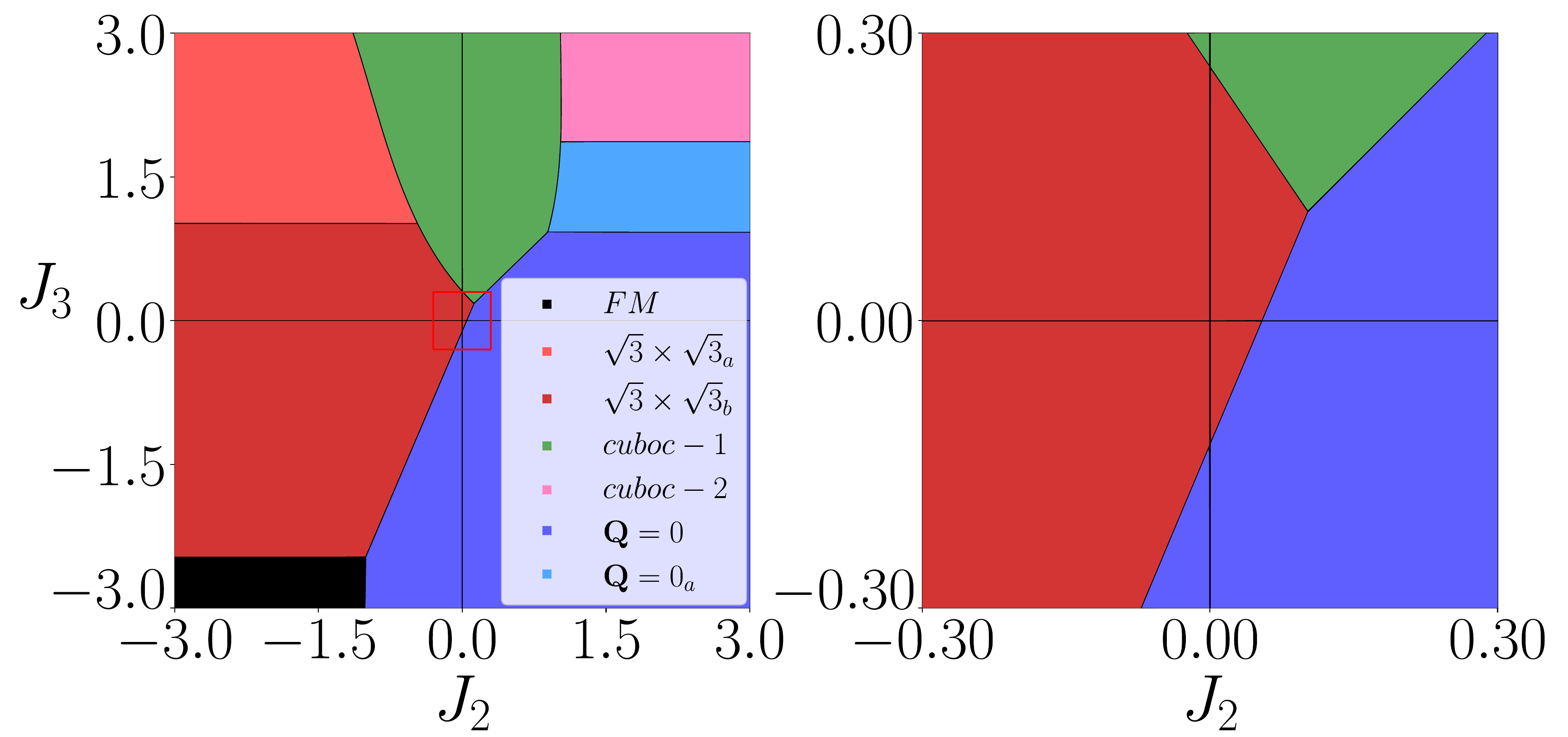}
	\caption{Classical phase diagram with DM angle $\phi=0.05$. On the left panel a bigger parameter space is shown while in the right one we zoom in the region close to the $J_2=J_3=0$ point.}
    \label{fig:classical_PD_005}
\end{figure}

Outside of these $SU(2)$ invariant points, the situation changes. The classical phase diagram for $\phi=0.05$ is shown in \cref{fig:classical_PD_005}. We first of all note that a finite value of $\phi$ lifts the degeneracy between the two vector chiralities of the $\sqrt{3}\times\sqrt{3}$ order. Also, the origin $J_2=J_3=0$ of the phase diagram is no longer a degenerate point of three orders and it becomes fully $\sqrt{3}\times\sqrt{3}$. By zooming in on the region close to the origin (right panel of \cref{fig:classical_PD_005}) we illustrate how the degeneracy is lifted leaving a clear $\sqrt{3}\times\sqrt{3}$ order at the origin. 

\subsection{Phase diagram of \texorpdfstring{$J_2=J_3=0$}{J2=J3=0}}     \label{subsec:SDM}
\begin{figure}
    \centering
    \includegraphics[width=0.45\textwidth]{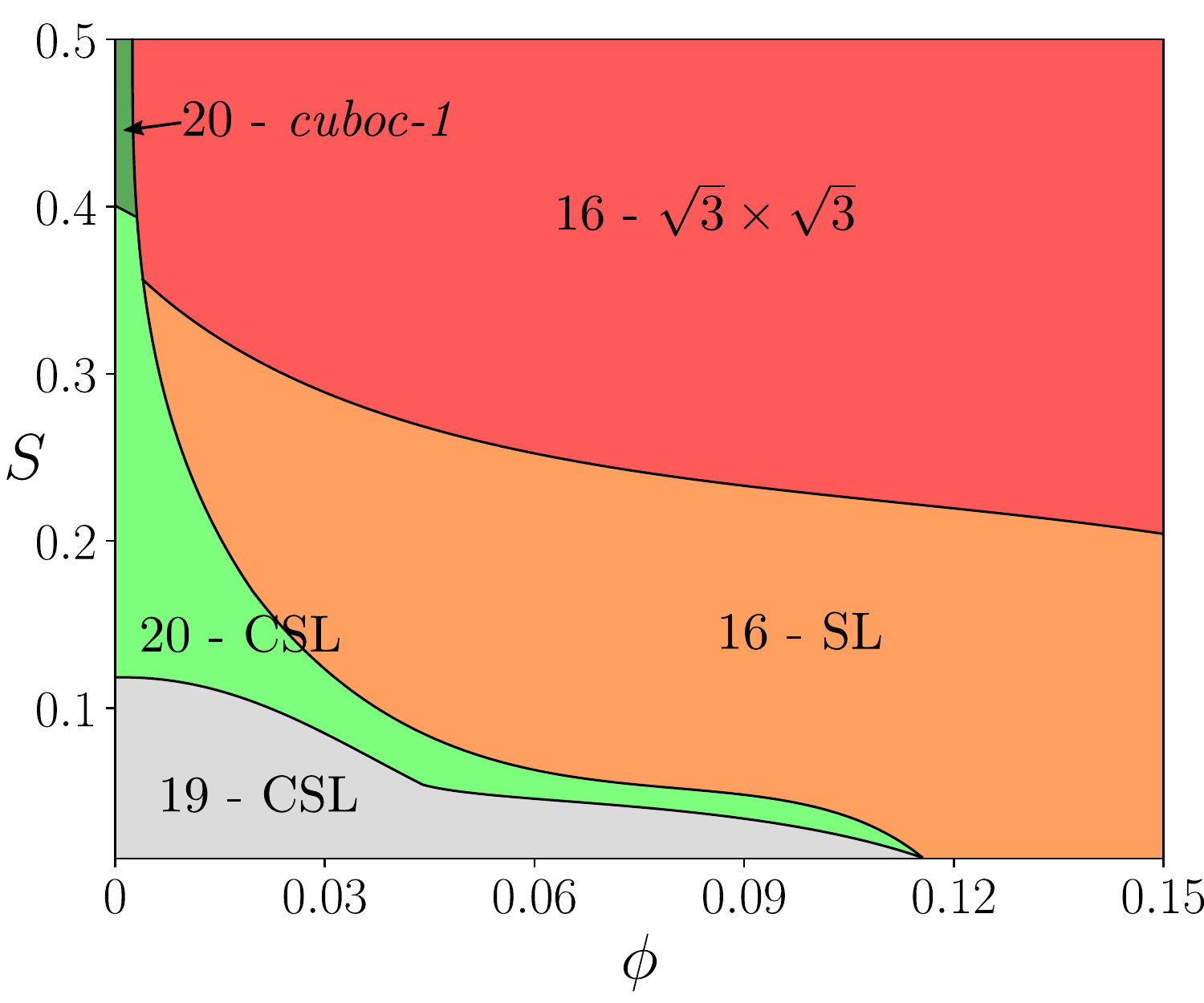}
	\caption{Phase diagram of $1$st n.n. Heisenberg model on kagome lattice with the addition of DM interaction as a function of spin $S$ and DM phase $\phi$. When gapped, solution $16$ is a spin liquid (SL) while solutions $19,20$ are chiral spin liquids (CSL). Ordered phases \textit{cuboc-1} and $\sqrt{3}\times\sqrt{3}$ are gapless solutions of ans\"atze $20$ and $16$, respectively.}
    \label{fig:SDM}
\end{figure}

Turning to the quantum case, let us start by computing the $J_2=J_3=0$ phase diagram in terms of the DM phase $\phi$ and spin $S$ along the same line as \cite{Messio2017,Huh2010,Mondal2017,Messio2010}, only with a uniform DM interaction instead. In this first calculation, we used a self-consistent method for finding the saddle points of the free energy and considered ans\"atze $15$ to $20$ in \cref{tab:allHeisTMD}. The mean field parameters thus are: $|A_1|,|B_1|,\phi_{B_1}$, with the addition of $\phi_{A_1}'$ for ans\"atze $19$ and $20$. The resulting phase diagram is reported in \cref{fig:SDM}.
First of all, we note that the chiral ground state at $0$ DM is rapidly substituted by a $\sqrt{3}\times\sqrt{3}$ type of phase as the DM phase increases. 
\begin{figure}[h!]
    \centering
    \includegraphics[width=0.35\textwidth]{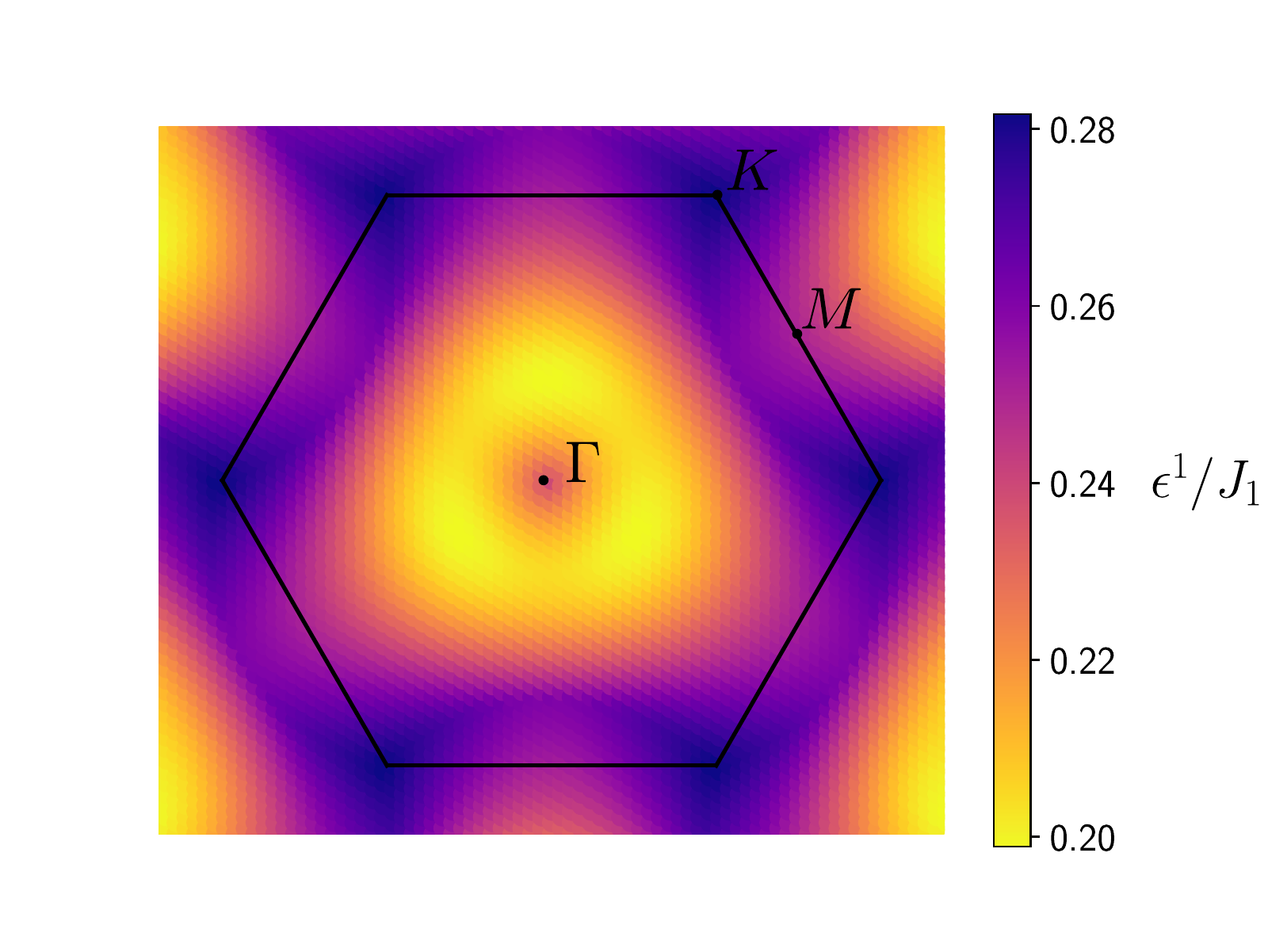}
	\caption{Lowest band spinon dispersion of ansatz $19$ for $S\sim0.11$ and $\phi=0$.}
    \label{fig:BZ_circle}
\end{figure}
\begin{figure*}
    \centering
    \includegraphics[width=\textwidth]{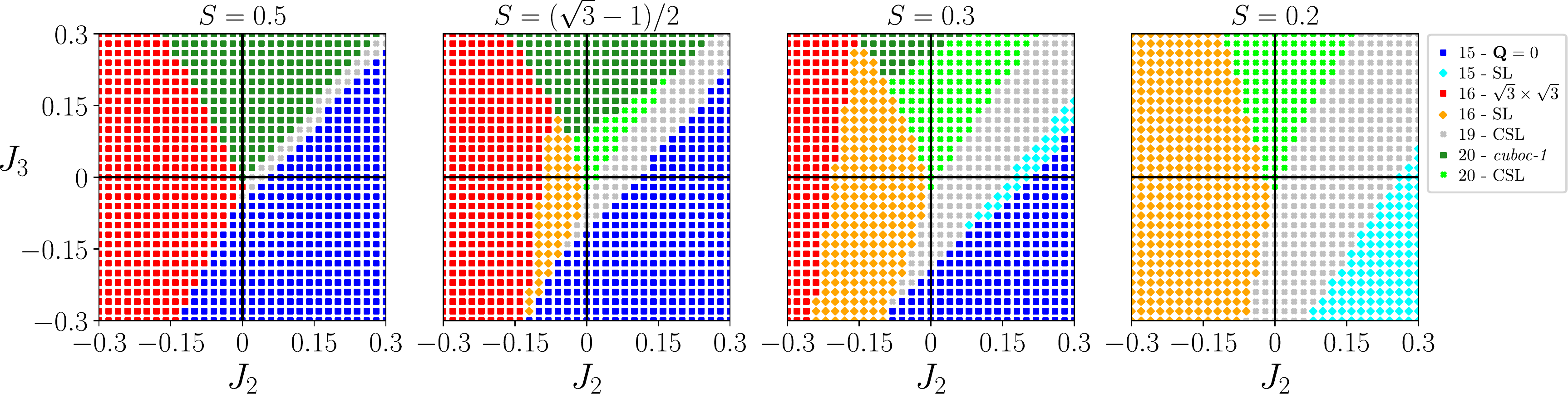}
	\caption{Phase diagrams of $J_1$-$J_2$-$J_3$ Heisenberg antiferromagnet at different spin values for DM phase $\phi=0$. The other $SU(2)$ invariant points result in equal diagrams. The orders appearing are the same as for the classical phase diagram, namely $\sqrt{3}\times\sqrt{3}$, $\mathbf{Q}=0$ and \textit{cuboc-1}. We find one additional solution ($19$) appearing at the border between $\mathbf{Q}=0$ and \textit{cuboc-1}, which is chiral and always gapped. We can also appreciate that already at $S=(\sqrt{3}-1)/2$ the ground state of the $J_2=J_3=0$ Heisenberg antiferromagnet is in a gapped chiral spin liquid phase. By decreasing further the spin value, the whole considered parameter space becomes gapped.}
    \label{fig:all_pd}
\end{figure*}
By decreasing the spin value, we see that LRO phases get substituted by SL ones, as expected since quantum fluctuations become stronger. Three of the six solutions of \cref{tab:allHeisTMD} appear in the phase diagram: $16$, $19$ and $20$. While $16$ and $20$ condense to planar $\sqrt{3}\times\sqrt{3}$ and chiral \textit{cuboc-1} orders, respectively, after gap closing, there is no known order corresponding to ansatz $19$. From the value of its phases, we determine that it is chiral, but it only appears as a SL in the phase diagram. This solution is the analog  of the phase $A_4(0,1)$ found in \cite{Messio2017} for uniform DM interactions. It also has the same circular minima in spinon spectrum, as shown in \cref{fig:BZ_circle}. 
In our simulations, in the entire parameter region in which it converges as a solution of the self-consistent equations, it remains gapped.

This phase diagram should coincide with the ones in \cite{Messio2017,Mondal2017} for DM phase $\phi=0$. However, we see a slight difference at small $S$ (around $S\lesssim0.1$), where we find solution $19$, while the \textit{cuboc-1} was reported up to very small spin values in \cite{Messio2017}. In our simulations, we took a precision of $10^{-7}$ for amplitudes and $10^{-5}$ for phases of the mean field parameters, normalizing with respect to the spin value in order to achieve a uniform precision. We also note that the phase $A_1(1,0,1)$ found in \cite{Messio2017}, which would correspond to our ansatz $17$, does not appear in our simulations. 

We also computed the spin structure factor
\begin{equation} \label{eq:ssf}
	\Xi(\vec{Q})=\frac{1}{\mathcal{N}}\sum_{i,j}e^{-i\vec{Q}\cdot(\vec{r}_i-\vec{r}_j)}\langle\vec{S}_i\cdot\vec{S}_j\rangle,
\end{equation}
to distinguish the phases. The details of the computation are given in \cref{App:ssf}. From the structure factors, we can see the peaks  which would show up in neutron scattering experiments in long-range orders, while we expect to see broader features in  the spin liquid phases . In general, we expect DM interactions to favour in-plane orders due to the out-of-plane direction of the DM vector (see \cref{fig:DM_directions}). Also, as argued in \cref{sec:model}, the presence of DM interactions reduces the symmetry of the model. This in turn reduces the effect of quantum fluctuations, thus favouring long-range order configurations. This general tendency can be observed in \cref{fig:SDM} by noticing that the transition line between $\sqrt{3}\times\sqrt{3}$ LRO to SL (ansatz $16$) happens at decreasing spin S values by increasing the DM strength.

Finally, we emphasize that the part of the phase diagram with multiple competing phases is restricted to DM angles $\phi\lesssim0.12$ with our choice of DM interactions. For larger DM angle, the only solution that remains is $16$, gapped (SL) and gapless ($\sqrt{3}\times\sqrt{3}$). In contrast, in the case of staggered DM interactions the phase diagram has multiple phases up to $\phi\sim0.3$, after which the $\mathbf{Q}=0$ order dominates.

\subsection{\texorpdfstring{$SU(2)$}{SU(2)} invariant points}

We now turn to the model with finite $J_2$ and $J_3$ parameters in the presence of uniform DM interactions of the form given in \cref{eq:DM_phases}. We start with the phase diagrams at the $SU(2)$ invariant points, i.e for DM phase equal to $\phi=0,\pi/3,2\pi/3$, in \cref{fig:all_pd}. We use a self-consistent method with precision $10^{-6}$ for amplitudes and $10^{-4}$ for phases, again normalized by the spin value in order to get a uniform precision in various plots. To distinguish between SL and LRO, we performed finite size scaling up to lattices of $2401$ (ansatz) unit cells and used a cutoff of $10^{-2}$ for the gap value, as discussed in detail in \cref{App:gap}.

The ans\"atze we considered are $15$ to $20$ in \cref{tab:allHeisTMD}, since they are more general and include the solutions $1$ to $14$, which are expected to be the ans\"atze for the $SU(2)$ invariant points. We computed the $J_1$-$J_2-J_3$ phase diagram for spin values $S=0.5,(\sqrt{3}-1)/2,0.3,0.2$. Four solutions mainly appear in the phase diagrams: $15$, $16$, $20$ and $19$. The first three condense, after gap closing to $\mathbf{Q}=0$, $\sqrt{3}\times\sqrt{3}$ and \textit{cuboc-1} orders, respectively. In particular, the solutions appearing in the diagrams correspond to ansatz $15$ with $p_2=p_3=1$, $16$ with $p_2=p_3=0$, $20$ with $p_2=p_3=1,\;p_4=0$, and finally $19$ with $p_2=p_3=1,\;p_4=p_5=0$. For ansatz $20$, the value of $p_5$ is not important since all the found solutions have $B_3\approx0$.

\begin{figure}
     \centering
     \begin{subfigure}[b]{0.45\columnwidth}
         \centering
         \includegraphics[width=\textwidth]{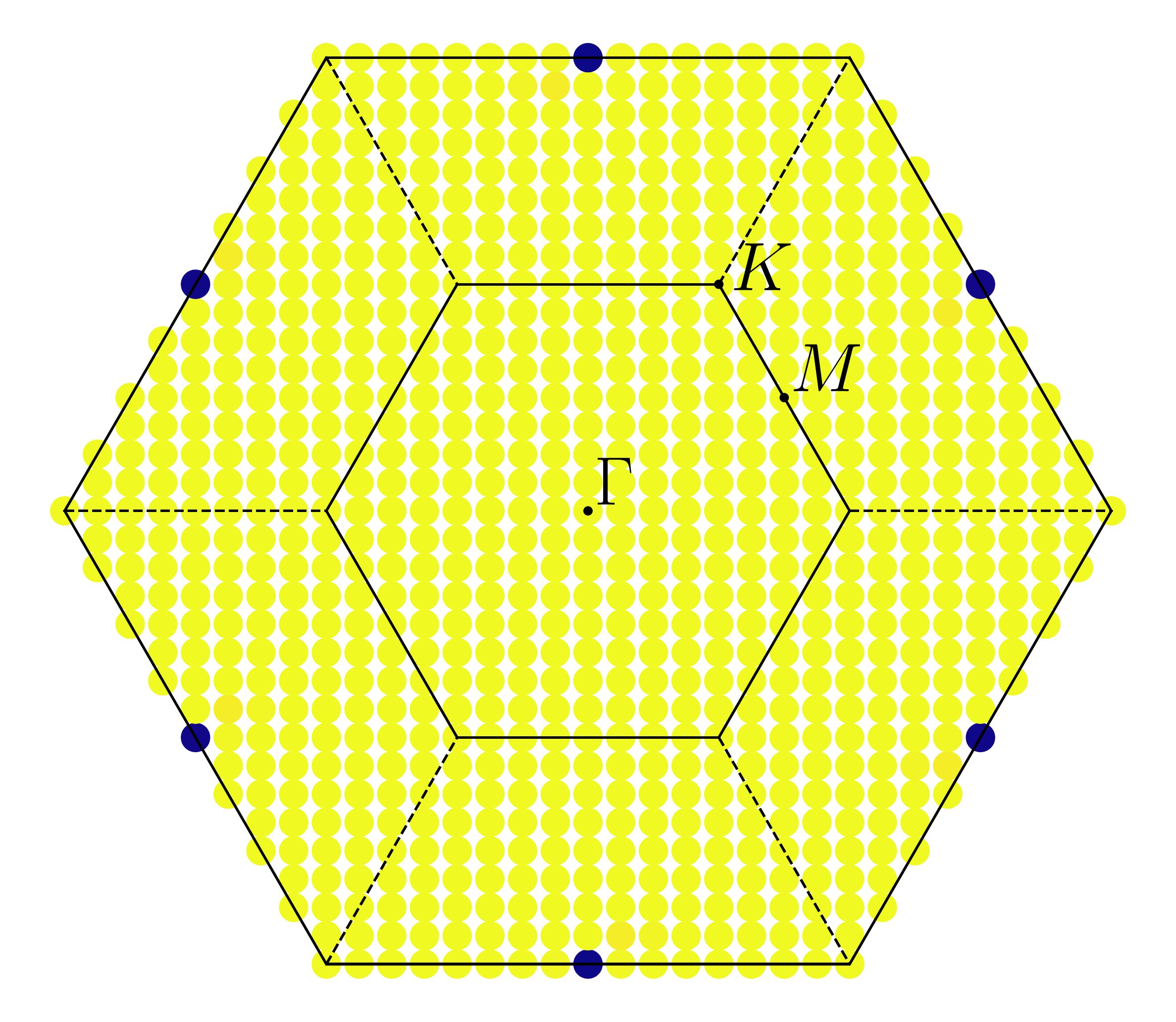}
         \caption{Ansatz $15$ at $J_2=0.3,\,J_3=0$ and \newline $\phi=0$.}
         \label{fig:15_000}
     \end{subfigure}
     \hfill
     \begin{subfigure}[b]{0.45\columnwidth}
         \centering
         \includegraphics[width=\textwidth]{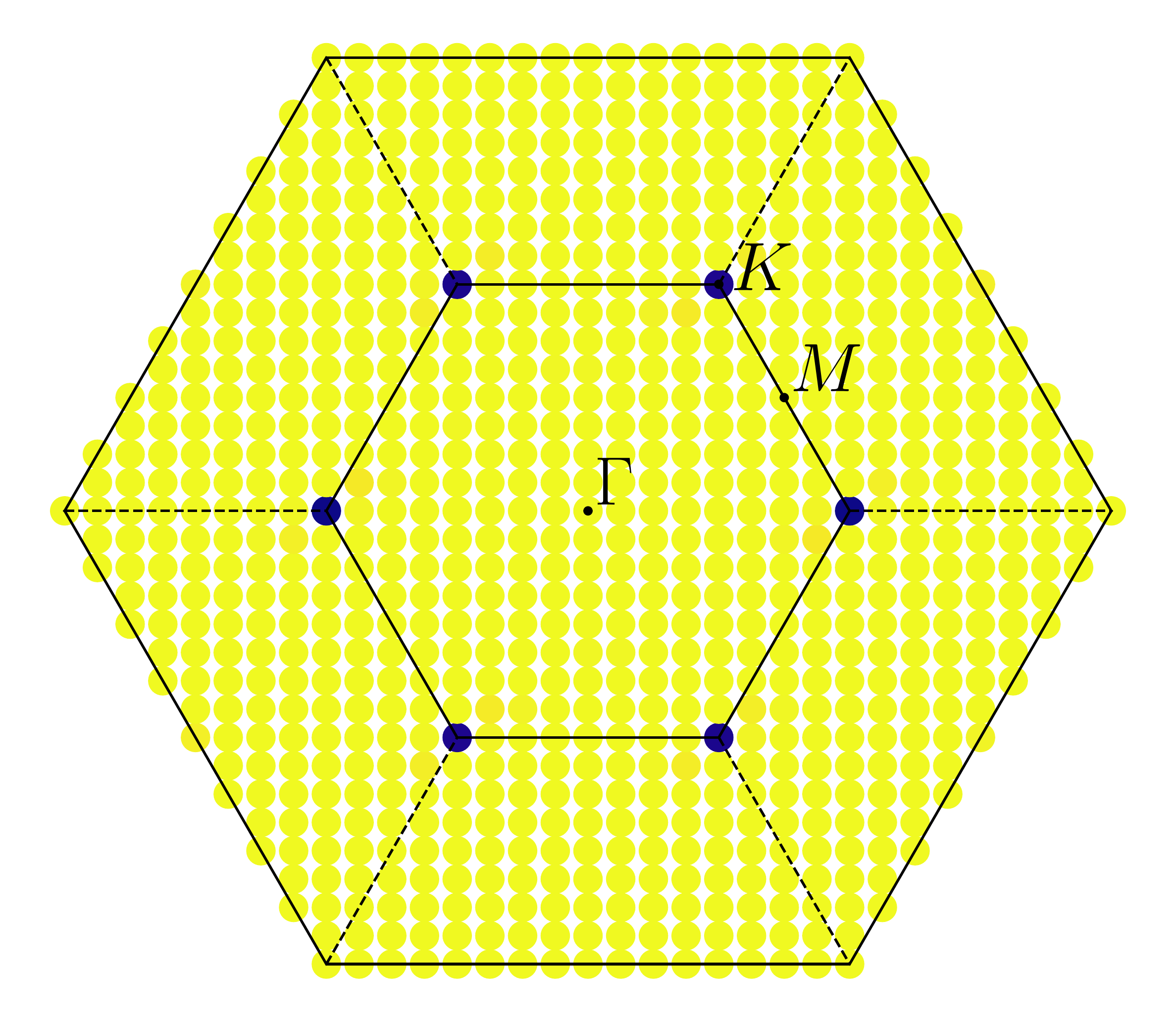}
         \caption{Ansatz $15$ at $J_2=0.3,\,J_3=0$ and \newline $\phi=2\pi/3$.}
         \label{fig:15_209}
     \end{subfigure}
     \vfill
     \begin{subfigure}[b]{0.45\columnwidth}
         \centering
         \includegraphics[width=\textwidth]{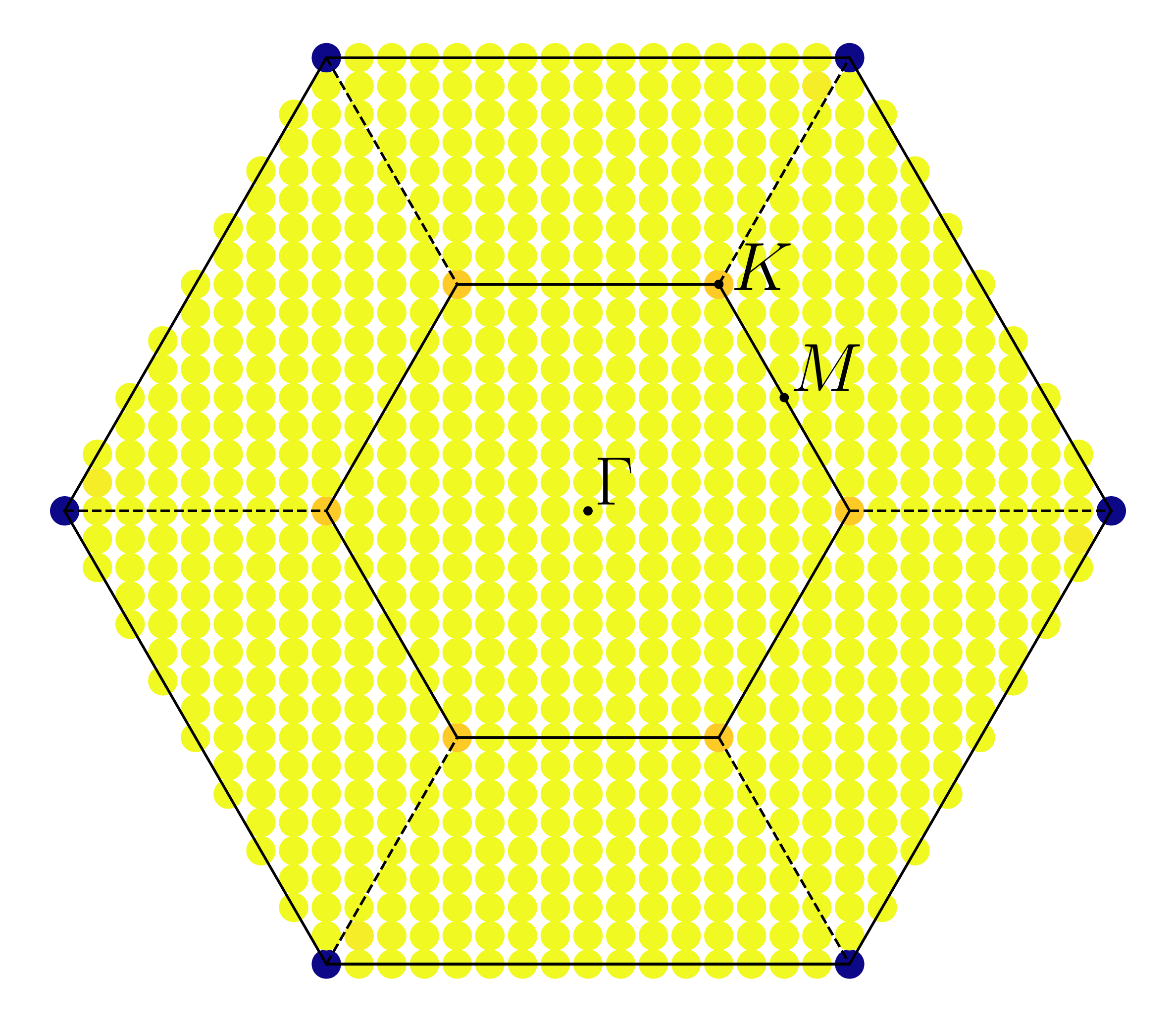}
         \caption{Ansatz $16$ at $J_2=-0.3,\,J_3=0$ and \newline $\phi=0$.}
         \label{fig:16_000}
     \end{subfigure}
     \hfill
     \begin{subfigure}[b]{0.45\columnwidth}
         \centering
         \includegraphics[width=\textwidth]{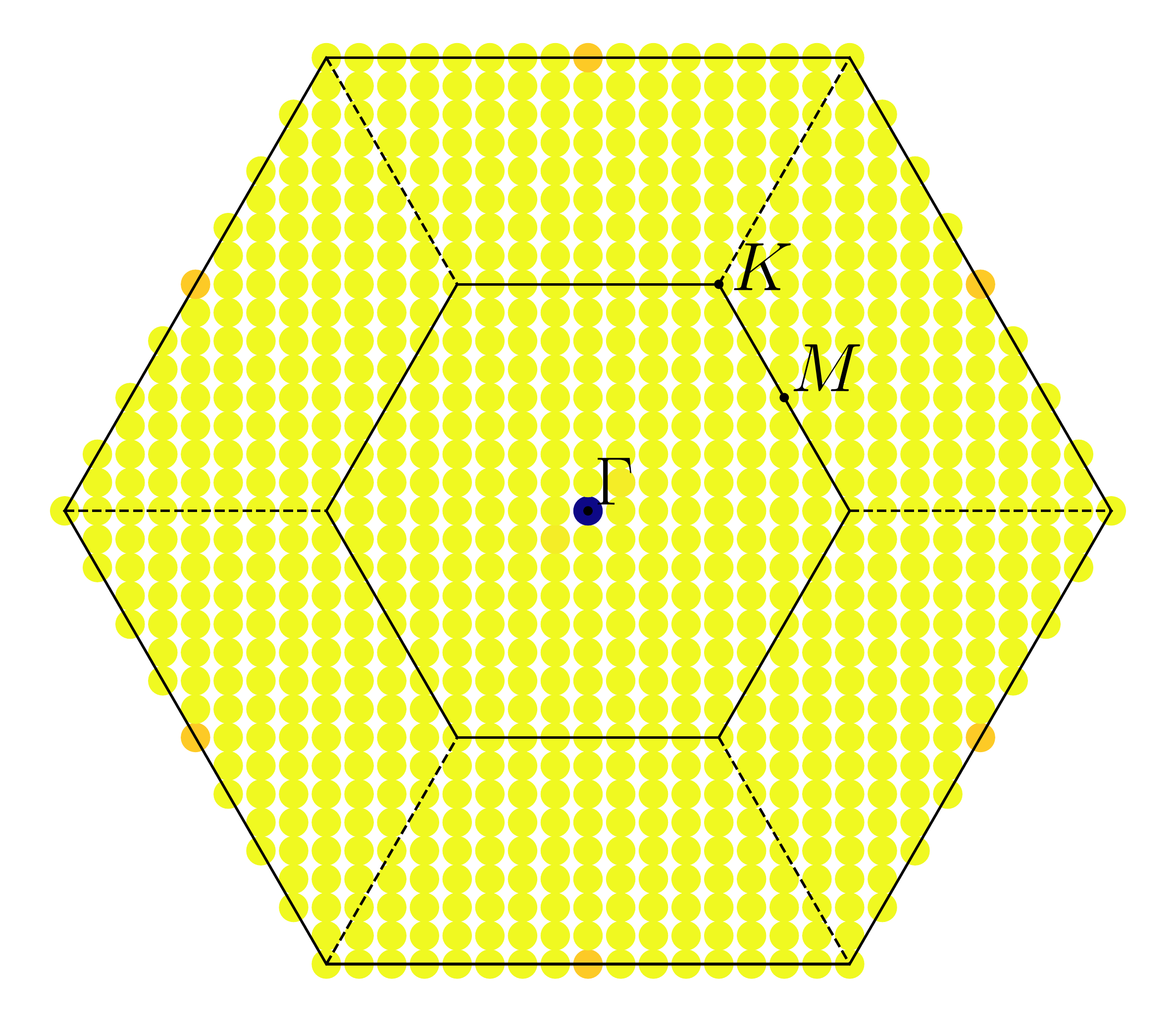}
         \caption{Ansatz $16$ at $J_2=-0.3,\,J_3=0$ and \newline $\phi=2\pi/3$.}
         \label{fig:16_209}
     \end{subfigure}
     \vfill
     \begin{subfigure}[b]{0.45\columnwidth}
         \centering
         \includegraphics[width=\textwidth]{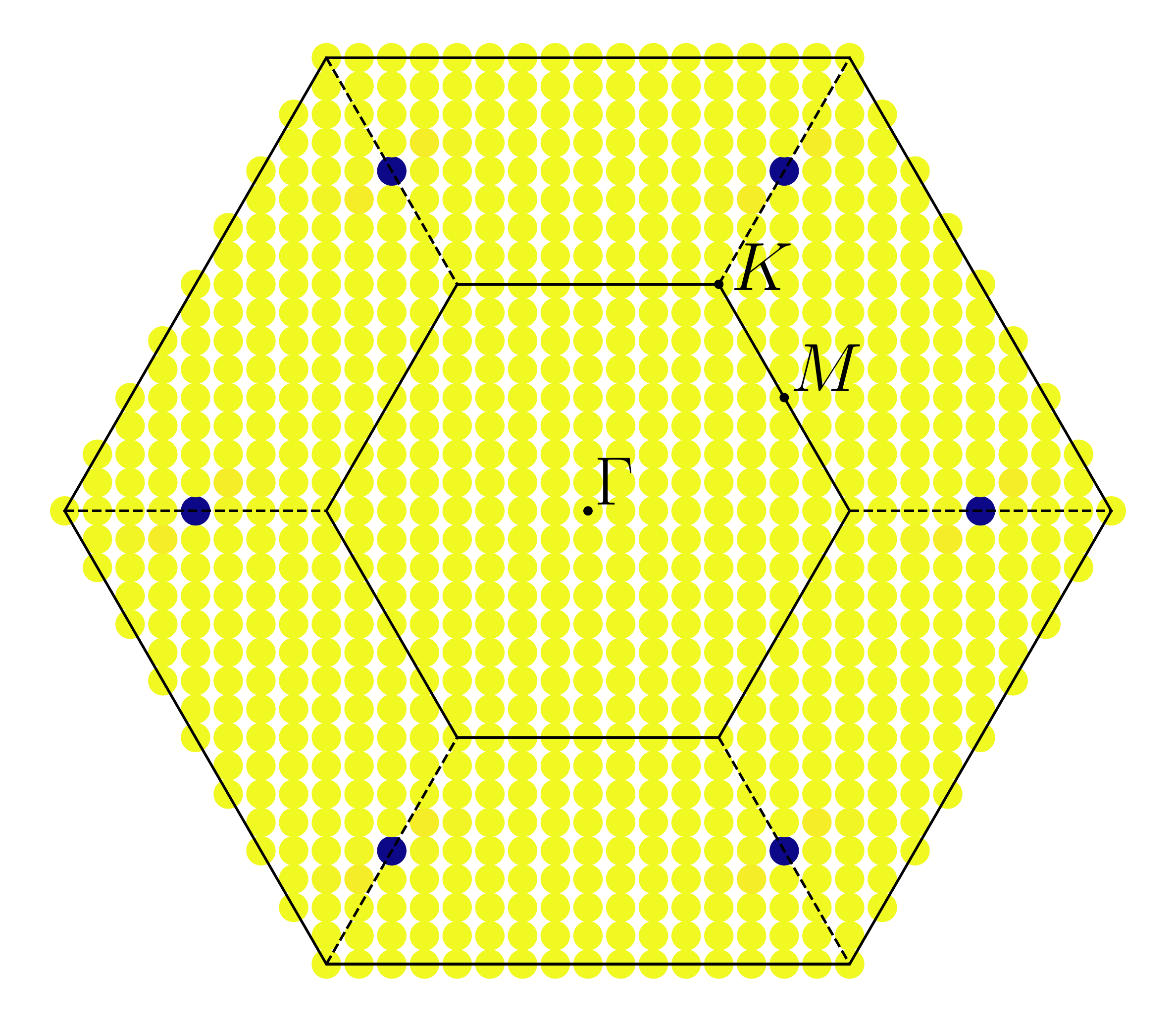}
         \caption{Ansatz $20$ at $J_2=0,\,J_3=0$ and \newline $\phi=0$.}
         \label{fig:20_000}
     \end{subfigure}
     \hfill
     \begin{subfigure}[b]{0.45\columnwidth}
         \centering
         \includegraphics[width=\textwidth]{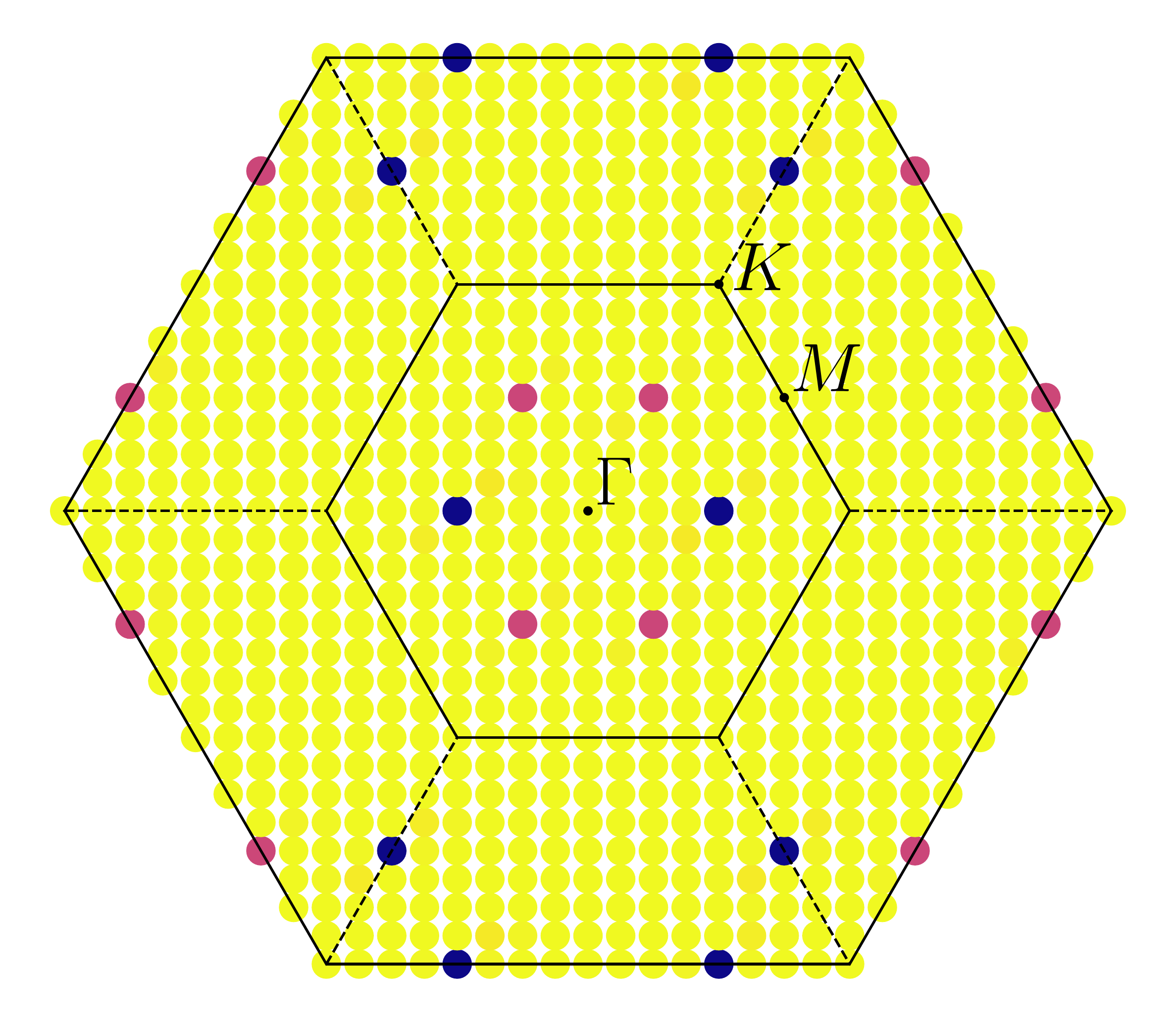}
         \caption{Ansatz $20$ at $J_2=0,\,J_3=0$ and \newline $\phi=2\pi/3$.}
         \label{fig:20_209}
     \end{subfigure}
     \vfill
     \centering
     \begin{subfigure}[b]{0.7\columnwidth}
         \centering
         \includegraphics[width=\textwidth]{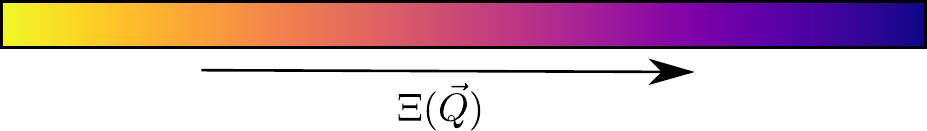}
     \end{subfigure}
    \caption{Spin structure factors evaluated using the method detailed in \cref{App:ssf_cond}, for spin $S=0.5$. Darker colors indicate a higher value of the structure factor.}
    \label{fig:SSF_LRO}
\end{figure}

As expected, the structures of the phase diagrams for the different $SU(2)$ invariant points are exactly the same. The only difference lies in the static spin structure factor of the LRO phases since the spin orientations are changed by the gauge transformation. We compute structure factors using the methods described in \cref{App:ssf} to compare the orders at different DM angles and spin values. In \cref{fig:SSF_LRO}, we report the SSF at $\phi=0,2\pi/3$ for points in the phase diagram in the regular LRO phases. These patterns coincide with the classical predictions for the respective regular orders. 

Let us now discuss the phase diagrams for $\phi=0$, \cref{fig:all_pd}. As expected, quantum effects become more relevant for decreasing spin values and from a mostly LRO phase diagram at $S=0.5$ we reach a completely SL one at $S=0.2$. The phase diagrams follow quite closely the classical prediction for higher spin values while they differ from it at lower $S$. As already reported in \cite{Messio2012}, we also see that the $J_2=J_3=0$ point has a chiral solution, which is a \textit{cuboc-1} order at high spin values and turns into a chiral spin liquid for $S<0.4$. More interestingly, for spin value $0.5$ we find an additional solution to the ones reported in \cite{Messio2012}, which is the further neighbor version of ansatz $19$, realising a chiral spin liquid phase. It is remarkable that this phase remains gapped even at spin $0.5$, a feature not observed in any SBMFT calculation on any lattice to the best of our knowledge. The region of the phase diagram where this new solution appears approximately coincides with the region identified in \cite{Gong2015} and \cite{Motruk2022} as chiral spin liquid. As mentioned above, a practice commonly used in SBMFT studies is to assume $S=(\sqrt{3}-1)/2\approx0.366$ as an effective spin value. In this case, the phase diagram is much richer since it features $\mathbf{Q}=0$,$\sqrt{3}\times\sqrt{3}$ and \textit{cuboc-1} long-range orders together with three different kinds of spin liquids. In particular, with solutions $19$ and $20$ being both chiral since $\phi_{A1}'\neq0,\pi$, we observe two different kinds of phases in the chiral spin liquid region identified in \cite{Gong2015,Motruk2022}.  The main difference in the form of the ans\"atze is $p_1$, thus, they have different unit cell size. By further decreasing the spin value, the chiral region increases and all phases turn towards gapped solutions.

\begin{figure}
     \centering
     \begin{subfigure}[b]{0.45\columnwidth}
         \centering
         \includegraphics[width=\textwidth]{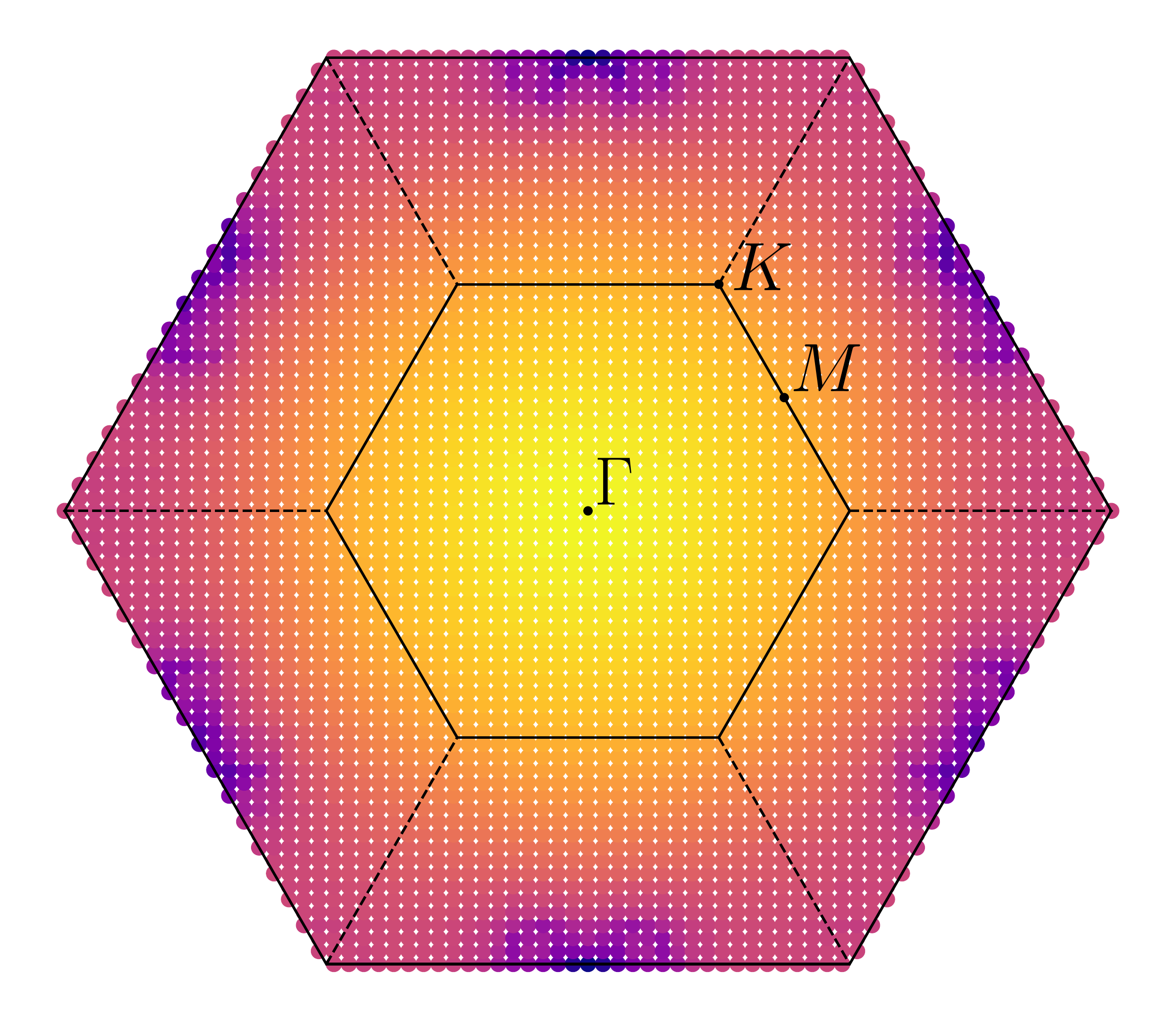}
         \caption{Ansatz $19$ at $J_2=0.06,\,J_3=0.02$, $\phi=0$ and spin $S=0.5$.}
         \label{fig:19_SL}
     \end{subfigure}
     \hfill
     \begin{subfigure}[b]{0.45\columnwidth}
         \centering
         \includegraphics[width=\textwidth]{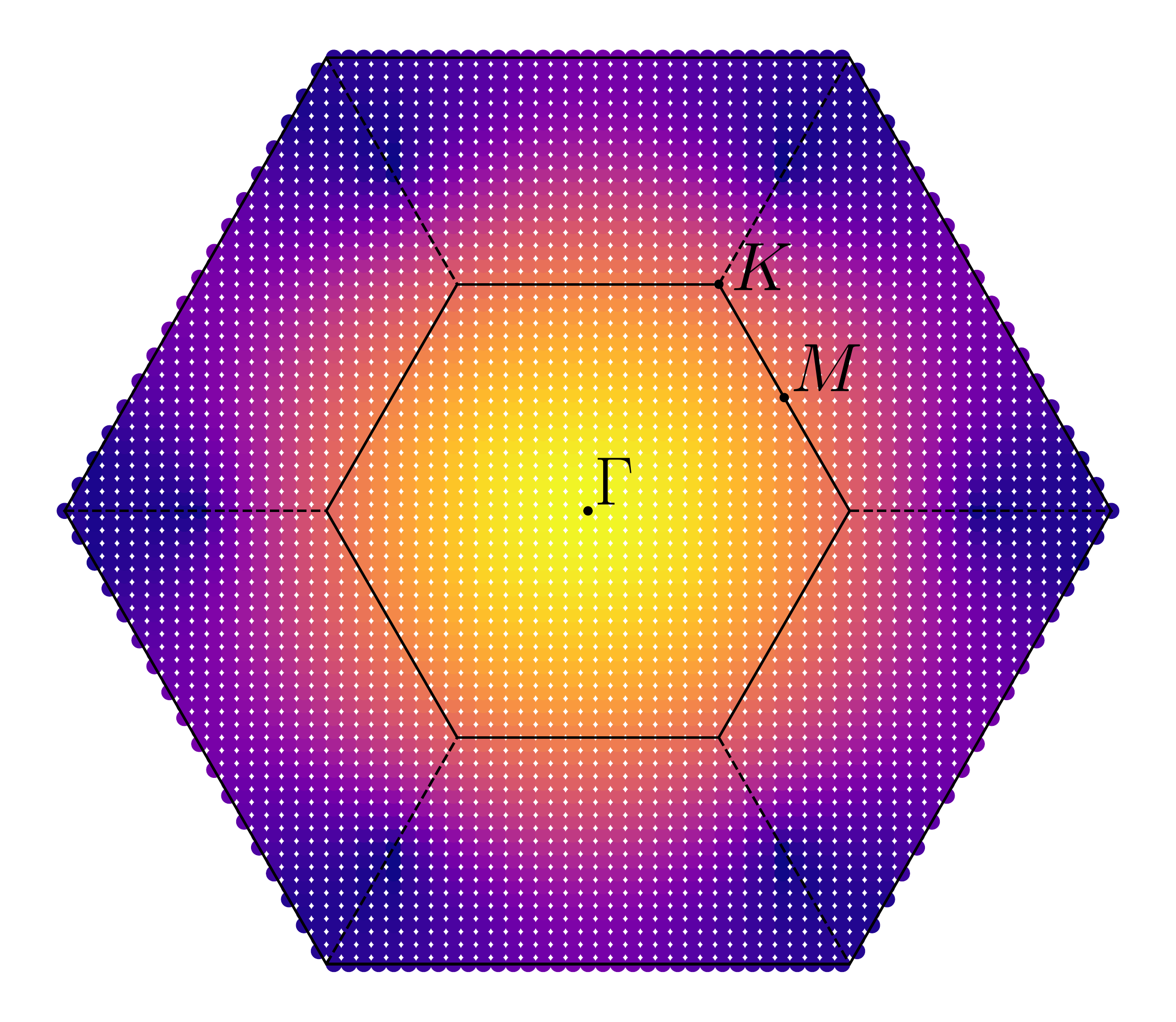}
         \caption{Ansatz $20$ at $J_2=0,\,J_3=0$, $\phi=0$ and spin $S=(\sqrt{3}-1)/2$.}
         \label{fig:20_SL}
     \end{subfigure}
    \caption{Spin structure factor of Ansatz 19 evaluated using the method detailed in \cref{App:ssf_gs}. As in \cref{fig:SSF_LRO}, darker colors refer to higher values.}
    \label{fig:SSF_SL}
\end{figure}

We compute the spin structure factor for ansatz $19$ and $20$ (\cref{fig:SSF_SL}) for values of the parameters $J_2,J_3,S$ where they appear as gapped SL, for comparison. For ansatz $20$ at $J_2=J_3=0$ and $S=(\sqrt{3}-1)/2$, the structure factor is more smeared out with the peaks remaining at the positions of the LRO ones (see \cref{fig:20_000}). More interesting is the case of ansatz $19$, where the structure factor shows a smeared pattern with peaks around the position expected for $\mathbf{Q}=0$, as reported in \cref{fig:15_000}. This is not surprising since the value of $\phi_{A_1}'$ of ansatz $19$ is around $0.3$, and this ansatz coincides with $15$ when $\phi_{A_1}'=0$ as shown in  \cref{tab:allHeisTMD}.
\begin{figure*}
    \centering
    \includegraphics[width=0.95\textwidth]{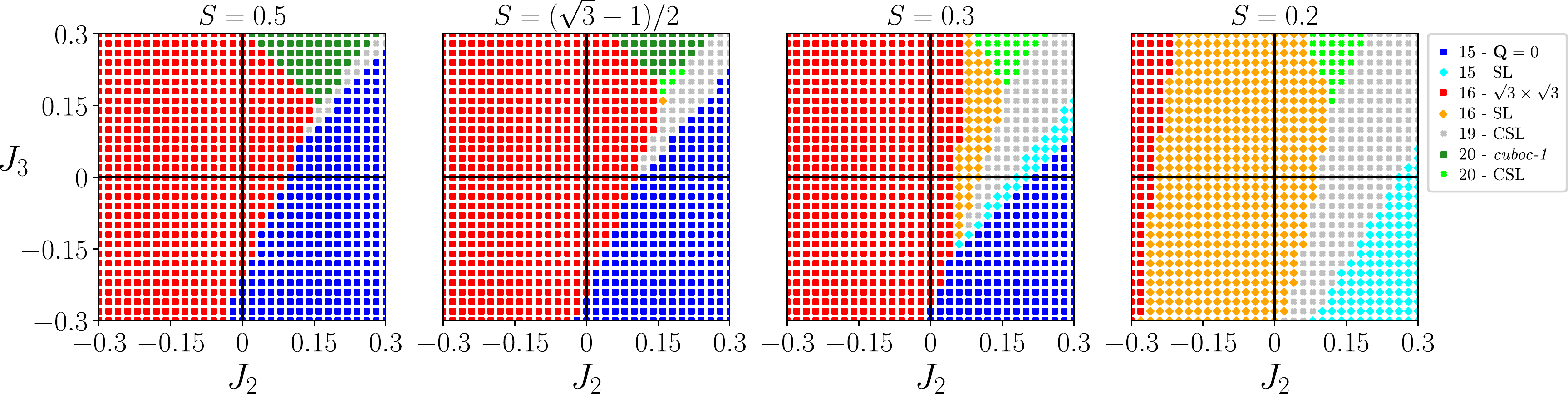}
	\caption{Phase diagrams of $J_1$-$J_2$-$J_3$ Heisenberg antiferromagnet with uniform DM interaction $\phi=0.05$. The orders appearing in these phase diagrams are the same as for the $SU(2)$ invariant points in \cref{fig:all_pd}. The DM interaction favours solution $16$ which becomes predominant in the considered parameter space. Also, it reduces the quantum fluctuations resulting in a phase diagram with more LRO gapless phases with respect to the same spin values in \cref{fig:all_pd}.}
    \label{fig:all_pd_dm}
\end{figure*}

\subsection{Finite DM angle}
We finally study the case with a finite offset of the DM value from the $SU(2)$ invariant point. In particular, we consider a small phase $\phi=0.05$ since we have seen how even a small value can strongly affect the phase diagram. We again take into account ans\"atze $15$ to $20$ in \cref{tab:allHeisTMD} and find a self-consistent solution, similar to the $SU(2)$-invariant case. The results are shown in \cref{fig:all_pd_dm}. 

The solutions at finite DM angle belong to the same phases which were found above without DM interaction. In principle, the system is free to converge towards more general solutions since there are fewer symmetries (and in turn constraints). Nevertheless, we find that all solutions which converge to a saddle point in the energy are more symmetric than their original Hamiltonian. In particular, we notice that there are no solutions with $|A_2|\neq|A_2'|$ or $|B_2|\neq|B_2'|$. This behaviour can actually be expected from the choice of DM interaction \cref{eq:DM_phases}. In fact, with the second nearest-neighbour phase being equal to zero,  we could have considered the $R_6$ symmetry in the PSG construction to be valid at the second-nearest neighbor level, yielding a constraint for these amplitudes.

We notice that the $J_2=J_3=0$ point remains $\sqrt{3}\times\sqrt{3}$ ordered up to much lower spin values with the introduction of the DM interaction. Also, the SL region is reduced greatly in general, as is evident from the two plots at $S=(\sqrt{3}-1)/2$. As argued in \cref{subsec:SDM}, this tendency to favour in-plane configurations and reduce quantum fluctuations is expected in the presence of DM interactions.  The results of this section show that the phases need to be kept close to the $SU(2)$ invariant points for spin liquid ground states to appear~\cite{Kiese2022}. This constrains the displacement fields for potential realizations of the kagome lattice in TMD homobilayers.

\section{Discussion}        \label{sec:conclusion}

In this work, we explored the phase diagram of the Heisenberg model on the kagome lattice taking into consideration further neighbor exchange and Dzyaloshinskii-Moriya interactions. Through the projective symmetry group approach, we classified all possible chiral and symmetric ans\"atze of the model both with and without DM interactions. Then, we employed Schwinger boson mean field theory to compute the ground state of the model across a wide range of parameters. We first studied how uniform (as opposed to staggered in Ref.~\cite{Messio2017}) DM interactions affect the ground states in the nearest-neighbor kagome model. We find that the uniform DM pattern destabilizes the \textit{cuboc-1} order and CSL at their respective effective spin values for even smaller finite DM angle than in the staggered case. This hints at a limited stability of potential spin liquid phases in TMD moir\'e bilayers away from the $SU(2)$ symmetric point, consistent with a functional renormalization group study on the triangular lattice~\cite{Kiese2022}.

We then investigated the model at the $SU(2)$ symmetric points including second and third neighbor interactions. The SBMFT accurately predicts the same phase boundaries for these points that can be related by local spin rotations in the $xy$ plane. Furthermore, it correctly captures the different coplanar magnetic orders related by this gauge transformation, e.g. the two vector chiralities of the $\sqrt{3} \times \sqrt{3}$ and the in-plane ferromagnet. Regarding spin liquids, surprisingly, we found a chiral spin liquid close to the $J_2=J_3$ line that remains stable up to a spin of $S=0.5$, described by an ansatz that had previously not been considered~\cite{Messio2012}. This is, to the best of our knowledge, the first time a spin liquid in SMBFT has been reported at such a high value of the spin. It appears around the region in which the CSL had been detected in DMRG simulations~\cite{Gong2015,Motruk2022} and could indicate an exceptional stability of this state. An open question is the relation of this CSL, which had been reported in the nearest-neighbor model in Ref.~\cite{Messio2017}, to the one of the quantum disordered version of the \textit{cuboc-1} state that emerges next to it for sufficiently small spin values. 

Finally, we added DM interactions away from the $SU(2)$ symmetric point also in the $J_1$-$J_2$-$J_3$ model and found that the spin liquids quickly destabilize in analogy with the nearest neighbor system. Additionally, the coplanar orders are favored over the chiral \textit{cuboc-1} state. Although the symmetry of the Hamiltonian has been reduced from $SU(2)$ to $U(1)$, we still only find magnetically ordered states with a classical analog among the regular magnetic orders of $O(3)$  symmetric Hamiltonians~\cite{Messio2011,Mondal2021}.

Overall, we can conclude that the mean-field calculations of the phase diagrams qualitatively agree well with numerically exact DMRG results in parameter regimes where the latter have been performed and for the phases that can be captured by the Schwinger boson ansatz. The spin value of $S=(\sqrt{3}-1)/2 \approx 0.366$ that is motivated by setting the onsite spin fluctuations equal to the quantum mechanical value, in particular predicts comparable parameter values for the transition lines between long-range orders and spin liquids. We thus demonstrated that SBMFT is an excellent approach to narrow down a large parameter space to regions that can potentially host spin liquid ground states on extended kagome lattice models, without the need to immediately resort to numerically more expensive methods. This might be of great use especially in the rapidly evolving understanding of the microscopic description of TMD moir\'e bilayers~\cite{Claassen2022,Reddy2023}.

\begin{acknowledgments}
The authors wish to thank L.\ Messio and Y.\ Iqbal for useful discussions. Support by the Swiss National Science Foundation (SNSF) and by the European Research Council (ERC) under the European Union’s Horizon 2020 research and innovation program (grant agreement No.~864597) is gratefully acknowledged. J.M. was supported by the SNSF Swiss Postdoctoral Fellowship grant 210478. L.R. is funded by the SNSF through Starting Grant 211296.
\end{acknowledgments}

\appendix
\section{Gap scaling}   \label{App:gap}
As mentioned in the main text, the difference between LRO and SL for a given ansatz can be identified in terms of the value of the Bogoliubov spinons' gap in the thermodynamic limit. If the gap closes the ground state gets a macroscopic occupation of spinons and this condensate leads to a LRO phase, since the global spin rotational invariance has been broken. On the other hand, if the gap remains finite then the system is in a $\mathds{Z}_2$ spin liquid phase. The value of the gap is meaningful just at the thermodynamic limit, so we need to perform finite size scaling in order to infer its behavior.

The mean field free energy that needs to be minimized in order to obtain the ground state is \cref{eq:min_H} and we can see that it contains a summation over the Brillouin Zone (BZ). For each value of $k$ we consider in the BZ we then need to construct the matrix $\mathcal{N}_k$ and diagonalize it with the Bogoliubov transformation in order to get Bogoliubov spinons' bands $\epsilon^\mu(k)$. Each value of $k$ that we take corresponds to a new unit cell in the real space that we are considering for the value of the energy. In the code, we consider a grid of values over the BZ, which contains $\Omega = N_k\times N_k$ $k$-points. As value of the gap is kept the smallest energy of the lowest band on the points of the grid. 

It is clear that in order to correctly estimate the gap value we need to consider grids where the gap closing point is very close to one of the points of the grid. The position in the BZ where the gap closes for different LRO can be derived exactly. For the $q=0$ order the gap closes at $(0,0)$-point in the BZ, so there will be two degenerate eigenvectors associated to the zero energy eigenvalue. For the $\sqrt{3}\times\sqrt{3}$ order the gap closes at $\vec{k}=(2\pi/3,2\pi/3)$ while for the \textit{cuboc-1} at $\vec{k}=(\pi/2,\sqrt{3}\pi/2)$. These values are all easily included in the grid because they are dependent of the size of the unit cell of the corresponding LRO. The $\mathbf{Q}=0$ has a three-site unit cell, meaning that any even \footnote{The grids consider both initial and final point of the BZ, so in the end $N_k$ is increased by $1$.} value of $N_k$ will contain it. The $\sqrt{3}\times\sqrt{3}$ order has nine sites in the unit cell, meaning $3$ kagome unit cells, so any value of $N_k$ multiple of three will do. Finally, the \textit{cuboc-1} has a $12$ sites unit cell, corresponding to $4$ kagome unit cells, so $N_k$ needs to be a multiple of $4$. Following this scheme, the first value of $N_k$ containing all the desired $k$ points is $12$, then $24$, $36$ etc. By plotting the value of the gap for increasing grid's thickness we can see how it scales towards the thermodynamic limit. 

Here I show some examples of typical behaviors. The gap should scale as $1/N_s$ with $N_s$ the number of sites. In terms of $N_k$ and of the unit cell size $m$ this corresponds to $N_s=mN_k^2$. We can fit with a function $a/N_s+b$ and check the value of $b$. Finally, we will have to introduce a cutoff on the value of the gap at the thermodynamic limit in order to distinguish between the two phases.

\begin{figure*}
    \centering
    \includegraphics[width=\textwidth]{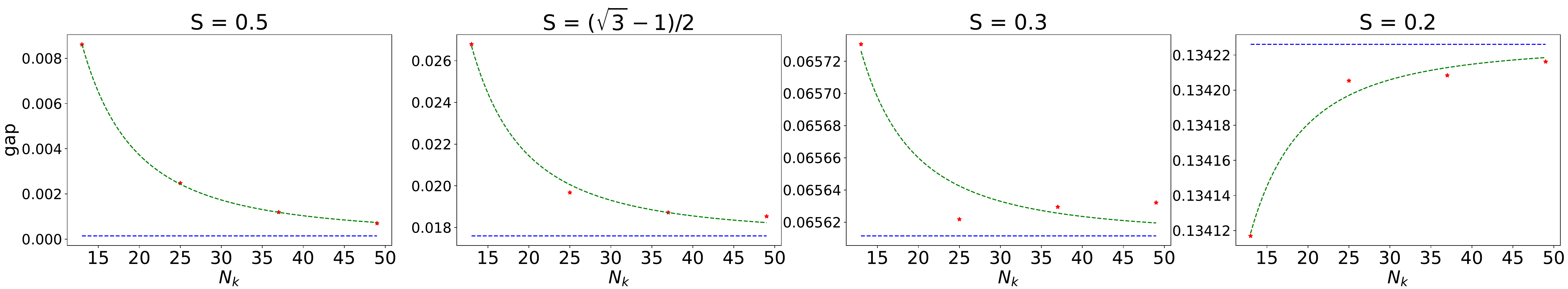}
	\caption{Gap scaling at different spin values for ansatz $20$ solutions at $J_2=J_3=0$. The fit in green is performed using a function of the type $a/N_s+b$ where $N_s$ is the number of sites in the system. In blue is reported the asymptotic value of the fit.}
    \label{fig:gaps_cb1}
\end{figure*}

In \cref{fig:gaps_cb1} we look at the gap scaling for ansatz $20$ for different spin values at $J_2=J_3=0$. The gap values go from being very well described by the fit at $S=0.50$, indicating a LRO, to being very far from it at lower spin values, indicating a SL. 

\section{Spin structure factor} \label{App:ssf}
\subsection{Structure factor using the condensate}  \label{App:ssf_cond}
In the case of LRO we can extract the spin structure factor from the solution of the minimization by looking at the shape of the condensate. The procedure to do so is explained in \cite{Sachdev1992,Wang2006}, here we report it with some additional practical details. 

Once the gap closes, the spinons condense and form a particular orientation of the spins. The information is encoded in the spinon condensate, which we can write as
\begin{equation}\label{eq:cond1}
	\chi(r)=\begin{pmatrix}
		\langle a_r \rangle	\\
		\langle b_r \rangle
	\end{pmatrix}.
\end{equation}
The first think to look at are points in the BZ where the gap closes. These can be degenerate or more than one. Once these have been identified, let us call them $\tilde{k}$, we need to extract the eigenvectors corresponding to the zero-energy bands. In particular, these will be the corresponding columns of $M_{\tilde{k}}$ as defined in \cref{eq:bogo}. Let us call them $\phi_{\tilde{k}}$. The presence of a condensate is evidenced by a non-zero expectation value in the ground state of the wavefunction, i.e.
\begin{equation}    \label{eq:cond2}
	\langle \psi_{\tilde{k}}\rangle = \begin{pmatrix}
	\langle a_{\tilde{k}} \rangle	\\
	\langle b^\dagger_{\tilde{k}} \rangle
	\end{pmatrix}	= 	c\phi_{\tilde{k}}.
\end{equation}
By Fourier transforming the spinons in \cref{eq:cond1} we get
\begin{equation}
	\chi(r)=\begin{pmatrix}
		\langle\sum_ke^{ikr}a_k\rangle	\\
		\langle\sum_ke^{ikr}b_k\rangle
	\end{pmatrix}=\begin{pmatrix}
	\sum_{\tilde{k}}\langle a_{\tilde{k}}\rangle e^{i\tilde{k}r}	\\
	\sum_{\tilde{k}}\langle b_{\tilde{k}}\rangle e^{i\tilde{k}r}	
\end{pmatrix},
\end{equation}
since the expectation value of the spinons on all other $k$'s is zero. The constants $c$ in \cref{eq:cond2} will determine the global spin orientations while leaving the relative angles unchanged. The final value of the spins' directions will be given by 
\begin{equation}
	\vec{S}(r)=\frac{1}{2}\chi^\dagger(r)\vec{\sigma}\chi(r).
\end{equation}
Once the directions of the spins have been determined, we compute the spin structure factor in the usual way as for the classical orders using the definition, \cref{eq:ssf}.

\subsection{Structure factor using the ground state}        \label{App:ssf_gs}
In the case of a SL phase in order to compute the spin structure factor we need to rely on a slightly more involved calculation \cite{Halimeh2016}. We can use the same method as for self consistency and write down the spin structure factor in terms of Bogoliubov spinons. Then we evaluate it on the ground state, which is the vacuum of spinons. 

Let us start by writing the spinons $a$ and $b$ in terms of thew Bogoliubov bosons $\alpha$ and $\beta$
\begin{subequations}
	\begin{align*}
	&\psi_k=\begin{pmatrix}
		a_k	\\
		b^\dagger_{-k}
	\end{pmatrix}=M_k\tilde{\psi}_k=\begin{pmatrix}
	U_k&X_k	\\
	V_k&Y_k
\end{pmatrix}\begin{pmatrix}
\alpha_k	\\
\beta^\dagger_{-k}
\end{pmatrix},		\\
	&\psi^\dagger_k=\begin{pmatrix}
		a^\dagger_k	& b_{-k}
	\end{pmatrix}=\tilde{\psi}^\dagger_kM^\dagger_k=\begin{pmatrix}
	\alpha^\dagger_k	&	\beta_{-k}
\end{pmatrix}\begin{pmatrix}
		U^\dagger_k&V^\dagger_k	\\
		X^\dagger_k&Y^\dagger_k
	\end{pmatrix}.
\end{align*}
\end{subequations}
We need to keep in mind that there is an additional unit cell index ranging from $1$ to $m$, where $m$ is the unit cell size of the ansatz. Let us consider in detail how to derive the $xx$ component of the spin structure factor \cref{eq:ssf}, i.e. just the terms $S_i^xS_j^x$. In fact, for $SU(2)$ invariant models it is enough to compute only one of the three components.
\begin{equation*}
\begin{split}
	\Xi(\vec{Q})^{xx} = \frac{1}{\mathcal{N}}\langle&\sum_{i}\sum_{k,k'}e^{-i\vec{r_i}\cdot\vec{Q}-i\vec{x_i}\cdot(\vec{k}-\vec{k'})}\left(a^\dagger_kb_{k'}+b^\dagger_ka_{k'}\right)  \\
 &\sum_{j}\sum_{t,t'}e^{i\vec{r_j}\cdot\vec{Q}-i\vec{x_i}\cdot(\vec{t}-\vec{t'})}\left(a^\dagger_tb_{t'}+b^\dagger_ta_{t'}\right)\rangle.
 \end{split}
\end{equation*}
Note that the position vector $r_i$ is the \emph{site} position whereas $\vec{x_i}$ in our convention is the ansatz unit cell position. We can decouple $r_i$ into unit cell position plus position of site in the unit cell, which will be denoted by $s_\mu$ ($\mu=1,\dots m$). By summing over the unit cells we get two delta functions of the form $\delta(\vec{Q}+k-k')$ and $\delta(-\vec{Q}+t-t')$ so that
\begin{equation*}
\begin{split}
	\Xi(\vec{Q})^{xx} = \frac{1}{\mathcal{N}}\langle\sum_{\mu,\gamma}\sum_{k,t}&\left(a^\dagger_{\mu,k}b_{\mu,k+Q}+b^\dagger_{\mu,k}a_{\mu,k+Q}\right)     \\
 &\left(a^\dagger_{\gamma,t}b_{\gamma,t-Q}+b^\dagger_{\gamma,t}a_{\gamma,t-Q}\right)\rangle.
 \end{split}
\end{equation*}
where we call $\mu,\gamma$ the unit cell index of sites $i,j$ respectively. Now, we can substitute Schwinger bosons $a,b$ with Bogoliubov bosons $\alpha,\beta$. We know these will respect the same commutation relations because of the construction of $M_k$ as detailed in \cref{sec:SBMFT}. By using the fact that the ground state is the vacuum of Bogoliubov bosons we arrive at the final form
\onecolumngrid
\vspace{\columnsep}
\begin{equation*}
\begin{split}
    \Xi(\vec{Q})^{xx}=\frac{1}{\mathcal{N}}\sum_{k,\mu,\gamma}e^{i\vec{Q}\cdot(s_\gamma-s_\mu)} 
    \Big[ &X_{\mu\nu}^*(k)Y_{\mu,\nu'}^*(-k-Q)\big(X_{\gamma\nu}(k)Y_{\gamma\nu'}(-k-Q)+ 
    Y_{\gamma\nu}(k)X_{\gamma\nu'}(-k-Q)\big)       \\
    &+V_{\mu\nu}(-k)U_{\mu\nu'}(k+Q)\big(V_{\gamma\nu}^*(-k)U_{\gamma\nu'}^*(k+Q) + 
    U_{\gamma\nu}^*(-k)V_{\gamma\nu'}^*(k+Q)\big)\Big].
    \end{split}
\end{equation*}
Similar expressions can be derived for $\Xi(\vec{Q})^{yy}$ and $\Xi(\vec{Q})^{zz}$.
\twocolumngrid

%


\begin{thebibliography}{102}%
\makeatletter
\providecommand \@ifxundefined [1]{%
 \@ifx{#1\undefined}
}%
\providecommand \@ifnum [1]{%
 \ifnum #1\expandafter \@firstoftwo
 \else \expandafter \@secondoftwo
 \fi
}%
\providecommand \@ifx [1]{%
 \ifx #1\expandafter \@firstoftwo
 \else \expandafter \@secondoftwo
 \fi
}%
\providecommand \natexlab [1]{#1}%
\providecommand \enquote  [1]{``#1''}%
\providecommand \bibnamefont  [1]{#1}%
\providecommand \bibfnamefont [1]{#1}%
\providecommand \citenamefont [1]{#1}%
\providecommand \href@noop [0]{\@secondoftwo}%
\providecommand \href [0]{\begingroup \@sanitize@url \@href}%
\providecommand \@href[1]{\@@startlink{#1}\@@href}%
\providecommand \@@href[1]{\endgroup#1\@@endlink}%
\providecommand \@sanitize@url [0]{\catcode `\\12\catcode `\$12\catcode
  `\&12\catcode `\#12\catcode `\^12\catcode `\_12\catcode `\%12\relax}%
\providecommand \@@startlink[1]{}%
\providecommand \@@endlink[0]{}%
\providecommand \url  [0]{\begingroup\@sanitize@url \@url }%
\providecommand \@url [1]{\endgroup\@href {#1}{\urlprefix }}%
\providecommand \urlprefix  [0]{URL }%
\providecommand \Eprint [0]{\href }%
\providecommand \doibase [0]{https://doi.org/}%
\providecommand \selectlanguage [0]{\@gobble}%
\providecommand \bibinfo  [0]{\@secondoftwo}%
\providecommand \bibfield  [0]{\@secondoftwo}%
\providecommand \translation [1]{[#1]}%
\providecommand \BibitemOpen [0]{}%
\providecommand \bibitemStop [0]{}%
\providecommand \bibitemNoStop [0]{.\EOS\space}%
\providecommand \EOS [0]{\spacefactor3000\relax}%
\providecommand \BibitemShut  [1]{\csname bibitem#1\endcsname}%
\let\auto@bib@innerbib\@empty
\bibitem [{\citenamefont {Norman}(2016)}]{Norman2016}%
  \BibitemOpen
  \bibfield  {author} {\bibinfo {author} {\bibfnamefont {M.~R.}\ \bibnamefont
  {Norman}},\ }\bibfield  {title} {\bibinfo {title} {Colloquium:
  Herbertsmithite and the search for the quantum spin liquid},\ }\href
  {https://doi.org/10.1103/RevModPhys.88.041002} {\bibfield  {journal}
  {\bibinfo  {journal} {Rev. Mod. Phys.}\ }\textbf {\bibinfo {volume} {88}},\
  \bibinfo {pages} {041002} (\bibinfo {year} {2016})}\BibitemShut {NoStop}%
\bibitem [{\citenamefont {Mendels}\ and\ \citenamefont
  {Bert}(2016)}]{Mendels2016}%
  \BibitemOpen
  \bibfield  {author} {\bibinfo {author} {\bibfnamefont {P.}~\bibnamefont
  {Mendels}}\ and\ \bibinfo {author} {\bibfnamefont {F.}~\bibnamefont {Bert}},\
  }\bibfield  {title} {\bibinfo {title} {Quantum kagome frustrated
  antiferromagnets: One route to quantum spin liquids},\ }\href
  {https://doi.org/https://doi.org/10.1016/j.crhy.2015.12.001} {\bibfield
  {journal} {\bibinfo  {journal} {Comptes Rendus Physique}\ }\textbf {\bibinfo
  {volume} {17}},\ \bibinfo {pages} {455} (\bibinfo {year} {2016})}\BibitemShut
  {NoStop}%
\bibitem [{\citenamefont {Savary}\ and\ \citenamefont
  {Balents}(2016)}]{Savary2016}%
  \BibitemOpen
  \bibfield  {author} {\bibinfo {author} {\bibfnamefont {L.}~\bibnamefont
  {Savary}}\ and\ \bibinfo {author} {\bibfnamefont {L.}~\bibnamefont
  {Balents}},\ }\bibfield  {title} {\bibinfo {title} {Quantum spin liquids: a
  review},\ }\href {https://doi.org/10.1088/0034-4885/80/1/016502} {\bibfield
  {journal} {\bibinfo  {journal} {Reports on Progress in Physics}\ }\textbf
  {\bibinfo {volume} {80}},\ \bibinfo {pages} {016502} (\bibinfo {year}
  {2016})}\BibitemShut {NoStop}%
\bibitem [{\citenamefont {Knolle}\ and\ \citenamefont
  {Moessner}(2019)}]{Knolle2019}%
  \BibitemOpen
  \bibfield  {author} {\bibinfo {author} {\bibfnamefont {J.}~\bibnamefont
  {Knolle}}\ and\ \bibinfo {author} {\bibfnamefont {R.}~\bibnamefont
  {Moessner}},\ }\bibfield  {title} {\bibinfo {title} {{A Field Guide to Spin
  Liquids}},\ }\href {https://doi.org/10.1146/annurev-conmatphys-031218-013401}
  {\bibfield  {journal} {\bibinfo  {journal} {Annual Review of Condensed Matter
  Physics}\ }\textbf {\bibinfo {volume} {10}},\ \bibinfo {pages} {451}
  (\bibinfo {year} {2019})}\BibitemShut {NoStop}%
\bibitem [{\citenamefont {Shores}\ \emph {et~al.}(2005)\citenamefont {Shores},
  \citenamefont {Nytko}, \citenamefont {Bartlett},\ and\ \citenamefont
  {Nocera}}]{Shores2005}%
  \BibitemOpen
  \bibfield  {author} {\bibinfo {author} {\bibfnamefont {M.~P.}\ \bibnamefont
  {Shores}}, \bibinfo {author} {\bibfnamefont {E.~A.}\ \bibnamefont {Nytko}},
  \bibinfo {author} {\bibfnamefont {B.~M.}\ \bibnamefont {Bartlett}},\ and\
  \bibinfo {author} {\bibfnamefont {D.~G.}\ \bibnamefont {Nocera}},\ }\bibfield
   {title} {\bibinfo {title} {{A Structurally Perfect S = 1/2 Kagomé
  Antiferromagnet}},\ }\href {https://doi.org/10.1021/ja053891p} {\bibfield
  {journal} {\bibinfo  {journal} {Journal of the American Chemical Society}\
  }\textbf {\bibinfo {volume} {127}},\ \bibinfo {pages} {13462} (\bibinfo
  {year} {2005})}\BibitemShut {NoStop}%
\bibitem [{\citenamefont {Helton}\ \emph {et~al.}(2007)\citenamefont {Helton},
  \citenamefont {Matan}, \citenamefont {Shores}, \citenamefont {Nytko},
  \citenamefont {Bartlett}, \citenamefont {Yoshida}, \citenamefont {Takano},
  \citenamefont {Suslov}, \citenamefont {Qiu}, \citenamefont {Chung},
  \citenamefont {Nocera},\ and\ \citenamefont {Lee}}]{Helton2007}%
  \BibitemOpen
  \bibfield  {author} {\bibinfo {author} {\bibfnamefont {J.~S.}\ \bibnamefont
  {Helton}}, \bibinfo {author} {\bibfnamefont {K.}~\bibnamefont {Matan}},
  \bibinfo {author} {\bibfnamefont {M.~P.}\ \bibnamefont {Shores}}, \bibinfo
  {author} {\bibfnamefont {E.~A.}\ \bibnamefont {Nytko}}, \bibinfo {author}
  {\bibfnamefont {B.~M.}\ \bibnamefont {Bartlett}}, \bibinfo {author}
  {\bibfnamefont {Y.}~\bibnamefont {Yoshida}}, \bibinfo {author} {\bibfnamefont
  {Y.}~\bibnamefont {Takano}}, \bibinfo {author} {\bibfnamefont
  {A.}~\bibnamefont {Suslov}}, \bibinfo {author} {\bibfnamefont
  {Y.}~\bibnamefont {Qiu}}, \bibinfo {author} {\bibfnamefont {J.-H.}\
  \bibnamefont {Chung}}, \bibinfo {author} {\bibfnamefont {D.~G.}\ \bibnamefont
  {Nocera}},\ and\ \bibinfo {author} {\bibfnamefont {Y.~S.}\ \bibnamefont
  {Lee}},\ }\bibfield  {title} {\bibinfo {title} {{Spin Dynamics of the
  Spin-$1/2$ Kagome Lattice Antiferromagnet
  ${\mathrm{ZnCu}}_{3}(\mathrm{OH}{)}_{6}{\mathrm{Cl}}_{2}$}},\ }\href
  {https://doi.org/10.1103/PhysRevLett.98.107204} {\bibfield  {journal}
  {\bibinfo  {journal} {Phys. Rev. Lett.}\ }\textbf {\bibinfo {volume} {98}},\
  \bibinfo {pages} {107204} (\bibinfo {year} {2007})}\BibitemShut {NoStop}%
\bibitem [{\citenamefont {Bert}\ \emph {et~al.}(2007)\citenamefont {Bert},
  \citenamefont {Nakamae}, \citenamefont {Ladieu}, \citenamefont {L'H\^ote},
  \citenamefont {Bonville}, \citenamefont {Duc}, \citenamefont {Trombe},\ and\
  \citenamefont {Mendels}}]{Bert2007}%
  \BibitemOpen
  \bibfield  {author} {\bibinfo {author} {\bibfnamefont {F.}~\bibnamefont
  {Bert}}, \bibinfo {author} {\bibfnamefont {S.}~\bibnamefont {Nakamae}},
  \bibinfo {author} {\bibfnamefont {F.}~\bibnamefont {Ladieu}}, \bibinfo
  {author} {\bibfnamefont {D.}~\bibnamefont {L'H\^ote}}, \bibinfo {author}
  {\bibfnamefont {P.}~\bibnamefont {Bonville}}, \bibinfo {author}
  {\bibfnamefont {F.}~\bibnamefont {Duc}}, \bibinfo {author} {\bibfnamefont
  {J.-C.}\ \bibnamefont {Trombe}},\ and\ \bibinfo {author} {\bibfnamefont
  {P.}~\bibnamefont {Mendels}},\ }\bibfield  {title} {\bibinfo {title} {{Low
  temperature magnetization of the $S=\frac{1}{2}$ kagome antiferromagnet
  $\mathrm{Zn}{\mathrm{Cu}}_{3}{(\mathrm{O}\mathrm{H})}_{6}{\mathrm{Cl}}_{2}$}},\
  }\href {https://doi.org/10.1103/PhysRevB.76.132411} {\bibfield  {journal}
  {\bibinfo  {journal} {Phys. Rev. B}\ }\textbf {\bibinfo {volume} {76}},\
  \bibinfo {pages} {132411} (\bibinfo {year} {2007})}\BibitemShut {NoStop}%
\bibitem [{\citenamefont {Mendels}\ \emph {et~al.}(2007)\citenamefont
  {Mendels}, \citenamefont {Bert}, \citenamefont {de~Vries}, \citenamefont
  {Olariu}, \citenamefont {Harrison}, \citenamefont {Duc}, \citenamefont
  {Trombe}, \citenamefont {Lord}, \citenamefont {Amato},\ and\ \citenamefont
  {Baines}}]{Mendels2007}%
  \BibitemOpen
  \bibfield  {author} {\bibinfo {author} {\bibfnamefont {P.}~\bibnamefont
  {Mendels}}, \bibinfo {author} {\bibfnamefont {F.}~\bibnamefont {Bert}},
  \bibinfo {author} {\bibfnamefont {M.~A.}\ \bibnamefont {de~Vries}}, \bibinfo
  {author} {\bibfnamefont {A.}~\bibnamefont {Olariu}}, \bibinfo {author}
  {\bibfnamefont {A.}~\bibnamefont {Harrison}}, \bibinfo {author}
  {\bibfnamefont {F.}~\bibnamefont {Duc}}, \bibinfo {author} {\bibfnamefont
  {J.~C.}\ \bibnamefont {Trombe}}, \bibinfo {author} {\bibfnamefont {J.~S.}\
  \bibnamefont {Lord}}, \bibinfo {author} {\bibfnamefont {A.}~\bibnamefont
  {Amato}},\ and\ \bibinfo {author} {\bibfnamefont {C.}~\bibnamefont
  {Baines}},\ }\bibfield  {title} {\bibinfo {title} {{Quantum Magnetism in the
  Paratacamite Family: Towards an Ideal Kagom\'e Lattice}},\ }\href
  {https://doi.org/10.1103/PhysRevLett.98.077204} {\bibfield  {journal}
  {\bibinfo  {journal} {Phys. Rev. Lett.}\ }\textbf {\bibinfo {volume} {98}},\
  \bibinfo {pages} {077204} (\bibinfo {year} {2007})}\BibitemShut {NoStop}%
\bibitem [{\citenamefont {Khuntia}\ \emph {et~al.}(2020)\citenamefont
  {Khuntia}, \citenamefont {Velazquez}, \citenamefont {Barth{\'e}lemy},
  \citenamefont {Bert}, \citenamefont {Kermarrec}, \citenamefont {Legros},
  \citenamefont {Bernu}, \citenamefont {Messio}, \citenamefont {Zorko},\ and\
  \citenamefont {Mendels}}]{Khuntia2020}%
  \BibitemOpen
  \bibfield  {author} {\bibinfo {author} {\bibfnamefont {P.}~\bibnamefont
  {Khuntia}}, \bibinfo {author} {\bibfnamefont {M.}~\bibnamefont {Velazquez}},
  \bibinfo {author} {\bibfnamefont {Q.}~\bibnamefont {Barth{\'e}lemy}},
  \bibinfo {author} {\bibfnamefont {F.}~\bibnamefont {Bert}}, \bibinfo {author}
  {\bibfnamefont {E.}~\bibnamefont {Kermarrec}}, \bibinfo {author}
  {\bibfnamefont {A.}~\bibnamefont {Legros}}, \bibinfo {author} {\bibfnamefont
  {B.}~\bibnamefont {Bernu}}, \bibinfo {author} {\bibfnamefont
  {L.}~\bibnamefont {Messio}}, \bibinfo {author} {\bibfnamefont
  {A.}~\bibnamefont {Zorko}},\ and\ \bibinfo {author} {\bibfnamefont
  {P.}~\bibnamefont {Mendels}},\ }\bibfield  {title} {\bibinfo {title}
  {{Gapless ground state in the archetypal quantum kagome antiferromagnet
  ZnCu$_3$(OH)$_6$Cl$_2$}},\ }\href {https://doi.org/10.1038/s41567-020-0792-1}
  {\bibfield  {journal} {\bibinfo  {journal} {Nature Physics}\ }\textbf
  {\bibinfo {volume} {16}},\ \bibinfo {pages} {469} (\bibinfo {year}
  {2020})}\BibitemShut {NoStop}%
\bibitem [{\citenamefont {Aidoudi}\ \emph {et~al.}(2011)\citenamefont
  {Aidoudi}, \citenamefont {Aldous}, \citenamefont {Goff}, \citenamefont
  {Slawin}, \citenamefont {Attfield}, \citenamefont {Morris},\ and\
  \citenamefont {Lightfoot}}]{Aidoudi2011}%
  \BibitemOpen
  \bibfield  {author} {\bibinfo {author} {\bibfnamefont {F.~H.}\ \bibnamefont
  {Aidoudi}}, \bibinfo {author} {\bibfnamefont {D.~W.}\ \bibnamefont {Aldous}},
  \bibinfo {author} {\bibfnamefont {R.~J.}\ \bibnamefont {Goff}}, \bibinfo
  {author} {\bibfnamefont {A.~M.~Z.}\ \bibnamefont {Slawin}}, \bibinfo {author}
  {\bibfnamefont {J.~P.}\ \bibnamefont {Attfield}}, \bibinfo {author}
  {\bibfnamefont {R.~E.}\ \bibnamefont {Morris}},\ and\ \bibinfo {author}
  {\bibfnamefont {P.}~\bibnamefont {Lightfoot}},\ }\bibfield  {title} {\bibinfo
  {title} {{An ionothermally prepared $S=1/2$ vanadium oxyfluoride kagome
  lattice}},\ }\href {https://doi.org/10.1038/nchem.1129} {\bibfield  {journal}
  {\bibinfo  {journal} {Nature Chemistry}\ }\textbf {\bibinfo {volume} {3}},\
  \bibinfo {pages} {801} (\bibinfo {year} {2011})}\BibitemShut {NoStop}%
\bibitem [{\citenamefont {F\aa{}k}\ \emph {et~al.}(2012)\citenamefont
  {F\aa{}k}, \citenamefont {Kermarrec}, \citenamefont {Messio}, \citenamefont
  {Bernu}, \citenamefont {Lhuillier}, \citenamefont {Bert}, \citenamefont
  {Mendels}, \citenamefont {Koteswararao}, \citenamefont {Bouquet},
  \citenamefont {Ollivier}, \citenamefont {Hillier}, \citenamefont {Amato},
  \citenamefont {Colman},\ and\ \citenamefont {Wills}}]{Faak2012}%
  \BibitemOpen
  \bibfield  {author} {\bibinfo {author} {\bibfnamefont {B.}~\bibnamefont
  {F\aa{}k}}, \bibinfo {author} {\bibfnamefont {E.}~\bibnamefont {Kermarrec}},
  \bibinfo {author} {\bibfnamefont {L.}~\bibnamefont {Messio}}, \bibinfo
  {author} {\bibfnamefont {B.}~\bibnamefont {Bernu}}, \bibinfo {author}
  {\bibfnamefont {C.}~\bibnamefont {Lhuillier}}, \bibinfo {author}
  {\bibfnamefont {F.}~\bibnamefont {Bert}}, \bibinfo {author} {\bibfnamefont
  {P.}~\bibnamefont {Mendels}}, \bibinfo {author} {\bibfnamefont
  {B.}~\bibnamefont {Koteswararao}}, \bibinfo {author} {\bibfnamefont
  {F.}~\bibnamefont {Bouquet}}, \bibinfo {author} {\bibfnamefont
  {J.}~\bibnamefont {Ollivier}}, \bibinfo {author} {\bibfnamefont {A.~D.}\
  \bibnamefont {Hillier}}, \bibinfo {author} {\bibfnamefont {A.}~\bibnamefont
  {Amato}}, \bibinfo {author} {\bibfnamefont {R.~H.}\ \bibnamefont {Colman}},\
  and\ \bibinfo {author} {\bibfnamefont {A.~S.}\ \bibnamefont {Wills}},\
  }\bibfield  {title} {\bibinfo {title} {{Kapellasite: A Kagome Quantum Spin
  Liquid with Competing Interactions}},\ }\href
  {https://doi.org/10.1103/PhysRevLett.109.037208} {\bibfield  {journal}
  {\bibinfo  {journal} {Phys. Rev. Lett.}\ }\textbf {\bibinfo {volume} {109}},\
  \bibinfo {pages} {037208} (\bibinfo {year} {2012})}\BibitemShut {NoStop}%
\bibitem [{\citenamefont {Clark}\ \emph
  {et~al.}(2013{\natexlab{a}})\citenamefont {Clark}, \citenamefont {Orain},
  \citenamefont {Bert}, \citenamefont {De~Vries}, \citenamefont {Aidoudi},
  \citenamefont {Morris}, \citenamefont {Lightfoot}, \citenamefont {Lord},
  \citenamefont {Telling}, \citenamefont {Bonville}, \citenamefont {Attfield},
  \citenamefont {Mendels},\ and\ \citenamefont {Harrison}}]{Clark2013a}%
  \BibitemOpen
  \bibfield  {author} {\bibinfo {author} {\bibfnamefont {L.}~\bibnamefont
  {Clark}}, \bibinfo {author} {\bibfnamefont {J.~C.}\ \bibnamefont {Orain}},
  \bibinfo {author} {\bibfnamefont {F.}~\bibnamefont {Bert}}, \bibinfo {author}
  {\bibfnamefont {M.~A.}\ \bibnamefont {De~Vries}}, \bibinfo {author}
  {\bibfnamefont {F.~H.}\ \bibnamefont {Aidoudi}}, \bibinfo {author}
  {\bibfnamefont {R.~E.}\ \bibnamefont {Morris}}, \bibinfo {author}
  {\bibfnamefont {P.}~\bibnamefont {Lightfoot}}, \bibinfo {author}
  {\bibfnamefont {J.~S.}\ \bibnamefont {Lord}}, \bibinfo {author}
  {\bibfnamefont {M.~T.~F.}\ \bibnamefont {Telling}}, \bibinfo {author}
  {\bibfnamefont {P.}~\bibnamefont {Bonville}}, \bibinfo {author}
  {\bibfnamefont {J.~P.}\ \bibnamefont {Attfield}}, \bibinfo {author}
  {\bibfnamefont {P.}~\bibnamefont {Mendels}},\ and\ \bibinfo {author}
  {\bibfnamefont {A.}~\bibnamefont {Harrison}},\ }\bibfield  {title} {\bibinfo
  {title} {{Gapless Spin Liquid Ground State in the $S\mathbf{=}1/2$ Vanadium
  Oxyfluoride Kagome Antiferromagnet
  $[{\mathrm{NH}}_{4}{]}_{2}[{\mathbf{C}}_{7}{\mathbf{H}}_{14}\mathbf{N}][{\mathbf{V}}_{7}{\mathbf{O}}_{6}{\mathbf{F}}_{18}]$}},\
  }\href {https://doi.org/10.1103/PhysRevLett.110.207208} {\bibfield  {journal}
  {\bibinfo  {journal} {Phys. Rev. Lett.}\ }\textbf {\bibinfo {volume} {110}},\
  \bibinfo {pages} {207208} (\bibinfo {year} {2013}{\natexlab{a}})}\BibitemShut
  {NoStop}%
\bibitem [{\citenamefont {Chen}\ \emph {et~al.}(2020)\citenamefont {Chen},
  \citenamefont {Huang}, \citenamefont {Pan},\ and\ \citenamefont
  {Mi}}]{Chen2020}%
  \BibitemOpen
  \bibfield  {author} {\bibinfo {author} {\bibfnamefont {X.-H.}\ \bibnamefont
  {Chen}}, \bibinfo {author} {\bibfnamefont {Y.-X.}\ \bibnamefont {Huang}},
  \bibinfo {author} {\bibfnamefont {Y.}~\bibnamefont {Pan}},\ and\ \bibinfo
  {author} {\bibfnamefont {J.-X.}\ \bibnamefont {Mi}},\ }\bibfield  {title}
  {\bibinfo {title} {{Quantum spin liquid candidate
  YCu$_3$(OH)$_6$Br$_2$[Br$_x$(OH)$_{1-x}$] (x$\simeq$0.51): With an almost
  perfect kagomé layer}},\ }\href
  {https://doi.org/https://doi.org/10.1016/j.jmmm.2020.167066} {\bibfield
  {journal} {\bibinfo  {journal} {Journal of Magnetism and Magnetic Materials}\
  }\textbf {\bibinfo {volume} {512}},\ \bibinfo {pages} {167066} (\bibinfo
  {year} {2020})}\BibitemShut {NoStop}%
\bibitem [{\citenamefont {Zeng}\ \emph {et~al.}(2022)\citenamefont {Zeng},
  \citenamefont {Ma}, \citenamefont {Wu}, \citenamefont {Li}, \citenamefont
  {Tao}, \citenamefont {Lu}, \citenamefont {Chen}, \citenamefont {Mi},
  \citenamefont {Song}, \citenamefont {Cao}, \citenamefont {Che}, \citenamefont
  {Li}, \citenamefont {Li}, \citenamefont {Luo}, \citenamefont {Meng},\ and\
  \citenamefont {Li}}]{Zeng2022}%
  \BibitemOpen
  \bibfield  {author} {\bibinfo {author} {\bibfnamefont {Z.}~\bibnamefont
  {Zeng}}, \bibinfo {author} {\bibfnamefont {X.}~\bibnamefont {Ma}}, \bibinfo
  {author} {\bibfnamefont {S.}~\bibnamefont {Wu}}, \bibinfo {author}
  {\bibfnamefont {H.-F.}\ \bibnamefont {Li}}, \bibinfo {author} {\bibfnamefont
  {Z.}~\bibnamefont {Tao}}, \bibinfo {author} {\bibfnamefont {X.}~\bibnamefont
  {Lu}}, \bibinfo {author} {\bibfnamefont {X.-h.}\ \bibnamefont {Chen}},
  \bibinfo {author} {\bibfnamefont {J.-X.}\ \bibnamefont {Mi}}, \bibinfo
  {author} {\bibfnamefont {S.-J.}\ \bibnamefont {Song}}, \bibinfo {author}
  {\bibfnamefont {G.-H.}\ \bibnamefont {Cao}}, \bibinfo {author} {\bibfnamefont
  {G.}~\bibnamefont {Che}}, \bibinfo {author} {\bibfnamefont {K.}~\bibnamefont
  {Li}}, \bibinfo {author} {\bibfnamefont {G.}~\bibnamefont {Li}}, \bibinfo
  {author} {\bibfnamefont {H.}~\bibnamefont {Luo}}, \bibinfo {author}
  {\bibfnamefont {Z.~Y.}\ \bibnamefont {Meng}},\ and\ \bibinfo {author}
  {\bibfnamefont {S.}~\bibnamefont {Li}},\ }\bibfield  {title} {\bibinfo
  {title} {{Possible Dirac quantum spin liquid in the kagome quantum
  antiferromagnet
  ${\mathrm{YCu}}_{3}{(\mathrm{OH})}_{6}{\mathrm{Br}}_{2}[{\mathrm{Br}}_{x}{(\mathrm{OH})}_{1\ensuremath{-}x}]$}},\
  }\href {https://doi.org/10.1103/PhysRevB.105.L121109} {\bibfield  {journal}
  {\bibinfo  {journal} {Phys. Rev. B}\ }\textbf {\bibinfo {volume} {105}},\
  \bibinfo {pages} {L121109} (\bibinfo {year} {2022})}\BibitemShut {NoStop}%
\bibitem [{\citenamefont {Liu}\ \emph {et~al.}(2022)\citenamefont {Liu},
  \citenamefont {Yuan}, \citenamefont {Li}, \citenamefont {Li}, \citenamefont
  {Zhao}, \citenamefont {Liao},\ and\ \citenamefont {Li}}]{Liu2022}%
  \BibitemOpen
  \bibfield  {author} {\bibinfo {author} {\bibfnamefont {J.}~\bibnamefont
  {Liu}}, \bibinfo {author} {\bibfnamefont {L.}~\bibnamefont {Yuan}}, \bibinfo
  {author} {\bibfnamefont {X.}~\bibnamefont {Li}}, \bibinfo {author}
  {\bibfnamefont {B.}~\bibnamefont {Li}}, \bibinfo {author} {\bibfnamefont
  {K.}~\bibnamefont {Zhao}}, \bibinfo {author} {\bibfnamefont {H.}~\bibnamefont
  {Liao}},\ and\ \bibinfo {author} {\bibfnamefont {Y.}~\bibnamefont {Li}},\
  }\bibfield  {title} {\bibinfo {title} {Gapless spin liquid behavior in a
  kagome {Heisenberg} antiferromagnet with randomly distributed hexagons of
  alternate bonds},\ }\href {https://doi.org/10.1103/PhysRevB.105.024418}
  {\bibfield  {journal} {\bibinfo  {journal} {Phys. Rev. B}\ }\textbf {\bibinfo
  {volume} {105}},\ \bibinfo {pages} {024418} (\bibinfo {year}
  {2022})}\BibitemShut {NoStop}%
\bibitem [{\citenamefont {Lu}\ \emph {et~al.}(2022)\citenamefont {Lu},
  \citenamefont {Yuan}, \citenamefont {Zhang}, \citenamefont {Li},
  \citenamefont {Luo},\ and\ \citenamefont {Li}}]{Lu2022}%
  \BibitemOpen
  \bibfield  {author} {\bibinfo {author} {\bibfnamefont {F.}~\bibnamefont
  {Lu}}, \bibinfo {author} {\bibfnamefont {L.}~\bibnamefont {Yuan}}, \bibinfo
  {author} {\bibfnamefont {J.}~\bibnamefont {Zhang}}, \bibinfo {author}
  {\bibfnamefont {B.}~\bibnamefont {Li}}, \bibinfo {author} {\bibfnamefont
  {Y.}~\bibnamefont {Luo}},\ and\ \bibinfo {author} {\bibfnamefont
  {Y.}~\bibnamefont {Li}},\ }\bibfield  {title} {\bibinfo {title} {{The
  observation of quantum fluctuations in a kagome Heisenberg
  antiferromagnet}},\ }\href {https://doi.org/10.1038/s42005-022-01053-4}
  {\bibfield  {journal} {\bibinfo  {journal} {Communications Physics}\ }\textbf
  {\bibinfo {volume} {5}},\ \bibinfo {pages} {272} (\bibinfo {year}
  {2022})}\BibitemShut {NoStop}%
\bibitem [{\citenamefont {Sachdev}(1992)}]{Sachdev1992}%
  \BibitemOpen
  \bibfield  {author} {\bibinfo {author} {\bibfnamefont {S.}~\bibnamefont
  {Sachdev}},\ }\bibfield  {title} {\bibinfo {title}
  {{Kagome\ifmmode\acute\else\textasciiacute\fi{}- and triangular-lattice
  Heisenberg antiferromagnets: Ordering from quantum fluctuations and
  quantum-disordered ground states with unconfined bosonic spinons}},\ }\href
  {https://doi.org/10.1103/PhysRevB.45.12377} {\bibfield  {journal} {\bibinfo
  {journal} {Phys. Rev. B}\ }\textbf {\bibinfo {volume} {45}},\ \bibinfo
  {pages} {12377} (\bibinfo {year} {1992})}\BibitemShut {NoStop}%
\bibitem [{\citenamefont {Ran}\ \emph {et~al.}(2007)\citenamefont {Ran},
  \citenamefont {Hermele}, \citenamefont {Lee},\ and\ \citenamefont
  {Wen}}]{Ran2007}%
  \BibitemOpen
  \bibfield  {author} {\bibinfo {author} {\bibfnamefont {Y.}~\bibnamefont
  {Ran}}, \bibinfo {author} {\bibfnamefont {M.}~\bibnamefont {Hermele}},
  \bibinfo {author} {\bibfnamefont {P.~A.}\ \bibnamefont {Lee}},\ and\ \bibinfo
  {author} {\bibfnamefont {X.-G.}\ \bibnamefont {Wen}},\ }\bibfield  {title}
  {\bibinfo {title} {{Projected-Wave-Function Study of the Spin-$1/2$
  Heisenberg Model on the Kagom\'e Lattice}},\ }\href
  {https://doi.org/10.1103/PhysRevLett.98.117205} {\bibfield  {journal}
  {\bibinfo  {journal} {Phys. Rev. Lett.}\ }\textbf {\bibinfo {volume} {98}},\
  \bibinfo {pages} {117205} (\bibinfo {year} {2007})}\BibitemShut {NoStop}%
\bibitem [{\citenamefont {Hermele}\ \emph {et~al.}(2008)\citenamefont
  {Hermele}, \citenamefont {Ran}, \citenamefont {Lee},\ and\ \citenamefont
  {Wen}}]{Hermele2008}%
  \BibitemOpen
  \bibfield  {author} {\bibinfo {author} {\bibfnamefont {M.}~\bibnamefont
  {Hermele}}, \bibinfo {author} {\bibfnamefont {Y.}~\bibnamefont {Ran}},
  \bibinfo {author} {\bibfnamefont {P.~A.}\ \bibnamefont {Lee}},\ and\ \bibinfo
  {author} {\bibfnamefont {X.-G.}\ \bibnamefont {Wen}},\ }\bibfield  {title}
  {\bibinfo {title} {Properties of an algebraic spin liquid on the kagome
  lattice},\ }\href {https://doi.org/10.1103/PhysRevB.77.224413} {\bibfield
  {journal} {\bibinfo  {journal} {Phys. Rev. B}\ }\textbf {\bibinfo {volume}
  {77}},\ \bibinfo {pages} {224413} (\bibinfo {year} {2008})}\BibitemShut
  {NoStop}%
\bibitem [{\citenamefont {Lu}\ \emph {et~al.}(2011)\citenamefont {Lu},
  \citenamefont {Ran},\ and\ \citenamefont {Lee}}]{Lu2011}%
  \BibitemOpen
  \bibfield  {author} {\bibinfo {author} {\bibfnamefont {Y.-M.}\ \bibnamefont
  {Lu}}, \bibinfo {author} {\bibfnamefont {Y.}~\bibnamefont {Ran}},\ and\
  \bibinfo {author} {\bibfnamefont {P.~A.}\ \bibnamefont {Lee}},\ }\bibfield
  {title} {\bibinfo {title} {{${\mathbb{Z}}_{2}$ spin liquids in the
  $S=\frac{1}{2}$ Heisenberg model on the kagome lattice: A projective
  symmetry-group study of Schwinger fermion mean-field states}},\ }\href
  {https://doi.org/10.1103/PhysRevB.83.224413} {\bibfield  {journal} {\bibinfo
  {journal} {Phys. Rev. B}\ }\textbf {\bibinfo {volume} {83}},\ \bibinfo
  {pages} {224413} (\bibinfo {year} {2011})}\BibitemShut {NoStop}%
\bibitem [{\citenamefont {Messio}\ \emph {et~al.}(2012)\citenamefont {Messio},
  \citenamefont {Bernu},\ and\ \citenamefont {Lhuillier}}]{Messio2012}%
  \BibitemOpen
  \bibfield  {author} {\bibinfo {author} {\bibfnamefont {L.}~\bibnamefont
  {Messio}}, \bibinfo {author} {\bibfnamefont {B.}~\bibnamefont {Bernu}},\ and\
  \bibinfo {author} {\bibfnamefont {C.}~\bibnamefont {Lhuillier}},\ }\bibfield
  {title} {\bibinfo {title} {{Kagome Antiferromagnet: A Chiral Topological Spin
  Liquid?}},\ }\href {https://doi.org/10.1103/PhysRevLett.108.207204}
  {\bibfield  {journal} {\bibinfo  {journal} {Phys. Rev. Lett.}\ }\textbf
  {\bibinfo {volume} {108}},\ \bibinfo {pages} {207204} (\bibinfo {year}
  {2012})}\BibitemShut {NoStop}%
\bibitem [{\citenamefont {Tay}\ and\ \citenamefont
  {Motrunich}(2011)}]{Tay2011}%
  \BibitemOpen
  \bibfield  {author} {\bibinfo {author} {\bibfnamefont {T.}~\bibnamefont
  {Tay}}\ and\ \bibinfo {author} {\bibfnamefont {O.~I.}\ \bibnamefont
  {Motrunich}},\ }\bibfield  {title} {\bibinfo {title} {{Variational study of
  ${J}_{1}$-${J}_{2}$ Heisenberg model on kagome lattice using projected
  Schwinger-boson wave functions}},\ }\href
  {https://doi.org/10.1103/PhysRevB.84.020404} {\bibfield  {journal} {\bibinfo
  {journal} {Phys. Rev. B}\ }\textbf {\bibinfo {volume} {84}},\ \bibinfo
  {pages} {020404(R)} (\bibinfo {year} {2011})}\BibitemShut {NoStop}%
\bibitem [{\citenamefont {Iqbal}\ \emph
  {et~al.}(2011{\natexlab{a}})\citenamefont {Iqbal}, \citenamefont {Becca},\
  and\ \citenamefont {Poilblanc}}]{Iqbal2011b}%
  \BibitemOpen
  \bibfield  {author} {\bibinfo {author} {\bibfnamefont {Y.}~\bibnamefont
  {Iqbal}}, \bibinfo {author} {\bibfnamefont {F.}~\bibnamefont {Becca}},\ and\
  \bibinfo {author} {\bibfnamefont {D.}~\bibnamefont {Poilblanc}},\ }\bibfield
  {title} {\bibinfo {title} {{Projected wave function study of
  ${\mathbb{Z}}_{2}$ spin liquids on the kagome lattice for the
  spin-$\frac{1}{2}$ quantum Heisenberg antiferromagnet}},\ }\href
  {https://doi.org/10.1103/PhysRevB.84.020407} {\bibfield  {journal} {\bibinfo
  {journal} {Phys. Rev. B}\ }\textbf {\bibinfo {volume} {84}},\ \bibinfo
  {pages} {020407(R)} (\bibinfo {year} {2011}{\natexlab{a}})}\BibitemShut
  {NoStop}%
\bibitem [{\citenamefont {Iqbal}\ \emph {et~al.}(2013)\citenamefont {Iqbal},
  \citenamefont {Becca}, \citenamefont {Sorella},\ and\ \citenamefont
  {Poilblanc}}]{Iqbal2013}%
  \BibitemOpen
  \bibfield  {author} {\bibinfo {author} {\bibfnamefont {Y.}~\bibnamefont
  {Iqbal}}, \bibinfo {author} {\bibfnamefont {F.}~\bibnamefont {Becca}},
  \bibinfo {author} {\bibfnamefont {S.}~\bibnamefont {Sorella}},\ and\ \bibinfo
  {author} {\bibfnamefont {D.}~\bibnamefont {Poilblanc}},\ }\bibfield  {title}
  {\bibinfo {title} {{Gapless spin-liquid phase in the kagome
  spin-$\frac{1}{2}$ Heisenberg antiferromagnet}},\ }\href
  {https://doi.org/10.1103/PhysRevB.87.060405} {\bibfield  {journal} {\bibinfo
  {journal} {Phys. Rev. B}\ }\textbf {\bibinfo {volume} {87}},\ \bibinfo
  {pages} {060405(R)} (\bibinfo {year} {2013})}\BibitemShut {NoStop}%
\bibitem [{\citenamefont {Clark}\ \emph
  {et~al.}(2013{\natexlab{b}})\citenamefont {Clark}, \citenamefont {Kinder},
  \citenamefont {Neuscamman}, \citenamefont {Chan},\ and\ \citenamefont
  {Lawler}}]{Clark2013b}%
  \BibitemOpen
  \bibfield  {author} {\bibinfo {author} {\bibfnamefont {B.~K.}\ \bibnamefont
  {Clark}}, \bibinfo {author} {\bibfnamefont {J.~M.}\ \bibnamefont {Kinder}},
  \bibinfo {author} {\bibfnamefont {E.}~\bibnamefont {Neuscamman}}, \bibinfo
  {author} {\bibfnamefont {G.~K.-L.}\ \bibnamefont {Chan}},\ and\ \bibinfo
  {author} {\bibfnamefont {M.~J.}\ \bibnamefont {Lawler}},\ }\bibfield  {title}
  {\bibinfo {title} {{Striped Spin Liquid Crystal Ground State Instability of
  Kagome Antiferromagnets}},\ }\href
  {https://doi.org/10.1103/PhysRevLett.111.187205} {\bibfield  {journal}
  {\bibinfo  {journal} {Phys. Rev. Lett.}\ }\textbf {\bibinfo {volume} {111}},\
  \bibinfo {pages} {187205} (\bibinfo {year} {2013}{\natexlab{b}})}\BibitemShut
  {NoStop}%
\bibitem [{\citenamefont {Iqbal}\ \emph {et~al.}(2014)\citenamefont {Iqbal},
  \citenamefont {Poilblanc},\ and\ \citenamefont {Becca}}]{Iqbal2014}%
  \BibitemOpen
  \bibfield  {author} {\bibinfo {author} {\bibfnamefont {Y.}~\bibnamefont
  {Iqbal}}, \bibinfo {author} {\bibfnamefont {D.}~\bibnamefont {Poilblanc}},\
  and\ \bibinfo {author} {\bibfnamefont {F.}~\bibnamefont {Becca}},\ }\bibfield
   {title} {\bibinfo {title} {{Vanishing spin gap in a competing spin-liquid
  phase in the kagome Heisenberg antiferromagnet}},\ }\href
  {https://doi.org/10.1103/PhysRevB.89.020407} {\bibfield  {journal} {\bibinfo
  {journal} {Phys. Rev. B}\ }\textbf {\bibinfo {volume} {89}},\ \bibinfo
  {pages} {020407(R)} (\bibinfo {year} {2014})}\BibitemShut {NoStop}%
\bibitem [{\citenamefont {Buessen}\ and\ \citenamefont
  {Trebst}(2016)}]{Buessen2016}%
  \BibitemOpen
  \bibfield  {author} {\bibinfo {author} {\bibfnamefont {F.~L.}\ \bibnamefont
  {Buessen}}\ and\ \bibinfo {author} {\bibfnamefont {S.}~\bibnamefont
  {Trebst}},\ }\bibfield  {title} {\bibinfo {title} {Competing magnetic orders
  and spin liquids in two- and three-dimensional kagome systems: Pseudofermion
  functional renormalization group perspective},\ }\href
  {https://doi.org/10.1103/PhysRevB.94.235138} {\bibfield  {journal} {\bibinfo
  {journal} {Phys. Rev. B}\ }\textbf {\bibinfo {volume} {94}},\ \bibinfo
  {pages} {235138} (\bibinfo {year} {2016})}\BibitemShut {NoStop}%
\bibitem [{\citenamefont {Hering}\ \emph {et~al.}(2019)\citenamefont {Hering},
  \citenamefont {Sonnenschein}, \citenamefont {Iqbal},\ and\ \citenamefont
  {Reuther}}]{Hering2019}%
  \BibitemOpen
  \bibfield  {author} {\bibinfo {author} {\bibfnamefont {M.}~\bibnamefont
  {Hering}}, \bibinfo {author} {\bibfnamefont {J.}~\bibnamefont
  {Sonnenschein}}, \bibinfo {author} {\bibfnamefont {Y.}~\bibnamefont
  {Iqbal}},\ and\ \bibinfo {author} {\bibfnamefont {J.}~\bibnamefont
  {Reuther}},\ }\bibfield  {title} {\bibinfo {title} {Characterization of
  quantum spin liquids and their spinon band structures via functional
  renormalization},\ }\href {https://doi.org/10.1103/PhysRevB.99.100405}
  {\bibfield  {journal} {\bibinfo  {journal} {Phys. Rev. B}\ }\textbf {\bibinfo
  {volume} {99}},\ \bibinfo {pages} {100405(R)} (\bibinfo {year}
  {2019})}\BibitemShut {NoStop}%
\bibitem [{\citenamefont {Thoenniss}\ \emph {et~al.}(2020)\citenamefont
  {Thoenniss}, \citenamefont {Ritter}, \citenamefont {Kugler}, \citenamefont
  {von Delft},\ and\ \citenamefont {Punk}}]{Thoenniss2020}%
  \BibitemOpen
  \bibfield  {author} {\bibinfo {author} {\bibfnamefont {J.}~\bibnamefont
  {Thoenniss}}, \bibinfo {author} {\bibfnamefont {M.~K.}\ \bibnamefont
  {Ritter}}, \bibinfo {author} {\bibfnamefont {F.~B.}\ \bibnamefont {Kugler}},
  \bibinfo {author} {\bibfnamefont {J.}~\bibnamefont {von Delft}},\ and\
  \bibinfo {author} {\bibfnamefont {M.}~\bibnamefont {Punk}},\ }\href@noop {}
  {\bibinfo {title} {Multiloop pseudofermion functional renormalization for
  quantum spin systems: Application to the spin-$\frac{1}{2}$ kagome heisenberg
  model}} (\bibinfo {year} {2020}),\ \Eprint {https://arxiv.org/abs/2011.01268}
  {arXiv:2011.01268} \BibitemShut {NoStop}%
\bibitem [{\citenamefont {Chalker}\ and\ \citenamefont
  {Eastmond}(1992)}]{Chalker1992}%
  \BibitemOpen
  \bibfield  {author} {\bibinfo {author} {\bibfnamefont {J.~T.}\ \bibnamefont
  {Chalker}}\ and\ \bibinfo {author} {\bibfnamefont {J.~F.~G.}\ \bibnamefont
  {Eastmond}},\ }\bibfield  {title} {\bibinfo {title} {{Ground-state disorder
  in the spin-1/2 kagom\'e Heisenberg antiferromagnet}},\ }\href
  {https://doi.org/10.1103/PhysRevB.46.14201} {\bibfield  {journal} {\bibinfo
  {journal} {Phys. Rev. B}\ }\textbf {\bibinfo {volume} {46}},\ \bibinfo
  {pages} {14201} (\bibinfo {year} {1992})}\BibitemShut {NoStop}%
\bibitem [{\citenamefont {Leung}\ and\ \citenamefont
  {Elser}(1993)}]{Leung1993}%
  \BibitemOpen
  \bibfield  {author} {\bibinfo {author} {\bibfnamefont {P.~W.}\ \bibnamefont
  {Leung}}\ and\ \bibinfo {author} {\bibfnamefont {V.}~\bibnamefont {Elser}},\
  }\bibfield  {title} {\bibinfo {title} {{Numerical studies of a 36-site
  \textit{kagome}´ antiferromagnet}},\ }\href
  {https://doi.org/10.1103/PhysRevB.47.5459} {\bibfield  {journal} {\bibinfo
  {journal} {Phys. Rev. B}\ }\textbf {\bibinfo {volume} {47}},\ \bibinfo
  {pages} {5459} (\bibinfo {year} {1993})}\BibitemShut {NoStop}%
\bibitem [{\citenamefont {Lecheminant}\ \emph {et~al.}(1997)\citenamefont
  {Lecheminant}, \citenamefont {Bernu}, \citenamefont {Lhuillier},
  \citenamefont {Pierre},\ and\ \citenamefont {Sindzingre}}]{Lecheminant1997}%
  \BibitemOpen
  \bibfield  {author} {\bibinfo {author} {\bibfnamefont {P.}~\bibnamefont
  {Lecheminant}}, \bibinfo {author} {\bibfnamefont {B.}~\bibnamefont {Bernu}},
  \bibinfo {author} {\bibfnamefont {C.}~\bibnamefont {Lhuillier}}, \bibinfo
  {author} {\bibfnamefont {L.}~\bibnamefont {Pierre}},\ and\ \bibinfo {author}
  {\bibfnamefont {P.}~\bibnamefont {Sindzingre}},\ }\bibfield  {title}
  {\bibinfo {title} {{Order versus disorder in the quantum Heisenberg
  antiferromagnet on the kagom\'e lattice using exact spectra analysis}},\
  }\href {https://doi.org/10.1103/PhysRevB.56.2521} {\bibfield  {journal}
  {\bibinfo  {journal} {Phys. Rev. B}\ }\textbf {\bibinfo {volume} {56}},\
  \bibinfo {pages} {2521} (\bibinfo {year} {1997})}\BibitemShut {NoStop}%
\bibitem [{\citenamefont {Waldtmann}\ \emph {et~al.}(1998)\citenamefont
  {Waldtmann}, \citenamefont {Everts}, \citenamefont {Bernu}, \citenamefont
  {Lhuillier}, \citenamefont {Sindzingre}, \citenamefont {Lecheminant},\ and\
  \citenamefont {Pierre}}]{Waldtmann1998}%
  \BibitemOpen
  \bibfield  {author} {\bibinfo {author} {\bibfnamefont {C.}~\bibnamefont
  {Waldtmann}}, \bibinfo {author} {\bibfnamefont {H.-U.}\ \bibnamefont
  {Everts}}, \bibinfo {author} {\bibfnamefont {B.}~\bibnamefont {Bernu}},
  \bibinfo {author} {\bibfnamefont {C.}~\bibnamefont {Lhuillier}}, \bibinfo
  {author} {\bibfnamefont {P.}~\bibnamefont {Sindzingre}}, \bibinfo {author}
  {\bibfnamefont {P.}~\bibnamefont {Lecheminant}},\ and\ \bibinfo {author}
  {\bibfnamefont {L.}~\bibnamefont {Pierre}},\ }\bibfield  {title} {\bibinfo
  {title} {{First excitations of the spin 1/2 Heisenberg antiferromagnet on the
  kagom{\'e} lattice}},\ }\href {https://doi.org/10.1007/s100510050274}
  {\bibfield  {journal} {\bibinfo  {journal} {The European Physical Journal B -
  Condensed Matter and Complex Systems}\ }\textbf {\bibinfo {volume} {2}},\
  \bibinfo {pages} {501} (\bibinfo {year} {1998})}\BibitemShut {NoStop}%
\bibitem [{\citenamefont {Jiang}\ \emph {et~al.}(2008)\citenamefont {Jiang},
  \citenamefont {Weng},\ and\ \citenamefont {Sheng}}]{Jiang2008}%
  \BibitemOpen
  \bibfield  {author} {\bibinfo {author} {\bibfnamefont {H.~C.}\ \bibnamefont
  {Jiang}}, \bibinfo {author} {\bibfnamefont {Z.~Y.}\ \bibnamefont {Weng}},\
  and\ \bibinfo {author} {\bibfnamefont {D.~N.}\ \bibnamefont {Sheng}},\
  }\bibfield  {title} {\bibinfo {title} {{Density Matrix Renormalization Group
  Numerical Study of the Kagome Antiferromagnet}},\ }\href
  {https://doi.org/10.1103/PhysRevLett.101.117203} {\bibfield  {journal}
  {\bibinfo  {journal} {Phys. Rev. Lett.}\ }\textbf {\bibinfo {volume} {101}},\
  \bibinfo {pages} {117203} (\bibinfo {year} {2008})}\BibitemShut {NoStop}%
\bibitem [{\citenamefont {Nakano}\ and\ \citenamefont
  {Sakai}(2011)}]{Nakano2011}%
  \BibitemOpen
  \bibfield  {author} {\bibinfo {author} {\bibfnamefont {H.}~\bibnamefont
  {Nakano}}\ and\ \bibinfo {author} {\bibfnamefont {T.}~\bibnamefont {Sakai}},\
  }\bibfield  {title} {\bibinfo {title} {{Numerical-Diagonalization Study of
  Spin Gap Issue of the Kagome Lattice Heisenberg Antiferromagnet}},\ }\href
  {https://doi.org/10.1143/JPSJ.80.053704} {\bibfield  {journal} {\bibinfo
  {journal} {Journal of the Physical Society of Japan}\ }\textbf {\bibinfo
  {volume} {80}},\ \bibinfo {pages} {053704} (\bibinfo {year}
  {2011})}\BibitemShut {NoStop}%
\bibitem [{\citenamefont {L\"auchli}\ \emph {et~al.}(2011)\citenamefont
  {L\"auchli}, \citenamefont {Sudan},\ and\ \citenamefont
  {S\o{}rensen}}]{Lauchli2011}%
  \BibitemOpen
  \bibfield  {author} {\bibinfo {author} {\bibfnamefont {A.~M.}\ \bibnamefont
  {L\"auchli}}, \bibinfo {author} {\bibfnamefont {J.}~\bibnamefont {Sudan}},\
  and\ \bibinfo {author} {\bibfnamefont {E.~S.}\ \bibnamefont {S\o{}rensen}},\
  }\bibfield  {title} {\bibinfo {title} {Ground-state energy and spin gap of
  spin-$\frac{1}{2}$ kagom\'e-heisenberg antiferromagnetic clusters:
  Large-scale exact diagonalization results},\ }\href
  {https://doi.org/10.1103/PhysRevB.83.212401} {\bibfield  {journal} {\bibinfo
  {journal} {Phys. Rev. B}\ }\textbf {\bibinfo {volume} {83}},\ \bibinfo
  {pages} {212401} (\bibinfo {year} {2011})}\BibitemShut {NoStop}%
\bibitem [{\citenamefont {Yan}\ \emph {et~al.}(2011)\citenamefont {Yan},
  \citenamefont {Huse},\ and\ \citenamefont {White}}]{Yan2011}%
  \BibitemOpen
  \bibfield  {author} {\bibinfo {author} {\bibfnamefont {S.}~\bibnamefont
  {Yan}}, \bibinfo {author} {\bibfnamefont {D.~A.}\ \bibnamefont {Huse}},\ and\
  \bibinfo {author} {\bibfnamefont {S.~R.}\ \bibnamefont {White}},\ }\bibfield
  {title} {\bibinfo {title} {{Spin-Liquid Ground State of the \textit{S} = 1/2
  Kagome Heisenberg Antiferromagnet}},\ }\href
  {https://doi.org/10.1126/science.1201080} {\bibfield  {journal} {\bibinfo
  {journal} {Science}\ }\textbf {\bibinfo {volume} {332}},\ \bibinfo {pages}
  {1173} (\bibinfo {year} {2011})}\BibitemShut {NoStop}%
\bibitem [{\citenamefont {Depenbrock}\ \emph {et~al.}(2012)\citenamefont
  {Depenbrock}, \citenamefont {McCulloch},\ and\ \citenamefont
  {Schollw\"ock}}]{Depenbrock2012}%
  \BibitemOpen
  \bibfield  {author} {\bibinfo {author} {\bibfnamefont {S.}~\bibnamefont
  {Depenbrock}}, \bibinfo {author} {\bibfnamefont {I.~P.}\ \bibnamefont
  {McCulloch}},\ and\ \bibinfo {author} {\bibfnamefont {U.}~\bibnamefont
  {Schollw\"ock}},\ }\bibfield  {title} {\bibinfo {title} {{Nature of the
  Spin-Liquid Ground State of the $S=1/2$ Heisenberg Model on the Kagome
  Lattice}},\ }\href {https://doi.org/10.1103/PhysRevLett.109.067201}
  {\bibfield  {journal} {\bibinfo  {journal} {Phys. Rev. Lett.}\ }\textbf
  {\bibinfo {volume} {109}},\ \bibinfo {pages} {067201} (\bibinfo {year}
  {2012})}\BibitemShut {NoStop}%
\bibitem [{\citenamefont {Jiang}\ \emph {et~al.}(2012)\citenamefont {Jiang},
  \citenamefont {Wang},\ and\ \citenamefont {Balents}}]{Jiang2012}%
  \BibitemOpen
  \bibfield  {author} {\bibinfo {author} {\bibfnamefont {H.-C.}\ \bibnamefont
  {Jiang}}, \bibinfo {author} {\bibfnamefont {Z.}~\bibnamefont {Wang}},\ and\
  \bibinfo {author} {\bibfnamefont {L.}~\bibnamefont {Balents}},\ }\bibfield
  {title} {\bibinfo {title} {Identifying topological order by entanglement
  entropy},\ }\href {https://doi.org/10.1038/nphys2465} {\bibfield  {journal}
  {\bibinfo  {journal} {Nature Physics}\ }\textbf {\bibinfo {volume} {8}},\
  \bibinfo {pages} {902} (\bibinfo {year} {2012})}\BibitemShut {NoStop}%
\bibitem [{\citenamefont {He}\ \emph {et~al.}(2017)\citenamefont {He},
  \citenamefont {Zaletel}, \citenamefont {Oshikawa},\ and\ \citenamefont
  {Pollmann}}]{He2017}%
  \BibitemOpen
  \bibfield  {author} {\bibinfo {author} {\bibfnamefont {Y.-C.}\ \bibnamefont
  {He}}, \bibinfo {author} {\bibfnamefont {M.~P.}\ \bibnamefont {Zaletel}},
  \bibinfo {author} {\bibfnamefont {M.}~\bibnamefont {Oshikawa}},\ and\
  \bibinfo {author} {\bibfnamefont {F.}~\bibnamefont {Pollmann}},\ }\bibfield
  {title} {\bibinfo {title} {{Signatures of Dirac Cones in a DMRG Study of the
  Kagome Heisenberg Model}},\ }\href
  {https://doi.org/10.1103/PhysRevX.7.031020} {\bibfield  {journal} {\bibinfo
  {journal} {Phys. Rev. X}\ }\textbf {\bibinfo {volume} {7}},\ \bibinfo {pages}
  {031020} (\bibinfo {year} {2017})}\BibitemShut {NoStop}%
\bibitem [{\citenamefont {Changlani}\ \emph {et~al.}(2018)\citenamefont
  {Changlani}, \citenamefont {Kochkov}, \citenamefont {Kumar}, \citenamefont
  {Clark},\ and\ \citenamefont {Fradkin}}]{Changlani2018}%
  \BibitemOpen
  \bibfield  {author} {\bibinfo {author} {\bibfnamefont {H.~J.}\ \bibnamefont
  {Changlani}}, \bibinfo {author} {\bibfnamefont {D.}~\bibnamefont {Kochkov}},
  \bibinfo {author} {\bibfnamefont {K.}~\bibnamefont {Kumar}}, \bibinfo
  {author} {\bibfnamefont {B.~K.}\ \bibnamefont {Clark}},\ and\ \bibinfo
  {author} {\bibfnamefont {E.}~\bibnamefont {Fradkin}},\ }\bibfield  {title}
  {\bibinfo {title} {{Macroscopically Degenerate Exactly Solvable Point in the
  Spin-$1/2$ Kagome Quantum Antiferromagnet}},\ }\href
  {https://doi.org/10.1103/PhysRevLett.120.117202} {\bibfield  {journal}
  {\bibinfo  {journal} {Phys. Rev. Lett.}\ }\textbf {\bibinfo {volume} {120}},\
  \bibinfo {pages} {117202} (\bibinfo {year} {2018})}\BibitemShut {NoStop}%
\bibitem [{\citenamefont {Zhu}\ \emph {et~al.}(2018)\citenamefont {Zhu},
  \citenamefont {Chen}, \citenamefont {He},\ and\ \citenamefont
  {Witczak-Krempa}}]{Zhu2018}%
  \BibitemOpen
  \bibfield  {author} {\bibinfo {author} {\bibfnamefont {W.}~\bibnamefont
  {Zhu}}, \bibinfo {author} {\bibfnamefont {X.}~\bibnamefont {Chen}}, \bibinfo
  {author} {\bibfnamefont {Y.-C.}\ \bibnamefont {He}},\ and\ \bibinfo {author}
  {\bibfnamefont {W.}~\bibnamefont {Witczak-Krempa}},\ }\bibfield  {title}
  {\bibinfo {title} {{Entanglement signatures of emergent Dirac fermions:
  Kagome spin liquid and quantum criticality}},\ }\href
  {https://doi.org/10.1126/sciadv.aat5535} {\bibfield  {journal} {\bibinfo
  {journal} {Science Advances}\ }\textbf {\bibinfo {volume} {4}},\ \bibinfo
  {pages} {eaat5535} (\bibinfo {year} {2018})}\BibitemShut {NoStop}%
\bibitem [{\citenamefont {L\"auchli}\ \emph {et~al.}(2019)\citenamefont
  {L\"auchli}, \citenamefont {Sudan},\ and\ \citenamefont
  {Moessner}}]{Lauchli2019}%
  \BibitemOpen
  \bibfield  {author} {\bibinfo {author} {\bibfnamefont {A.~M.}\ \bibnamefont
  {L\"auchli}}, \bibinfo {author} {\bibfnamefont {J.}~\bibnamefont {Sudan}},\
  and\ \bibinfo {author} {\bibfnamefont {R.}~\bibnamefont {Moessner}},\
  }\bibfield  {title} {\bibinfo {title} {{$S=\frac{1}{2}$ kagome Heisenberg
  antiferromagnet revisited}},\ }\href
  {https://doi.org/10.1103/PhysRevB.100.155142} {\bibfield  {journal} {\bibinfo
   {journal} {Phys. Rev. B}\ }\textbf {\bibinfo {volume} {100}},\ \bibinfo
  {pages} {155142} (\bibinfo {year} {2019})}\BibitemShut {NoStop}%
\bibitem [{\citenamefont {Xie}\ \emph {et~al.}(2014)\citenamefont {Xie},
  \citenamefont {Chen}, \citenamefont {Yu}, \citenamefont {Kong}, \citenamefont
  {Normand},\ and\ \citenamefont {Xiang}}]{Xie2014}%
  \BibitemOpen
  \bibfield  {author} {\bibinfo {author} {\bibfnamefont {Z.~Y.}\ \bibnamefont
  {Xie}}, \bibinfo {author} {\bibfnamefont {J.}~\bibnamefont {Chen}}, \bibinfo
  {author} {\bibfnamefont {J.~F.}\ \bibnamefont {Yu}}, \bibinfo {author}
  {\bibfnamefont {X.}~\bibnamefont {Kong}}, \bibinfo {author} {\bibfnamefont
  {B.}~\bibnamefont {Normand}},\ and\ \bibinfo {author} {\bibfnamefont
  {T.}~\bibnamefont {Xiang}},\ }\bibfield  {title} {\bibinfo {title} {{Tensor
  Renormalization of Quantum Many-Body Systems Using Projected Entangled
  Simplex States}},\ }\href {https://doi.org/10.1103/PhysRevX.4.011025}
  {\bibfield  {journal} {\bibinfo  {journal} {Phys. Rev. X}\ }\textbf {\bibinfo
  {volume} {4}},\ \bibinfo {pages} {011025} (\bibinfo {year}
  {2014})}\BibitemShut {NoStop}%
\bibitem [{\citenamefont {Mei}\ \emph {et~al.}(2017)\citenamefont {Mei},
  \citenamefont {Chen}, \citenamefont {He},\ and\ \citenamefont
  {Wen}}]{Mei2017}%
  \BibitemOpen
  \bibfield  {author} {\bibinfo {author} {\bibfnamefont {J.-W.}\ \bibnamefont
  {Mei}}, \bibinfo {author} {\bibfnamefont {J.-Y.}\ \bibnamefont {Chen}},
  \bibinfo {author} {\bibfnamefont {H.}~\bibnamefont {He}},\ and\ \bibinfo
  {author} {\bibfnamefont {X.-G.}\ \bibnamefont {Wen}},\ }\bibfield  {title}
  {\bibinfo {title} {{Gapped spin liquid with ${\mathbb{Z}}_{2}$ topological
  order for the kagome Heisenberg model}},\ }\href
  {https://doi.org/10.1103/PhysRevB.95.235107} {\bibfield  {journal} {\bibinfo
  {journal} {Phys. Rev. B}\ }\textbf {\bibinfo {volume} {95}},\ \bibinfo
  {pages} {235107} (\bibinfo {year} {2017})}\BibitemShut {NoStop}%
\bibitem [{\citenamefont {Liao}\ \emph {et~al.}(2017)\citenamefont {Liao},
  \citenamefont {Xie}, \citenamefont {Chen}, \citenamefont {Liu}, \citenamefont
  {Xie}, \citenamefont {Huang}, \citenamefont {Normand},\ and\ \citenamefont
  {Xiang}}]{Liao2017}%
  \BibitemOpen
  \bibfield  {author} {\bibinfo {author} {\bibfnamefont {H.~J.}\ \bibnamefont
  {Liao}}, \bibinfo {author} {\bibfnamefont {Z.~Y.}\ \bibnamefont {Xie}},
  \bibinfo {author} {\bibfnamefont {J.}~\bibnamefont {Chen}}, \bibinfo {author}
  {\bibfnamefont {Z.~Y.}\ \bibnamefont {Liu}}, \bibinfo {author} {\bibfnamefont
  {H.~D.}\ \bibnamefont {Xie}}, \bibinfo {author} {\bibfnamefont {R.~Z.}\
  \bibnamefont {Huang}}, \bibinfo {author} {\bibfnamefont {B.}~\bibnamefont
  {Normand}},\ and\ \bibinfo {author} {\bibfnamefont {T.}~\bibnamefont
  {Xiang}},\ }\bibfield  {title} {\bibinfo {title} {{Gapless Spin-Liquid Ground
  State in the $S=1/2$ Kagome Antiferromagnet}},\ }\href
  {https://doi.org/10.1103/PhysRevLett.118.137202} {\bibfield  {journal}
  {\bibinfo  {journal} {Phys. Rev. Lett.}\ }\textbf {\bibinfo {volume} {118}},\
  \bibinfo {pages} {137202} (\bibinfo {year} {2017})}\BibitemShut {NoStop}%
\bibitem [{\citenamefont {Jiang}\ \emph {et~al.}(2019)\citenamefont {Jiang},
  \citenamefont {Kim}, \citenamefont {Han},\ and\ \citenamefont
  {Ran}}]{Jiang2019}%
  \BibitemOpen
  \bibfield  {author} {\bibinfo {author} {\bibfnamefont {S.}~\bibnamefont
  {Jiang}}, \bibinfo {author} {\bibfnamefont {P.}~\bibnamefont {Kim}}, \bibinfo
  {author} {\bibfnamefont {J.~H.}\ \bibnamefont {Han}},\ and\ \bibinfo {author}
  {\bibfnamefont {Y.}~\bibnamefont {Ran}},\ }\bibfield  {title} {\bibinfo
  {title} {{Competing Spin Liquid Phases in the S=$\frac{1}{2}$ Heisenberg
  Model on the Kagome Lattice}},\ }\href
  {https://doi.org/10.21468/SciPostPhys.7.1.006} {\bibfield  {journal}
  {\bibinfo  {journal} {SciPost Phys.}\ }\textbf {\bibinfo {volume} {7}},\
  \bibinfo {pages} {006} (\bibinfo {year} {2019})}\BibitemShut {NoStop}%
\bibitem [{\citenamefont {Suttner}\ \emph {et~al.}(2014)\citenamefont
  {Suttner}, \citenamefont {Platt}, \citenamefont {Reuther},\ and\
  \citenamefont {Thomale}}]{Suttner2014}%
  \BibitemOpen
  \bibfield  {author} {\bibinfo {author} {\bibfnamefont {R.}~\bibnamefont
  {Suttner}}, \bibinfo {author} {\bibfnamefont {C.}~\bibnamefont {Platt}},
  \bibinfo {author} {\bibfnamefont {J.}~\bibnamefont {Reuther}},\ and\ \bibinfo
  {author} {\bibfnamefont {R.}~\bibnamefont {Thomale}},\ }\bibfield  {title}
  {\bibinfo {title} {{Renormalization group analysis of competing quantum
  phases in the ${J}_{1}$-${J}_{2}$ Heisenberg model on the kagome lattice}},\
  }\href {https://doi.org/10.1103/PhysRevB.89.020408} {\bibfield  {journal}
  {\bibinfo  {journal} {Phys. Rev. B}\ }\textbf {\bibinfo {volume} {89}},\
  \bibinfo {pages} {020408(R)} (\bibinfo {year} {2014})}\BibitemShut {NoStop}%
\bibitem [{\citenamefont {Iqbal}\ \emph {et~al.}(2015)\citenamefont {Iqbal},
  \citenamefont {Poilblanc},\ and\ \citenamefont {Becca}}]{Iqbal2015}%
  \BibitemOpen
  \bibfield  {author} {\bibinfo {author} {\bibfnamefont {Y.}~\bibnamefont
  {Iqbal}}, \bibinfo {author} {\bibfnamefont {D.}~\bibnamefont {Poilblanc}},\
  and\ \bibinfo {author} {\bibfnamefont {F.}~\bibnamefont {Becca}},\ }\bibfield
   {title} {\bibinfo {title} {{Spin-$\frac{1}{2}$ Heisenberg
  ${J}_{1}\text{\ensuremath{-}}{J}_{2}$ antiferromagnet on the kagome
  lattice}},\ }\href {https://doi.org/10.1103/PhysRevB.91.020402} {\bibfield
  {journal} {\bibinfo  {journal} {Phys. Rev. B}\ }\textbf {\bibinfo {volume}
  {91}},\ \bibinfo {pages} {020402(R)} (\bibinfo {year} {2015})}\BibitemShut
  {NoStop}%
\bibitem [{\citenamefont {Kolley}\ \emph {et~al.}(2015)\citenamefont {Kolley},
  \citenamefont {Depenbrock}, \citenamefont {McCulloch}, \citenamefont
  {Schollw\"ock},\ and\ \citenamefont {Alba}}]{Kolley2015}%
  \BibitemOpen
  \bibfield  {author} {\bibinfo {author} {\bibfnamefont {F.}~\bibnamefont
  {Kolley}}, \bibinfo {author} {\bibfnamefont {S.}~\bibnamefont {Depenbrock}},
  \bibinfo {author} {\bibfnamefont {I.~P.}\ \bibnamefont {McCulloch}}, \bibinfo
  {author} {\bibfnamefont {U.}~\bibnamefont {Schollw\"ock}},\ and\ \bibinfo
  {author} {\bibfnamefont {V.}~\bibnamefont {Alba}},\ }\bibfield  {title}
  {\bibinfo {title} {{Phase diagram of the
  ${J}_{1}\text{\ensuremath{-}}{J}_{2}$ Heisenberg model on the kagome
  lattice}},\ }\href {https://doi.org/10.1103/PhysRevB.91.104418} {\bibfield
  {journal} {\bibinfo  {journal} {Phys. Rev. B}\ }\textbf {\bibinfo {volume}
  {91}},\ \bibinfo {pages} {104418} (\bibinfo {year} {2015})}\BibitemShut
  {NoStop}%
\bibitem [{\citenamefont {Gong}\ \emph {et~al.}(2014)\citenamefont {Gong},
  \citenamefont {Zhu},\ and\ \citenamefont {Sheng}}]{Gong2014}%
  \BibitemOpen
  \bibfield  {author} {\bibinfo {author} {\bibfnamefont {S.-S.}\ \bibnamefont
  {Gong}}, \bibinfo {author} {\bibfnamefont {W.}~\bibnamefont {Zhu}},\ and\
  \bibinfo {author} {\bibfnamefont {D.~N.}\ \bibnamefont {Sheng}},\ }\bibfield
  {title} {\bibinfo {title} {{Emergent Chiral Spin Liquid: Fractional Quantum
  Hall Effect in a Kagome Heisenberg Model}},\ }\href
  {https://doi.org/10.1038/srep06317} {\bibfield  {journal} {\bibinfo
  {journal} {Scientific Reports}\ }\textbf {\bibinfo {volume} {4}},\ \bibinfo
  {pages} {6317} (\bibinfo {year} {2014})}\BibitemShut {NoStop}%
\bibitem [{\citenamefont {Gong}\ \emph {et~al.}(2015)\citenamefont {Gong},
  \citenamefont {Zhu}, \citenamefont {Balents},\ and\ \citenamefont
  {Sheng}}]{Gong2015}%
  \BibitemOpen
  \bibfield  {author} {\bibinfo {author} {\bibfnamefont {S.-S.}\ \bibnamefont
  {Gong}}, \bibinfo {author} {\bibfnamefont {W.}~\bibnamefont {Zhu}}, \bibinfo
  {author} {\bibfnamefont {L.}~\bibnamefont {Balents}},\ and\ \bibinfo {author}
  {\bibfnamefont {D.~N.}\ \bibnamefont {Sheng}},\ }\bibfield  {title} {\bibinfo
  {title} {Global phase diagram of competing ordered and quantum spin-liquid
  phases on the kagome lattice},\ }\href
  {https://doi.org/10.1103/PhysRevB.91.075112} {\bibfield  {journal} {\bibinfo
  {journal} {Phys. Rev. B}\ }\textbf {\bibinfo {volume} {91}},\ \bibinfo
  {pages} {075112} (\bibinfo {year} {2015})}\BibitemShut {NoStop}%
\bibitem [{\citenamefont {Iqbal}\ \emph {et~al.}(2021)\citenamefont {Iqbal},
  \citenamefont {Ferrari}, \citenamefont {Chauhan}, \citenamefont {Parola},
  \citenamefont {Poilblanc},\ and\ \citenamefont {Becca}}]{Iqbal2021}%
  \BibitemOpen
  \bibfield  {author} {\bibinfo {author} {\bibfnamefont {Y.}~\bibnamefont
  {Iqbal}}, \bibinfo {author} {\bibfnamefont {F.}~\bibnamefont {Ferrari}},
  \bibinfo {author} {\bibfnamefont {A.}~\bibnamefont {Chauhan}}, \bibinfo
  {author} {\bibfnamefont {A.}~\bibnamefont {Parola}}, \bibinfo {author}
  {\bibfnamefont {D.}~\bibnamefont {Poilblanc}},\ and\ \bibinfo {author}
  {\bibfnamefont {F.}~\bibnamefont {Becca}},\ }\bibfield  {title} {\bibinfo
  {title} {{Gutzwiller projected states for the ${J}_{1}\ensuremath{-}{J}_{2}$
  Heisenberg model on the Kagome lattice: Achievements and pitfalls}},\ }\href
  {https://doi.org/10.1103/PhysRevB.104.144406} {\bibfield  {journal} {\bibinfo
   {journal} {Phys. Rev. B}\ }\textbf {\bibinfo {volume} {104}},\ \bibinfo
  {pages} {144406} (\bibinfo {year} {2021})}\BibitemShut {NoStop}%
\bibitem [{\citenamefont {Sun}\ \emph {et~al.}(2022)\citenamefont {Sun},
  \citenamefont {Jin}, \citenamefont {Tu},\ and\ \citenamefont
  {Zhou}}]{Sun2022}%
  \BibitemOpen
  \bibfield  {author} {\bibinfo {author} {\bibfnamefont {R.-Y.}\ \bibnamefont
  {Sun}}, \bibinfo {author} {\bibfnamefont {H.-K.}\ \bibnamefont {Jin}},
  \bibinfo {author} {\bibfnamefont {H.-H.}\ \bibnamefont {Tu}},\ and\ \bibinfo
  {author} {\bibfnamefont {Y.}~\bibnamefont {Zhou}},\ }\href@noop {} {\bibinfo
  {title} {{Possible chiral spin liquid state in the $S=1/2$ kagome Heisenberg
  model}}} (\bibinfo {year} {2022}),\ \Eprint
  {https://arxiv.org/abs/2203.07321} {arXiv:2203.07321} \BibitemShut {NoStop}%
\bibitem [{\citenamefont {Dzyaloshinsky}(1958)}]{Dzyaloshinsky1958}%
  \BibitemOpen
  \bibfield  {author} {\bibinfo {author} {\bibfnamefont {I.}~\bibnamefont
  {Dzyaloshinsky}},\ }\bibfield  {title} {\bibinfo {title} {A thermodynamic
  theory of “weak” ferromagnetism of antiferromagnetics},\ }\href
  {https://doi.org/https://doi.org/10.1016/0022-3697(58)90076-3} {\bibfield
  {journal} {\bibinfo  {journal} {Journal of Physics and Chemistry of Solids}\
  }\textbf {\bibinfo {volume} {4}},\ \bibinfo {pages} {241} (\bibinfo {year}
  {1958})}\BibitemShut {NoStop}%
\bibitem [{\citenamefont {Moriya}(1960)}]{Moriya1960}%
  \BibitemOpen
  \bibfield  {author} {\bibinfo {author} {\bibfnamefont {T.}~\bibnamefont
  {Moriya}},\ }\bibfield  {title} {\bibinfo {title} {{Anisotropic Superexchange
  Interaction and Weak Ferromagnetism}},\ }\href
  {https://doi.org/10.1103/PhysRev.120.91} {\bibfield  {journal} {\bibinfo
  {journal} {Phys. Rev.}\ }\textbf {\bibinfo {volume} {120}},\ \bibinfo {pages}
  {91} (\bibinfo {year} {1960})}\BibitemShut {NoStop}%
\bibitem [{\citenamefont {C\'epas}\ \emph {et~al.}(2008)\citenamefont
  {C\'epas}, \citenamefont {Fong}, \citenamefont {Leung},\ and\ \citenamefont
  {Lhuillier}}]{Cepas2008}%
  \BibitemOpen
  \bibfield  {author} {\bibinfo {author} {\bibfnamefont {O.}~\bibnamefont
  {C\'epas}}, \bibinfo {author} {\bibfnamefont {C.~M.}\ \bibnamefont {Fong}},
  \bibinfo {author} {\bibfnamefont {P.~W.}\ \bibnamefont {Leung}},\ and\
  \bibinfo {author} {\bibfnamefont {C.}~\bibnamefont {Lhuillier}},\ }\bibfield
  {title} {\bibinfo {title} {{Quantum phase transition induced by
  Dzyaloshinskii-Moriya interactions in the kagome antiferromagnet}},\ }\href
  {https://doi.org/10.1103/PhysRevB.78.140405} {\bibfield  {journal} {\bibinfo
  {journal} {Phys. Rev. B}\ }\textbf {\bibinfo {volume} {78}},\ \bibinfo
  {pages} {140405(R)} (\bibinfo {year} {2008})}\BibitemShut {NoStop}%
\bibitem [{\citenamefont {Rousochatzakis}\ \emph {et~al.}(2009)\citenamefont
  {Rousochatzakis}, \citenamefont {Manmana}, \citenamefont {L\"auchli},
  \citenamefont {Normand},\ and\ \citenamefont {Mila}}]{Rousochatzakis2009}%
  \BibitemOpen
  \bibfield  {author} {\bibinfo {author} {\bibfnamefont {I.}~\bibnamefont
  {Rousochatzakis}}, \bibinfo {author} {\bibfnamefont {S.~R.}\ \bibnamefont
  {Manmana}}, \bibinfo {author} {\bibfnamefont {A.~M.}\ \bibnamefont
  {L\"auchli}}, \bibinfo {author} {\bibfnamefont {B.}~\bibnamefont {Normand}},\
  and\ \bibinfo {author} {\bibfnamefont {F.}~\bibnamefont {Mila}},\ }\bibfield
  {title} {\bibinfo {title} {{Dzyaloshinskii-Moriya anisotropy and nonmagnetic
  impurities in the $s=\frac{1}{2}$ kagome system
  ${\text{ZnCu}}_{3}{(\text{OH})}_{6}{\text{Cl}}_{2}$}},\ }\href
  {https://doi.org/10.1103/PhysRevB.79.214415} {\bibfield  {journal} {\bibinfo
  {journal} {Phys. Rev. B}\ }\textbf {\bibinfo {volume} {79}},\ \bibinfo
  {pages} {214415} (\bibinfo {year} {2009})}\BibitemShut {NoStop}%
\bibitem [{\citenamefont {Messio}\ \emph {et~al.}(2010)\citenamefont {Messio},
  \citenamefont {C\'epas},\ and\ \citenamefont {Lhuillier}}]{Messio2010}%
  \BibitemOpen
  \bibfield  {author} {\bibinfo {author} {\bibfnamefont {L.}~\bibnamefont
  {Messio}}, \bibinfo {author} {\bibfnamefont {O.}~\bibnamefont {C\'epas}},\
  and\ \bibinfo {author} {\bibfnamefont {C.}~\bibnamefont {Lhuillier}},\
  }\bibfield  {title} {\bibinfo {title} {{Schwinger-boson approach to the
  kagome antiferromagnet with Dzyaloshinskii-Moriya interactions: Phase diagram
  and dynamical structure factors}},\ }\href
  {https://doi.org/10.1103/PhysRevB.81.064428} {\bibfield  {journal} {\bibinfo
  {journal} {Phys. Rev. B}\ }\textbf {\bibinfo {volume} {81}},\ \bibinfo
  {pages} {064428} (\bibinfo {year} {2010})}\BibitemShut {NoStop}%
\bibitem [{\citenamefont {Messio}\ \emph {et~al.}(2017)\citenamefont {Messio},
  \citenamefont {Bieri}, \citenamefont {Lhuillier},\ and\ \citenamefont
  {Bernu}}]{Messio2017}%
  \BibitemOpen
  \bibfield  {author} {\bibinfo {author} {\bibfnamefont {L.}~\bibnamefont
  {Messio}}, \bibinfo {author} {\bibfnamefont {S.}~\bibnamefont {Bieri}},
  \bibinfo {author} {\bibfnamefont {C.}~\bibnamefont {Lhuillier}},\ and\
  \bibinfo {author} {\bibfnamefont {B.}~\bibnamefont {Bernu}},\ }\bibfield
  {title} {\bibinfo {title} {{Chiral Spin Liquid on a Kagome Antiferromagnet
  Induced by the Dzyaloshinskii-Moriya Interaction}},\ }\href
  {https://doi.org/10.1103/PhysRevLett.118.267201} {\bibfield  {journal}
  {\bibinfo  {journal} {Phys. Rev. Lett.}\ }\textbf {\bibinfo {volume} {118}},\
  \bibinfo {pages} {267201} (\bibinfo {year} {2017})}\BibitemShut {NoStop}%
\bibitem [{\citenamefont {Mondal}\ and\ \citenamefont
  {Kadolkar}(2017)}]{Mondal2017}%
  \BibitemOpen
  \bibfield  {author} {\bibinfo {author} {\bibfnamefont {K.}~\bibnamefont
  {Mondal}}\ and\ \bibinfo {author} {\bibfnamefont {C.}~\bibnamefont
  {Kadolkar}},\ }\bibfield  {title} {\bibinfo {title} {{Schwinger boson
  mean-field theory of the kagome Heisenberg antiferromagnet with
  Dzyaloshinskii-Moriya interactions}},\ }\href
  {https://doi.org/10.1103/PhysRevB.95.134404} {\bibfield  {journal} {\bibinfo
  {journal} {Phys. Rev. B}\ }\textbf {\bibinfo {volume} {95}},\ \bibinfo
  {pages} {134404} (\bibinfo {year} {2017})}\BibitemShut {NoStop}%
\bibitem [{\citenamefont {Lee}\ \emph {et~al.}(2018)\citenamefont {Lee},
  \citenamefont {Normand},\ and\ \citenamefont {Kao}}]{Lee2018}%
  \BibitemOpen
  \bibfield  {author} {\bibinfo {author} {\bibfnamefont {C.-Y.}\ \bibnamefont
  {Lee}}, \bibinfo {author} {\bibfnamefont {B.}~\bibnamefont {Normand}},\ and\
  \bibinfo {author} {\bibfnamefont {Y.-J.}\ \bibnamefont {Kao}},\ }\bibfield
  {title} {\bibinfo {title} {{Gapless spin liquid in the kagome Heisenberg
  antiferromagnet with Dzyaloshinskii-Moriya interactions}},\ }\href
  {https://doi.org/10.1103/PhysRevB.98.224414} {\bibfield  {journal} {\bibinfo
  {journal} {Phys. Rev. B}\ }\textbf {\bibinfo {volume} {98}},\ \bibinfo
  {pages} {224414} (\bibinfo {year} {2018})}\BibitemShut {NoStop}%
\bibitem [{\citenamefont {Hering}\ and\ \citenamefont
  {Reuther}(2017)}]{Hering2017}%
  \BibitemOpen
  \bibfield  {author} {\bibinfo {author} {\bibfnamefont {M.}~\bibnamefont
  {Hering}}\ and\ \bibinfo {author} {\bibfnamefont {J.}~\bibnamefont
  {Reuther}},\ }\bibfield  {title} {\bibinfo {title} {{Functional
  renormalization group analysis of Dzyaloshinsky-Moriya and Heisenberg spin
  interactions on the kagome lattice}},\ }\href
  {https://doi.org/10.1103/PhysRevB.95.054418} {\bibfield  {journal} {\bibinfo
  {journal} {Phys. Rev. B}\ }\textbf {\bibinfo {volume} {95}},\ \bibinfo
  {pages} {054418} (\bibinfo {year} {2017})}\BibitemShut {NoStop}%
\bibitem [{\citenamefont {Buessen}\ \emph {et~al.}(2019)\citenamefont
  {Buessen}, \citenamefont {Noculak}, \citenamefont {Trebst},\ and\
  \citenamefont {Reuther}}]{Buessen2019}%
  \BibitemOpen
  \bibfield  {author} {\bibinfo {author} {\bibfnamefont {F.~L.}\ \bibnamefont
  {Buessen}}, \bibinfo {author} {\bibfnamefont {V.}~\bibnamefont {Noculak}},
  \bibinfo {author} {\bibfnamefont {S.}~\bibnamefont {Trebst}},\ and\ \bibinfo
  {author} {\bibfnamefont {J.}~\bibnamefont {Reuther}},\ }\bibfield  {title}
  {\bibinfo {title} {Functional renormalization group for frustrated magnets
  with nondiagonal spin interactions},\ }\href
  {https://doi.org/10.1103/PhysRevB.100.125164} {\bibfield  {journal} {\bibinfo
   {journal} {Phys. Rev. B}\ }\textbf {\bibinfo {volume} {100}},\ \bibinfo
  {pages} {125164} (\bibinfo {year} {2019})}\BibitemShut {NoStop}%
\bibitem [{\citenamefont {Kalmeyer}\ and\ \citenamefont
  {Laughlin}(1987)}]{Kalmeyer1987}%
  \BibitemOpen
  \bibfield  {author} {\bibinfo {author} {\bibfnamefont {V.}~\bibnamefont
  {Kalmeyer}}\ and\ \bibinfo {author} {\bibfnamefont {R.~B.}\ \bibnamefont
  {Laughlin}},\ }\bibfield  {title} {\bibinfo {title} {{Equivalence of the
  resonating-valence-bond and fractional quantum Hall states}},\ }\href
  {https://doi.org/10.1103/PhysRevLett.59.2095} {\bibfield  {journal} {\bibinfo
   {journal} {Phys. Rev. Lett.}\ }\textbf {\bibinfo {volume} {59}},\ \bibinfo
  {pages} {2095} (\bibinfo {year} {1987})}\BibitemShut {NoStop}%
\bibitem [{\citenamefont {Schroeter}\ \emph {et~al.}(2007)\citenamefont
  {Schroeter}, \citenamefont {Kapit}, \citenamefont {Thomale},\ and\
  \citenamefont {Greiter}}]{Schroeter2007}%
  \BibitemOpen
  \bibfield  {author} {\bibinfo {author} {\bibfnamefont {D.~F.}\ \bibnamefont
  {Schroeter}}, \bibinfo {author} {\bibfnamefont {E.}~\bibnamefont {Kapit}},
  \bibinfo {author} {\bibfnamefont {R.}~\bibnamefont {Thomale}},\ and\ \bibinfo
  {author} {\bibfnamefont {M.}~\bibnamefont {Greiter}},\ }\bibfield  {title}
  {\bibinfo {title} {{Spin Hamiltonian for which the Chiral Spin Liquid is the
  Exact Ground State}},\ }\href {https://doi.org/10.1103/physrevlett.99.097202}
  {\bibfield  {journal} {\bibinfo  {journal} {Phys. Rev. Lett.}\ }\textbf
  {\bibinfo {volume} {99}},\ \bibinfo {pages} {097202} (\bibinfo {year}
  {2007})}\BibitemShut {NoStop}%
\bibitem [{\citenamefont {Thomale}\ \emph {et~al.}(2009)\citenamefont
  {Thomale}, \citenamefont {Kapit}, \citenamefont {Schroeter},\ and\
  \citenamefont {Greiter}}]{Thomale2009}%
  \BibitemOpen
  \bibfield  {author} {\bibinfo {author} {\bibfnamefont {R.}~\bibnamefont
  {Thomale}}, \bibinfo {author} {\bibfnamefont {E.}~\bibnamefont {Kapit}},
  \bibinfo {author} {\bibfnamefont {D.~F.}\ \bibnamefont {Schroeter}},\ and\
  \bibinfo {author} {\bibfnamefont {M.}~\bibnamefont {Greiter}},\ }\bibfield
  {title} {\bibinfo {title} {Parent hamiltonian for the chiral spin liquid},\
  }\href {https://doi.org/10.1103/PhysRevB.80.104406} {\bibfield  {journal}
  {\bibinfo  {journal} {Phys. Rev. B}\ }\textbf {\bibinfo {volume} {80}},\
  \bibinfo {pages} {104406} (\bibinfo {year} {2009})}\BibitemShut {NoStop}%
\bibitem [{\citenamefont {Iqbal}\ \emph
  {et~al.}(2011{\natexlab{b}})\citenamefont {Iqbal}, \citenamefont {Becca},\
  and\ \citenamefont {Poilblanc}}]{Iqbal2011a}%
  \BibitemOpen
  \bibfield  {author} {\bibinfo {author} {\bibfnamefont {Y.}~\bibnamefont
  {Iqbal}}, \bibinfo {author} {\bibfnamefont {F.}~\bibnamefont {Becca}},\ and\
  \bibinfo {author} {\bibfnamefont {D.}~\bibnamefont {Poilblanc}},\ }\bibfield
  {title} {\bibinfo {title} {{Valence-bond crystal in the extended kagome
  spin-$\frac{1}{2}$ quantum Heisenberg antiferromagnet: A variational Monte
  Carlo approach}},\ }\href {https://doi.org/10.1103/PhysRevB.83.100404}
  {\bibfield  {journal} {\bibinfo  {journal} {Phys. Rev. B}\ }\textbf {\bibinfo
  {volume} {83}},\ \bibinfo {pages} {100404(R)} (\bibinfo {year}
  {2011}{\natexlab{b}})}\BibitemShut {NoStop}%
\bibitem [{\citenamefont {He}\ \emph {et~al.}(2014)\citenamefont {He},
  \citenamefont {Sheng},\ and\ \citenamefont {Chen}}]{He2014}%
  \BibitemOpen
  \bibfield  {author} {\bibinfo {author} {\bibfnamefont {Y.-C.}\ \bibnamefont
  {He}}, \bibinfo {author} {\bibfnamefont {D.~N.}\ \bibnamefont {Sheng}},\ and\
  \bibinfo {author} {\bibfnamefont {Y.}~\bibnamefont {Chen}},\ }\bibfield
  {title} {\bibinfo {title} {{Chiral Spin Liquid in a Frustrated Anisotropic
  Kagome Heisenberg Model}},\ }\href
  {https://doi.org/10.1103/PhysRevLett.112.137202} {\bibfield  {journal}
  {\bibinfo  {journal} {Phys. Rev. Lett.}\ }\textbf {\bibinfo {volume} {112}},\
  \bibinfo {pages} {137202} (\bibinfo {year} {2014})}\BibitemShut {NoStop}%
\bibitem [{\citenamefont {He}\ and\ \citenamefont {Chen}(2015)}]{He2015}%
  \BibitemOpen
  \bibfield  {author} {\bibinfo {author} {\bibfnamefont {Y.-C.}\ \bibnamefont
  {He}}\ and\ \bibinfo {author} {\bibfnamefont {Y.}~\bibnamefont {Chen}},\
  }\bibfield  {title} {\bibinfo {title} {{Distinct Spin Liquids and Their
  Transitions in Spin-$1/2$ $XXZ$ Kagome Antiferromagnets}},\ }\href
  {https://doi.org/10.1103/PhysRevLett.114.037201} {\bibfield  {journal}
  {\bibinfo  {journal} {Phys. Rev. Lett.}\ }\textbf {\bibinfo {volume} {114}},\
  \bibinfo {pages} {037201} (\bibinfo {year} {2015})}\BibitemShut {NoStop}%
\bibitem [{\citenamefont {Wietek}\ and\ \citenamefont
  {L\"auchli}(2020)}]{Wietek2020}%
  \BibitemOpen
  \bibfield  {author} {\bibinfo {author} {\bibfnamefont {A.}~\bibnamefont
  {Wietek}}\ and\ \bibinfo {author} {\bibfnamefont {A.~M.}\ \bibnamefont
  {L\"auchli}},\ }\bibfield  {title} {\bibinfo {title} {Valence bond solid and
  possible deconfined quantum criticality in an extended kagome lattice
  {Heisenberg} antiferromagnet},\ }\href
  {https://doi.org/10.1103/PhysRevB.102.020411} {\bibfield  {journal} {\bibinfo
   {journal} {Phys. Rev. B}\ }\textbf {\bibinfo {volume} {102}},\ \bibinfo
  {pages} {020411(R)} (\bibinfo {year} {2020})}\BibitemShut {NoStop}%
\bibitem [{\citenamefont {Wu}\ \emph {et~al.}(2018)\citenamefont {Wu},
  \citenamefont {Lovorn}, \citenamefont {Tutuc},\ and\ \citenamefont
  {MacDonald}}]{Wu2018}%
  \BibitemOpen
  \bibfield  {author} {\bibinfo {author} {\bibfnamefont {F.}~\bibnamefont
  {Wu}}, \bibinfo {author} {\bibfnamefont {T.}~\bibnamefont {Lovorn}}, \bibinfo
  {author} {\bibfnamefont {E.}~\bibnamefont {Tutuc}},\ and\ \bibinfo {author}
  {\bibfnamefont {A.~H.}\ \bibnamefont {MacDonald}},\ }\bibfield  {title}
  {\bibinfo {title} {{Hubbard Model Physics in Transition Metal Dichalcogenide
  Moir\'e Bands}},\ }\href {https://doi.org/10.1103/PhysRevLett.121.026402}
  {\bibfield  {journal} {\bibinfo  {journal} {Phys. Rev. Lett.}\ }\textbf
  {\bibinfo {volume} {121}},\ \bibinfo {pages} {026402} (\bibinfo {year}
  {2018})}\BibitemShut {NoStop}%
\bibitem [{\citenamefont {Wu}\ \emph {et~al.}(2019)\citenamefont {Wu},
  \citenamefont {Lovorn}, \citenamefont {Tutuc}, \citenamefont {Martin},\ and\
  \citenamefont {MacDonald}}]{Wu2019}%
  \BibitemOpen
  \bibfield  {author} {\bibinfo {author} {\bibfnamefont {F.}~\bibnamefont
  {Wu}}, \bibinfo {author} {\bibfnamefont {T.}~\bibnamefont {Lovorn}}, \bibinfo
  {author} {\bibfnamefont {E.}~\bibnamefont {Tutuc}}, \bibinfo {author}
  {\bibfnamefont {I.}~\bibnamefont {Martin}},\ and\ \bibinfo {author}
  {\bibfnamefont {A.~H.}\ \bibnamefont {MacDonald}},\ }\bibfield  {title}
  {\bibinfo {title} {{Topological Insulators in Twisted Transition Metal
  Dichalcogenide Homobilayers}},\ }\href
  {https://doi.org/10.1103/PhysRevLett.122.086402} {\bibfield  {journal}
  {\bibinfo  {journal} {Phys. Rev. Lett.}\ }\textbf {\bibinfo {volume} {122}},\
  \bibinfo {pages} {086402} (\bibinfo {year} {2019})}\BibitemShut {NoStop}%
\bibitem [{\citenamefont {Xu}\ \emph {et~al.}(2020)\citenamefont {Xu},
  \citenamefont {Liu}, \citenamefont {Rhodes}, \citenamefont {Watanabe},
  \citenamefont {Taniguchi}, \citenamefont {Hone}, \citenamefont {Elser},
  \citenamefont {Mak},\ and\ \citenamefont {Shan}}]{Xu2020}%
  \BibitemOpen
  \bibfield  {author} {\bibinfo {author} {\bibfnamefont {Y.}~\bibnamefont
  {Xu}}, \bibinfo {author} {\bibfnamefont {S.}~\bibnamefont {Liu}}, \bibinfo
  {author} {\bibfnamefont {D.~A.}\ \bibnamefont {Rhodes}}, \bibinfo {author}
  {\bibfnamefont {K.}~\bibnamefont {Watanabe}}, \bibinfo {author}
  {\bibfnamefont {T.}~\bibnamefont {Taniguchi}}, \bibinfo {author}
  {\bibfnamefont {J.}~\bibnamefont {Hone}}, \bibinfo {author} {\bibfnamefont
  {V.}~\bibnamefont {Elser}}, \bibinfo {author} {\bibfnamefont {K.~F.}\
  \bibnamefont {Mak}},\ and\ \bibinfo {author} {\bibfnamefont {J.}~\bibnamefont
  {Shan}},\ }\bibfield  {title} {\bibinfo {title} {{Correlated insulating
  states at fractional fillings of moir\'e superlattices}},\ }\href
  {https://doi.org/10.1038/s41586-020-2868-6} {\bibfield  {journal} {\bibinfo
  {journal} {Nature}\ }\textbf {\bibinfo {volume} {587}},\ \bibinfo {pages}
  {214} (\bibinfo {year} {2020})}\BibitemShut {NoStop}%
\bibitem [{\citenamefont {Huang}\ \emph {et~al.}(2021)\citenamefont {Huang},
  \citenamefont {Wang}, \citenamefont {Miao}, \citenamefont {Wang},
  \citenamefont {Li}, \citenamefont {Lian}, \citenamefont {Taniguchi},
  \citenamefont {Watanabe}, \citenamefont {Okamoto}, \citenamefont {Xiao},
  \citenamefont {Shi},\ and\ \citenamefont {Cui}}]{Huang2021}%
  \BibitemOpen
  \bibfield  {author} {\bibinfo {author} {\bibfnamefont {X.}~\bibnamefont
  {Huang}}, \bibinfo {author} {\bibfnamefont {T.}~\bibnamefont {Wang}},
  \bibinfo {author} {\bibfnamefont {S.}~\bibnamefont {Miao}}, \bibinfo {author}
  {\bibfnamefont {C.}~\bibnamefont {Wang}}, \bibinfo {author} {\bibfnamefont
  {Z.}~\bibnamefont {Li}}, \bibinfo {author} {\bibfnamefont {Z.}~\bibnamefont
  {Lian}}, \bibinfo {author} {\bibfnamefont {T.}~\bibnamefont {Taniguchi}},
  \bibinfo {author} {\bibfnamefont {K.}~\bibnamefont {Watanabe}}, \bibinfo
  {author} {\bibfnamefont {S.}~\bibnamefont {Okamoto}}, \bibinfo {author}
  {\bibfnamefont {D.}~\bibnamefont {Xiao}}, \bibinfo {author} {\bibfnamefont
  {S.-F.}\ \bibnamefont {Shi}},\ and\ \bibinfo {author} {\bibfnamefont {Y.-T.}\
  \bibnamefont {Cui}},\ }\bibfield  {title} {\bibinfo {title} {{Correlated
  insulating states at fractional fillings of the WS$_2$/WSe$_2$ moir\'e
  lattice}},\ }\href {https://doi.org/10.1038/s41567-021-01171-w} {\bibfield
  {journal} {\bibinfo  {journal} {Nature Physics}\ }\textbf {\bibinfo {volume}
  {17}},\ \bibinfo {pages} {715 } (\bibinfo {year} {2021})}\BibitemShut
  {NoStop}%
\bibitem [{\citenamefont {Motruk}\ \emph {et~al.}(2023)\citenamefont {Motruk},
  \citenamefont {Rossi}, \citenamefont {Abanin},\ and\ \citenamefont
  {Rademaker}}]{Motruk2022}%
  \BibitemOpen
  \bibfield  {author} {\bibinfo {author} {\bibfnamefont {J.}~\bibnamefont
  {Motruk}}, \bibinfo {author} {\bibfnamefont {D.}~\bibnamefont {Rossi}},
  \bibinfo {author} {\bibfnamefont {D.~A.}\ \bibnamefont {Abanin}},\ and\
  \bibinfo {author} {\bibfnamefont {L.}~\bibnamefont {Rademaker}},\ }\bibfield
  {title} {\bibinfo {title} {Kagome chiral spin liquid in transition metal
  dichalcogenide moir\'e bilayers},\ }\href
  {https://doi.org/10.1103/PhysRevResearch.5.L022049} {\bibfield  {journal}
  {\bibinfo  {journal} {Phys. Rev. Res.}\ }\textbf {\bibinfo {volume} {5}},\
  \bibinfo {pages} {L022049} (\bibinfo {year} {2023})}\BibitemShut {NoStop}%
\bibitem [{\citenamefont {Pan}\ \emph {et~al.}(2020{\natexlab{a}})\citenamefont
  {Pan}, \citenamefont {Wu},\ and\ \citenamefont {Das~Sarma}}]{Pan2020_1}%
  \BibitemOpen
  \bibfield  {author} {\bibinfo {author} {\bibfnamefont {H.}~\bibnamefont
  {Pan}}, \bibinfo {author} {\bibfnamefont {F.}~\bibnamefont {Wu}},\ and\
  \bibinfo {author} {\bibfnamefont {S.}~\bibnamefont {Das~Sarma}},\ }\bibfield
  {title} {\bibinfo {title} {{Band topology, Hubbard model, Heisenberg model,
  and Dzyaloshinskii-Moriya interaction in twisted bilayer
  ${\mathrm{WSe}}_{2}$}},\ }\href
  {https://doi.org/10.1103/PhysRevResearch.2.033087} {\bibfield  {journal}
  {\bibinfo  {journal} {Phys. Rev. Research}\ }\textbf {\bibinfo {volume}
  {2}},\ \bibinfo {pages} {033087} (\bibinfo {year}
  {2020}{\natexlab{a}})}\BibitemShut {NoStop}%
\bibitem [{\citenamefont {Pan}\ \emph {et~al.}(2020{\natexlab{b}})\citenamefont
  {Pan}, \citenamefont {Wu},\ and\ \citenamefont {Das~Sarma}}]{Pan2020_2}%
  \BibitemOpen
  \bibfield  {author} {\bibinfo {author} {\bibfnamefont {H.}~\bibnamefont
  {Pan}}, \bibinfo {author} {\bibfnamefont {F.}~\bibnamefont {Wu}},\ and\
  \bibinfo {author} {\bibfnamefont {S.}~\bibnamefont {Das~Sarma}},\ }\bibfield
  {title} {\bibinfo {title} {{Quantum phase diagram of a Moir\'e-Hubbard
  model}},\ }\href {https://doi.org/10.1103/PhysRevB.102.201104} {\bibfield
  {journal} {\bibinfo  {journal} {Phys. Rev. B}\ }\textbf {\bibinfo {volume}
  {102}},\ \bibinfo {pages} {201104(R)} (\bibinfo {year}
  {2020}{\natexlab{b}})}\BibitemShut {NoStop}%
\bibitem [{\citenamefont {Zang}\ \emph {et~al.}(2021)\citenamefont {Zang},
  \citenamefont {Wang}, \citenamefont {Cano},\ and\ \citenamefont
  {Millis}}]{Zang2021}%
  \BibitemOpen
  \bibfield  {author} {\bibinfo {author} {\bibfnamefont {J.}~\bibnamefont
  {Zang}}, \bibinfo {author} {\bibfnamefont {J.}~\bibnamefont {Wang}}, \bibinfo
  {author} {\bibfnamefont {J.}~\bibnamefont {Cano}},\ and\ \bibinfo {author}
  {\bibfnamefont {A.~J.}\ \bibnamefont {Millis}},\ }\bibfield  {title}
  {\bibinfo {title} {{Hartree-Fock study of the moir\'e Hubbard model for
  twisted bilayer transition metal dichalcogenides}},\ }\href
  {https://doi.org/10.1103/PhysRevB.104.075150} {\bibfield  {journal} {\bibinfo
   {journal} {Phys. Rev. B}\ }\textbf {\bibinfo {volume} {104}},\ \bibinfo
  {pages} {075150} (\bibinfo {year} {2021})}\BibitemShut {NoStop}%
\bibitem [{\citenamefont {Wietek}\ \emph {et~al.}(2022)\citenamefont {Wietek},
  \citenamefont {Wang}, \citenamefont {Zang}, \citenamefont {Cano},
  \citenamefont {Georges},\ and\ \citenamefont {Millis}}]{Wietek2022}%
  \BibitemOpen
  \bibfield  {author} {\bibinfo {author} {\bibfnamefont {A.}~\bibnamefont
  {Wietek}}, \bibinfo {author} {\bibfnamefont {J.}~\bibnamefont {Wang}},
  \bibinfo {author} {\bibfnamefont {J.}~\bibnamefont {Zang}}, \bibinfo {author}
  {\bibfnamefont {J.}~\bibnamefont {Cano}}, \bibinfo {author} {\bibfnamefont
  {A.}~\bibnamefont {Georges}},\ and\ \bibinfo {author} {\bibfnamefont
  {A.}~\bibnamefont {Millis}},\ }\bibfield  {title} {\bibinfo {title} {{Tunable
  stripe order and weak superconductivity in the Moir\'e Hubbard model}},\
  }\href {https://doi.org/10.1103/PhysRevResearch.4.043048} {\bibfield
  {journal} {\bibinfo  {journal} {Phys. Rev. Research}\ }\textbf {\bibinfo
  {volume} {4}},\ \bibinfo {pages} {043048} (\bibinfo {year}
  {2022})}\BibitemShut {NoStop}%
\bibitem [{\citenamefont {Kiese}\ \emph {et~al.}(2022)\citenamefont {Kiese},
  \citenamefont {He}, \citenamefont {Hickey}, \citenamefont {Rubio},\ and\
  \citenamefont {Kennes}}]{Kiese2022}%
  \BibitemOpen
  \bibfield  {author} {\bibinfo {author} {\bibfnamefont {D.}~\bibnamefont
  {Kiese}}, \bibinfo {author} {\bibfnamefont {Y.}~\bibnamefont {He}}, \bibinfo
  {author} {\bibfnamefont {C.}~\bibnamefont {Hickey}}, \bibinfo {author}
  {\bibfnamefont {A.}~\bibnamefont {Rubio}},\ and\ \bibinfo {author}
  {\bibfnamefont {D.~M.}\ \bibnamefont {Kennes}},\ }\bibfield  {title}
  {\bibinfo {title} {{TMDs as a platform for spin liquid physics: A strong
  coupling study of twisted bilayer WSe$_2$}},\ }\href
  {https://doi.org/10.1063/5.0077901} {\bibfield  {journal} {\bibinfo
  {journal} {APL Materials}\ }\textbf {\bibinfo {volume} {10}},\ \bibinfo
  {pages} {031113} (\bibinfo {year} {2022})}\BibitemShut {NoStop}%
\bibitem [{\citenamefont {Rademaker}(2022)}]{Rademaker2022}%
  \BibitemOpen
  \bibfield  {author} {\bibinfo {author} {\bibfnamefont {L.}~\bibnamefont
  {Rademaker}},\ }\bibfield  {title} {\bibinfo {title} {Spin-orbit coupling in
  transition metal dichalcogenide heterobilayer flat bands},\ }\href
  {https://doi.org/10.1103/PhysRevB.105.195428} {\bibfield  {journal} {\bibinfo
   {journal} {Phys. Rev. B}\ }\textbf {\bibinfo {volume} {105}},\ \bibinfo
  {pages} {195428} (\bibinfo {year} {2022})}\BibitemShut {NoStop}%
\bibitem [{Note1()}]{Note1}%
  \BibitemOpen
  \bibinfo {note} {There is a DM vector for each bond, the blue dot in \protect
  \cref {fig:DM_directions} indicates the direction for all the bonds
  neighboring the triangle.}\BibitemShut {Stop}%
\bibitem [{\citenamefont {Huh}\ \emph {et~al.}(2010)\citenamefont {Huh},
  \citenamefont {Fritz},\ and\ \citenamefont {Sachdev}}]{Huh2010}%
  \BibitemOpen
  \bibfield  {author} {\bibinfo {author} {\bibfnamefont {Y.}~\bibnamefont
  {Huh}}, \bibinfo {author} {\bibfnamefont {L.}~\bibnamefont {Fritz}},\ and\
  \bibinfo {author} {\bibfnamefont {S.}~\bibnamefont {Sachdev}},\ }\bibfield
  {title} {\bibinfo {title} {{Quantum criticality of the kagome antiferromagnet
  with Dzyaloshinskii-Moriya interactions}},\ }\href
  {https://doi.org/10.1103/PhysRevB.81.144432} {\bibfield  {journal} {\bibinfo
  {journal} {Phys. Rev. B}\ }\textbf {\bibinfo {volume} {81}},\ \bibinfo
  {pages} {144432} (\bibinfo {year} {2010})}\BibitemShut {NoStop}%
\bibitem [{\citenamefont {Auerbach}(1998)}]{Auerbach1998}%
  \BibitemOpen
  \bibfield  {author} {\bibinfo {author} {\bibfnamefont {A.}~\bibnamefont
  {Auerbach}},\ }\href@noop {} {\emph {\bibinfo {title} {Interacting electrons
  and quantum magnetism}}}\ (\bibinfo  {publisher} {Springer, New York},\
  \bibinfo {year} {1998})\BibitemShut {NoStop}%
\bibitem [{\citenamefont {Wen}(2002)}]{Wen2002}%
  \BibitemOpen
  \bibfield  {author} {\bibinfo {author} {\bibfnamefont {X.-G.}\ \bibnamefont
  {Wen}},\ }\bibfield  {title} {\bibinfo {title} {Quantum orders and symmetric
  spin liquids},\ }\href {https://doi.org/10.1103/PhysRevB.65.165113}
  {\bibfield  {journal} {\bibinfo  {journal} {Phys. Rev. B}\ }\textbf {\bibinfo
  {volume} {65}},\ \bibinfo {pages} {165113} (\bibinfo {year}
  {2002})}\BibitemShut {NoStop}%
\bibitem [{\citenamefont {Auerbach}\ and\ \citenamefont
  {Arovas}(1988)}]{Auerbach1988a}%
  \BibitemOpen
  \bibfield  {author} {\bibinfo {author} {\bibfnamefont {A.}~\bibnamefont
  {Auerbach}}\ and\ \bibinfo {author} {\bibfnamefont {D.~P.}\ \bibnamefont
  {Arovas}},\ }\bibfield  {title} {\bibinfo {title} {{Spin Dynamics in the
  Square-Lattice Antiferromagnet}},\ }\href
  {https://doi.org/10.1103/PhysRevLett.61.617} {\bibfield  {journal} {\bibinfo
  {journal} {Phys. Rev. Lett.}\ }\textbf {\bibinfo {volume} {61}},\ \bibinfo
  {pages} {617} (\bibinfo {year} {1988})}\BibitemShut {NoStop}%
\bibitem [{\citenamefont {Messio}\ \emph {et~al.}(2013)\citenamefont {Messio},
  \citenamefont {Lhuillier},\ and\ \citenamefont {Misguich}}]{Messio2013}%
  \BibitemOpen
  \bibfield  {author} {\bibinfo {author} {\bibfnamefont {L.}~\bibnamefont
  {Messio}}, \bibinfo {author} {\bibfnamefont {C.}~\bibnamefont {Lhuillier}},\
  and\ \bibinfo {author} {\bibfnamefont {G.}~\bibnamefont {Misguich}},\
  }\bibfield  {title} {\bibinfo {title} {Time reversal symmetry breaking chiral
  spin liquids: Projective symmetry group approach of bosonic mean-field
  theories},\ }\href {https://doi.org/10.1103/PhysRevB.87.125127} {\bibfield
  {journal} {\bibinfo  {journal} {Phys. Rev. B}\ }\textbf {\bibinfo {volume}
  {87}},\ \bibinfo {pages} {125127} (\bibinfo {year} {2013})}\BibitemShut
  {NoStop}%
\bibitem [{\citenamefont {Colpa}(1978)}]{Colpa1978}%
  \BibitemOpen
  \bibfield  {author} {\bibinfo {author} {\bibfnamefont {J.}~\bibnamefont
  {Colpa}},\ }\bibfield  {title} {\bibinfo {title} {Diagonalization of the
  quadratic boson hamiltonian},\ }\href
  {https://doi.org/https://doi.org/10.1016/0378-4371(78)90160-7} {\bibfield
  {journal} {\bibinfo  {journal} {Physica A: Statistical Mechanics and its
  Applications}\ }\textbf {\bibinfo {volume} {93}},\ \bibinfo {pages} {327}
  (\bibinfo {year} {1978})}\BibitemShut {NoStop}%
\bibitem [{Note2()}]{Note2}%
  \BibitemOpen
  \bibinfo {note} {This is due to the fact that there is no analytic expression
  for the eigenvalues of a $6\times 6$ matrix.}\BibitemShut {Stop}%
\bibitem [{Note3()}]{Note3}%
  \BibitemOpen
  \bibinfo {note} {Since the mean field parameters are in general complex
  valued, we also need to determine their phase. It turns out that in most
  cases they respect the same Hessian sign as their corresponding
  amplitude.}\BibitemShut {Stop}%
\bibitem [{\citenamefont {Misguich}(2012)}]{Misguich2012}%
  \BibitemOpen
  \bibfield  {author} {\bibinfo {author} {\bibfnamefont {G.}~\bibnamefont
  {Misguich}},\ }\bibfield  {title} {\bibinfo {title} {Schwinger boson
  mean-field theory: Numerics for the energy landscape and gauge excitations in
  two-dimensional antiferromagnets},\ }\href
  {https://doi.org/10.1103/PhysRevB.86.245132} {\bibfield  {journal} {\bibinfo
  {journal} {Phys. Rev. B}\ }\textbf {\bibinfo {volume} {86}},\ \bibinfo
  {pages} {245132} (\bibinfo {year} {2012})}\BibitemShut {NoStop}%
\bibitem [{\citenamefont {Wang}\ and\ \citenamefont
  {Vishwanath}(2006)}]{Wang2006}%
  \BibitemOpen
  \bibfield  {author} {\bibinfo {author} {\bibfnamefont {F.}~\bibnamefont
  {Wang}}\ and\ \bibinfo {author} {\bibfnamefont {A.}~\bibnamefont
  {Vishwanath}},\ }\bibfield  {title} {\bibinfo {title} {{Spin-liquid states on
  the triangular and Kagom\'e lattices: A projective-symmetry-group analysis of
  Schwinger boson states}},\ }\href
  {https://doi.org/10.1103/PhysRevB.74.174423} {\bibfield  {journal} {\bibinfo
  {journal} {Phys. Rev. B}\ }\textbf {\bibinfo {volume} {74}},\ \bibinfo
  {pages} {174423} (\bibinfo {year} {2006})}\BibitemShut {NoStop}%
\bibitem [{\citenamefont {Halimeh}\ and\ \citenamefont
  {Punk}(2016)}]{Halimeh2016}%
  \BibitemOpen
  \bibfield  {author} {\bibinfo {author} {\bibfnamefont {J.~C.}\ \bibnamefont
  {Halimeh}}\ and\ \bibinfo {author} {\bibfnamefont {M.}~\bibnamefont {Punk}},\
  }\bibfield  {title} {\bibinfo {title} {Spin structure factors of chiral
  quantum spin liquids on the kagome lattice},\ }\href
  {https://doi.org/10.1103/PhysRevB.94.104413} {\bibfield  {journal} {\bibinfo
  {journal} {Phys. Rev. B}\ }\textbf {\bibinfo {volume} {94}},\ \bibinfo
  {pages} {104413} (\bibinfo {year} {2016})}\BibitemShut {NoStop}%
\bibitem [{Note4()}]{Note4}%
  \BibitemOpen
  \bibinfo {note} {Note the ambiguity in the definition of staggered and
  uniform DM interactions \cite {GomezAlbarracin2018}. In this work we stick to
  the notation of \cite {Messio2010}.}\BibitemShut {Stop}%
\bibitem [{\citenamefont {Messio}\ \emph {et~al.}(2011)\citenamefont {Messio},
  \citenamefont {Lhuillier},\ and\ \citenamefont {Misguich}}]{Messio2011}%
  \BibitemOpen
  \bibfield  {author} {\bibinfo {author} {\bibfnamefont {L.}~\bibnamefont
  {Messio}}, \bibinfo {author} {\bibfnamefont {C.}~\bibnamefont {Lhuillier}},\
  and\ \bibinfo {author} {\bibfnamefont {G.}~\bibnamefont {Misguich}},\
  }\bibfield  {title} {\bibinfo {title} {Lattice symmetries and regular
  magnetic orders in classical frustrated antiferromagnets},\ }\href
  {https://doi.org/10.1103/PhysRevB.83.184401} {\bibfield  {journal} {\bibinfo
  {journal} {Phys. Rev. B}\ }\textbf {\bibinfo {volume} {83}},\ \bibinfo
  {pages} {184401} (\bibinfo {year} {2011})}\BibitemShut {NoStop}%
\bibitem [{Note5()}]{Note5}%
  \BibitemOpen
  \bibinfo {note} {The general unit cell has $6$ sites because of the parameter
  $p_1$ found in the PSG classification. By considering only a $3$-site unit
  cell, orders like \protect \textit {cuboc-1}, \protect \textit {cuboc-2} and
  \protect \textit {octahedral} are excluded.}\BibitemShut {Stop}%
\bibitem [{\citenamefont {Mondal}\ and\ \citenamefont
  {Kadolkar}(2021)}]{Mondal2021}%
  \BibitemOpen
  \bibfield  {author} {\bibinfo {author} {\bibfnamefont {K.}~\bibnamefont
  {Mondal}}\ and\ \bibinfo {author} {\bibfnamefont {C.}~\bibnamefont
  {Kadolkar}},\ }\bibfield  {title} {\bibinfo {title} {Regular magnetic orders
  in triangular and kagome lattices},\ }\href
  {https://doi.org/10.1088/1361-648X/ac27d9} {\bibfield  {journal} {\bibinfo
  {journal} {Journal of Physics: Condensed Matter}\ }\textbf {\bibinfo {volume}
  {33}},\ \bibinfo {pages} {505801} (\bibinfo {year} {2021})}\BibitemShut
  {NoStop}%
\bibitem [{\citenamefont {Claassen}\ \emph {et~al.}(2022)\citenamefont
  {Claassen}, \citenamefont {Xian}, \citenamefont {Kennes},\ and\ \citenamefont
  {Rubio}}]{Claassen2022}%
  \BibitemOpen
  \bibfield  {author} {\bibinfo {author} {\bibfnamefont {M.}~\bibnamefont
  {Claassen}}, \bibinfo {author} {\bibfnamefont {L.}~\bibnamefont {Xian}},
  \bibinfo {author} {\bibfnamefont {D.~M.}\ \bibnamefont {Kennes}},\ and\
  \bibinfo {author} {\bibfnamefont {A.}~\bibnamefont {Rubio}},\ }\bibfield
  {title} {\bibinfo {title} {Ultra-strong spin--orbit coupling and topological
  moir{\'e} engineering in twisted {ZrS$_2$} bilayers},\ }\href
  {https://doi.org/10.1038/s41467-022-31604-w} {\bibfield  {journal} {\bibinfo
  {journal} {Nature Communications}\ }\textbf {\bibinfo {volume} {13}},\
  \bibinfo {pages} {4915} (\bibinfo {year} {2022})}\BibitemShut {NoStop}%
\bibitem [{\citenamefont {Reddy}\ \emph {et~al.}(2023)\citenamefont {Reddy},
  \citenamefont {Devakul},\ and\ \citenamefont {Fu}}]{Reddy2023}%
  \BibitemOpen
  \bibfield  {author} {\bibinfo {author} {\bibfnamefont {A.~P.}\ \bibnamefont
  {Reddy}}, \bibinfo {author} {\bibfnamefont {T.}~\bibnamefont {Devakul}},\
  and\ \bibinfo {author} {\bibfnamefont {L.}~\bibnamefont {Fu}},\ }\href@noop
  {} {\bibinfo {title} {Moir\'e alchemy: artificial atoms, {Wigner} molecules,
  and emergent {Kagome} lattice}} (\bibinfo {year} {2023}),\ \Eprint
  {https://arxiv.org/abs/2301.00799} {arXiv:2301.00799} \BibitemShut {NoStop}%
\bibitem [{Note6()}]{Note6}%
  \BibitemOpen
  \bibinfo {note} {The grids consider both initial and final point of the BZ,
  so in the end $N_k$ is increased by $1$.}\BibitemShut {Stop}%
\bibitem [{\citenamefont {Gomez~Albarracin}\ and\ \citenamefont
  {Pujol}(2018)}]{GomezAlbarracin2018}%
  \BibitemOpen
  \bibfield  {author} {\bibinfo {author} {\bibfnamefont {F.~A.}\ \bibnamefont
  {Gomez~Albarracin}}\ and\ \bibinfo {author} {\bibfnamefont {P.}~\bibnamefont
  {Pujol}},\ }\bibfield  {title} {\bibinfo {title} {{Degenerate and chiral
  states in the extended Heisenberg model on the kagome lattice}},\ }\href
  {https://doi.org/10.1103/PhysRevB.97.104419} {\bibfield  {journal} {\bibinfo
  {journal} {Phys. Rev. B}\ }\textbf {\bibinfo {volume} {97}},\ \bibinfo
  {pages} {104419} (\bibinfo {year} {2018})}\BibitemShut {NoStop}%
\end{thebibliography}
\end{document}